\documentclass[11pt]{article}

\usepackage{amsfonts,amsmath,amssymb,chet,dsfont,enumitem,graphicx,mathrsfs,pgfplots,simpler-wick,tikz}
\usetikzlibrary{calc}
\pgfplotsset{compat=1.14}
\allowdisplaybreaks

\newcommand{\1}{\mathds{1}}
\newcommand{\bs}[1]{{\boldsymbol{#1}}}
\newcommand{\h}[1]{{\hat{#1}}}
\newcommand{\m}[1]{{\mathcal{#1}}}
\newcommand{\ee}[2]{\eta_{#1}\cdot\eta_{#2}}

\newcommand{\Op}[3]{\m{O}_{#1}^{#2}(\eta_{#3})}
\newcommand{\cOPE}[3]{{}_{#1}c_{#2}^{\phantom{#2}#3}}
\newcommand{\tOPE}[4]{{}_{#1}t_{#2}^{#4#3}}
\newcommand{\Vev}[1]{\left\langle{#1}\right\rangle}
\newcommand{\cCF}[2]{{}_{#1}c_{#2}}
\newcommand{\tCF}[3]{{}_{#1}t_{#2}^{#3}}

\DeclareMathOperator*{\Sym}{Sym}
\DeclareMathOperator*{\Asym}{Asym}

\title{\vspace{-3cm}Conformal Bootstrap Equations from the Embedding Space Operator Product Expansion}

\author{Jean-Fran\c{c}ois Fortin$^{\ast,}$\email{jean-francois.fortin@phy.ulaval.ca}, Wen-Jie Ma$^{\dagger,\$}$\email{wenjia.ma@bimsa.cn}, 
Valentina Prilepina$^{\S,}$\email{valentina.prilepina.1@ulaval.ca}, and Witold Skiba$^{\ddag,}$\email{witold.skiba@yale.edu}}

\affiliation{
$^\ast$D\'epartement de Physique, de G\'enie Physique et d'Optique,\\Universit\'e Laval, Qu\'ebec, QC G1V 0A6, Canada\\
$^\dagger$Beijing Institute of Mathematical Sciences and Applications (BIMSA), Beijing, 101408, China\\
$^\$$Yau Mathematical Sciences Center (YMSC), Tsinghua University, Beijing, 100084, China\\
$^\S$Perimeter Institute for Theoretical Physics, Waterloo, ON N2L 2Y5, Canada\\
$^\ddag$Department of Physics, Yale University, New Haven, CT 06520, USA
}%Choices for affiliations $^{\ast,\dagger,\$,\S,\ddag,}$

\abstract{We describe how to implement the conformal bootstrap program in the context of the embedding space OPE formalism introduced in previous work.  To take maximal advantage of the known properties of the scalar conformal blocks for symmetric-traceless exchange, we construct tensorial generalizations of the three-point and four-point scalar conformal blocks that have many nice properties.  Further, we present a special basis of tensor structures for three-point correlation functions endowed with the remarkable simplifying property that it does not mix under permutations of the external quasi-primary operators.  We find that in this approach, we can write the $M$-point conformal bootstrap equations explicitly in terms of the standard position space cross-ratios without the need to project back to position space, thus effectively deriving all conformal bootstrap equations directly from the embedding space.  Finally, we lay out an algorithm for generating the conformal bootstrap equations in this formalism.  Collectively, the tensorial generalizations, the new basis of tensor structures, as well as the procedure for deriving the conformal bootstrap equations lead to four-point bootstrap equations for quasi-primary operators in arbitrary Lorentz representations expressed as linear combinations of the standard scalar conformal blocks for spin-$\ell$ exchange, with finite $\ell$-independent terms.  Moreover, the OPE coefficients in these equations conveniently feature trivial symmetry properties.  The only inputs necessary are the relevant projection operators and tensor structures, which are all fixed by group theory.  To illustrate the procedure, we present one nontrivial example involving scalars $S$ and vectors $V$, namely $\Vev{SSSV}$.}

\date{March 2023} %Uncomment this line for month to be fixed

\begin{document}

\maketitle

\toc

%%%%%%%%%%%%%%%%%%%%%%%%%%%%%%%%%%%%%%%%%%%%%%%%%%
%%%%%%%%%%%%%%%%%%%%%%%%%%%%%%%%%%%%%%%%%%%%%%%%%%

\section{Introduction}\label{SecIntro}

The usual Poincar\'e symmetry group $SO(1,d-1)$ of quantum field theory (QFT) is extended to the broader conformal symmetry group $SO(2,d)$ at fixed points of the renormalization group flow.  Such fixed points are hence described by conformal field theory (CFT).  In contrast to QFT, the enhanced symmetry group of CFT lends a non-perturbative handle to confront computations of correlation functions.  CFTs thus bring with them the hope of effectively solving some higher-dimensional QFTs without needing to resort to supersymmetry.  Aside from furthering this promising endeavor, these remarkable theories are also important on their own.  In Minkowski signature, CFTs describe the universal behavior of scale invariant QFTs with infinite correlation length and, as such, they probe the space of QFTs.  In Euclidean signature, CFTs represent second order phase transitions of systems in statistical physics.  From their extensive implications for QFT and condensed matter systems to their appearance in quantum gravity through the AdS/CFT correspondence and holography, which authorizes the use of CFT methods for the study of black hole systems, it is evident that a deep understanding of CFTs is of paramount importance.

The seminal work of \cite{Rattazzi:2008pe} led to a renewed interest in non-perturbative CFT methods, which culminated in impressive results on the numerical values of some CFT data.  The technique relies on the original strategy of the conformal bootstrap approach \cite{Ferrara:1973yt,Polyakov:1974gs}, which harnesses the full power of symmetries and consistency conditions of the correlation functions to explore the parameter space of CFTs non-perturbatively.  It would be too lengthy to provide a comprehensive list of references on this vast subject, but the conformal bootstrap literature has been summarized in several reviews and lecture notes where they can be found (for a small set of reviews and lecture notes, see \cite{Rychkov:2016iqz,Simmons-Duffin:2016gjk,Poland:2018epd,Chester:2019wfx} and references therein).

Several approaches aiming to implement the bootstrap program, either numerically or analytically, have been proposed in the literature.  All methods necessitate the knowledge of the four-point conformal blocks, which are at the heart of four-point conformal correlation functions, and thus the conformal bootstrap.  Conformal blocks encode the different contributions to specific four-point correlation functions in terms of the allowed exchanged representations as prescribed by the operator product expansion (OPE).  Early works on conformal blocks include \cite{Ferrara:1973vz,Ferrara:1974nf,Dobrev:1977qv,Exton_1995,Dolan:2000ut,Dolan:2003hv}, and novel methods or improvements to established ones can be found in \textit{e.g.}\ \cite{Costa:2011mg,Dolan:2011dv,Costa:2011dw,SimmonsDuffin:2012uy,Costa:2014rya,Echeverri:2015rwa,Iliesiu:2015qra,Hijano:2015zsa,Rejon-Barrera:2015bpa,Penedones:2015aga,Iliesiu:2015akf,Alkalaev:2015fbw,Echeverri:2016dun,Isachenkov:2016gim,Fortin:2016lmf,Costa:2016hju,Costa:2016xah,Chen:2016bxc,Nishida:2016vds,Cordova:2016emh,Schomerus:2016epl,Fortin:2016dlj,Kravchuk:2016qvl,Gliozzi:2017hni,Castro:2017hpx,Dyer:2017zef,Sleight:2017fpc,Chen:2017yia,Pasterski:2017kqt,Cardoso:2017qmj,Karateev:2017jgd,Kravchuk:2017dzd,Schomerus:2017eny,Isachenkov:2017qgn,Faller:2017hyt,Chen:2017xdz,Sleight:2018epi,Costa:2018mcg,Kobayashi:2018okw,Bhatta:2018gjb,Lauria:2018klo,Liu:2018jhs,Gromov:2018hut,Rosenhaus:2018zqn,Zhou:2018sfz,Kazakov:2018gcy,Karateev:2019pvw,Comeau:2019xco,Fortin:2019fvx,Fortin:2019dnq,Li:2019dix,Goncalves:2019znr,Jepsen:2019svc,Fortin:2019xyr,Fortin:2019pep,Fortin:2019gck,Karlsson:2019dbd,Buric:2019dfk,Paulos:2019gtx,Reehorst:2019pzi,Fortin:2020ncr,Reehorst:2021ykw,Ghosh:2021ruh,Liendo:2021egi,Bissi:2022mrs,Laio:2022ayq,Buric:2022ucg,Li:2023whn,He:2023ewx}.

While the action of the conformal group is nontrivial in $d$-dimensional position space, it is linear in ($d+2$)-dimensional embedding space \cite{Dirac:1936fq,Mack:1969rr,Weinberg:2010fx,Weinberg:2012mz}, which effectively allows us to treat the conformal group as the Lorentz group.  It is therefore natural to work in the embedding space; in particular, it is beneficial to construct the OPE that is at the basis of all conformal correlation functions directly in the embedding space \cite{Ferrara:1971vh,Ferrara:1971zy,Ferrara:1972cq,Ferrara:1973eg,Dobrev:1975ru,Mack:1976pa}.  Recent works on a modified embedding---where all irreducible representations of the Lorentz group are treated on the same footing \cite{Fortin:2016lmf,Fortin:2016dlj,Comeau:2019xco,Fortin:2019fvx,Fortin:2019dnq} and where a smooth link to position space exists \cite{Fortin:2020des}---led to relatively straightforward algorithms for computing four-point conformal blocks in arbitrary correlation functions modulo inputs from group theory, namely the projection operators of the exchanged quasi-primary operators and the tensor structures \cite{Fortin:2019xyr,Fortin:2019pep,Fortin:2019gck,Fortin:2020ncr}.  Up to this point the method yields explicit results for four-point conformal blocks, but in terms of spin-$\ell$-dependent finite sums of scalar conformal blocks corresponding to scalar exchange, which is a significant drawback.  Our goal in this paper is to resolve this vital issue by re-expressing the resulting conformal blocks in terms of scalar conformal blocks corresponding to symmetric-traceless (spin-$\ell$), \textit{without} projecting back to position space.  We stress that contrary to the customary embedding space formalism where projection to position space results in a loss of (non-physical) information owing to the existence of a ``gauge invariance'', such a concern is not present in the modified uplift, where there is only one embedding space object per position space analog (informally, our method supplies a ``natural gauge fixing'').  The method hinges on the repeated application of contiguous relations that transform tensorial generalizations of the three- and four-point conformal blocks into standard blocks for spin-$\ell$ exchange, modulo functions of the standard conformal cross-ratios $u$ and $v$.  Here we redefine the tensor structure basis introduced in \cite{Fortin:2020des} to obtain a special basis that does not mix under permutations of the three quasi-primary operators appearing in the three-point correlation function of interest, effectively simplifying conformal bootstrap equations involving the same quasi-primary operators.  As it is sufficiently general, the method presented here can be extended to higher-point correlation functions with minor modifications.

The overall strategy of the method consists of acting with the embedding space OPE on two-point correlation functions once (twice) to generate three- (four)-point correlation functions with minimal input from group theory.  Computing the three-point correlation functions from the OPE determines the initial tensor structure basis which can then be recast in terms of the special basis mentioned above via multiplication by appropriate rotation matrices.  The computation of the four-point correlation functions can subsequently be re-expressed in this convenient basis and ultimately exploited directly in the conformal bootstrap equations through the repeated use of the contiguous relations.  The method thus relies on the knowledge of the projection operators (all approaches to the bootstrap need that information in one form or another), and the (somewhat complicated) contiguous relations.

After presenting a summary of the embedding space with our new uplift in Section \ref{SecReview}, with small notational improvements to build an even more univocal relationship with position space, this paper delves into the computations of two-, three-, and four-point correlation functions in the context of the embedding space formalism in Section \ref{SecCF}.  In that section, the three- and four-point correlation functions, which are obtained with the help of the appropriate projection operators, are explicitly written in terms of the so-called ``tensorial generalizations'' of the three- and four-point conformal blocks.  These objects carry extra indices that must later be contracted, \textit{e.g.}\ with the tensor structures.  Section \ref{SecTensorGeneralizations} studies their behaviors under contractions with the embedding space metrics and coordinates, which results in a set of contiguous relations.  Other techniques are also presented to relate the tensorial generalizations to the standard conformal blocks.  Next, Section \ref{SecBootstrap} harnesses the power of the contiguous relations to obtain fully-scalar conformal bootstrap equations directly from embedding space.  Indeed, by fully contracting the free embedding space indices appearing in the bootstrap equations with the available embedding space coordinates (packaged into natural objects determined by group theory), through the repeated use of the contiguous relations, the initial tensorial bootstrap equations are ultimately transformed into independent scalar bootstrap equations.  These equations feature linear combinations of scalar conformal blocks for spin-$\ell$ exchange with coefficients that are functions of the conformal cross-ratios, thus bypassing the need to project back to position space.  As a byproduct, it is shown in the case of three-point correlation functions that there exists a very convenient basis of tensor structures that does not mix under permutations of the external quasi-primary operators.  Section \ref{SecAlg} presents the full algorithm for generating the conformal blocks, rotation matrices, and conformal bootstrap equations in this formalism, along with a simple yet non-trivial example detailing the steps needed to arrive at the fully-scalar conformal bootstrap equations.  The example is not new (in particular, it involves three scalars and one vector operator), but it illustrates the technique while avoiding the pitfall of very lengthy equations.  Section \ref{SecConc} concludes with a summary of the method, its advantages and shortcomings, and a discussion of future work.  Finally, several appendices display the proofs of the contiguous relations (Appendix \ref{SAppProofs}), provide a technical result to simplify the computation of rotation matrix elements (Appendix \ref{SAppRM}), furnish the projection operators relevant for the examples discussed in the paper (Appendix \ref{SAppProj}), and supply the input data required for a more complicated example (Appendix \ref{SAppEx}).

%%%%%%%%%%%%%%%%%%%%%%%%%%%%%%%%%%%%%%%%%%%%%%%%%%
%%%%%%%%%%%%%%%%%%%%%%%%%%%%%%%%%%%%%%%%%%%%%%%%%%

\section{Embedding Space Formalism}\label{SecReview}

In this section we review the embedding space OPE formalism \cite{Fortin:2019fvx,Fortin:2019dnq}.  In the interest of making the already natural relationship between embedding space quantities and their position space counterparts even more transparent, we introduce small notational changes with respect to the original notation established in \cite{Fortin:2019fvx,Fortin:2019dnq}.

%%%%%%%%%%%%%%%%%%%%%%%%%%%%%%%%%%%%%%%%%%%%%%%%%%

\subsection{Irreducible Representations in Position Space}

First, we discuss the irreducible representations of $SO(1,d-1)$ (with rank $r=\lfloor d/2\rfloor$) relevant for position space quasi-primary operators $\m{O}^{(x)}$.  Let us restrict attention to odd spacetime dimensions for simplicity.\footnote{As shown in \cite{Fortin:2019dnq}, the formalism works for any spacetime dimension.  In even spacetime dimensions, there are slight complications due to the existence of two inequivalent irreducible spinor representations.}  In the language of this formalism, quasi-primary operators are expressed purely in terms of embedding space spinor indices.  Their behavior under Lorentz transformations is encoded via the position space half-projection operators $\m{T}$ such that $\m{O}^{(x)}\sim\m{T}$.  Here $\sim$ indicates that the spinor indices of both quantities transform similarly under the Lorentz group.

For arbitrary irreducible representations $\bs{N}=\sum_{i=1}^rN_i\bs{e}_i$, where $N_i$  and $(\bs{e}_i)_j=\delta_{ij}$ denote the Dynkin indices and the unit vectors, respectively, the position space half-projectors are given by
\eqna{
(\m{T}^{\bs{N}})_{\alpha_1\cdots\alpha_n}^{\mu_1\cdots\mu_{n_v}\delta}&=\left((\m{T}^{\bs{e}_1})^{N_1}\cdots(\m{T}^{\bs{e}_{r-1}})^{N_{r-1}}(\m{T}^{2\bs{e}_r})^{\lfloor N_r/2\rfloor}(\m{T}^{\bs{e}_r})^{N_r-2\lfloor N_r/2\rfloor}\right)_{\alpha_1\cdots\alpha_n}^{\mu'_1\cdots\mu'_{n_v}\delta'}\\
&\phantom{=}\qquad\times(\h{\m{P}}^{\bs{N}})_{\delta'\mu'_{n_v}\cdots\mu'_1}^{\phantom{\delta'\mu'_{n_v}\cdots\mu'_1}\mu_1\cdots\mu_{n_v}\delta}\\
&=(\m{P}_{\bs{N}})_{\alpha_1\cdots\alpha_n}^{\phantom{\alpha_1\cdots\alpha_n}\alpha'_n\cdots\alpha'_1}\left((\m{T}^{\bs{e}_1})^{N_1}\cdots(\m{T}^{\bs{e}_{r-1}})^{N_{r-1}}(\m{T}^{2\bs{e}_r})^{\lfloor N_r/2\rfloor}(\m{T}^{\bs{e}_r})^{N_r-2\lfloor N_r/2\rfloor}\right)_{\alpha'_1\cdots\alpha'_n}^{\mu_1\cdots\mu_{n_v}\delta},
}[EqTPS]
where $n=2\sum_{i=1}^{r-1}N_i+N_r=2S$ is twice the spin $S$ of the irreducible representation $\bs{N}$.

The half-projectors serve to translate the spinor indices carried by each operator to dummy vector and spinor indices that must then be properly contracted.  These objects carry two types of position space indices, namely the spinor and vector indices.  In particular, the lower set denotes the position space spinor indices $\alpha_1,\ldots,\alpha_n$ in \eqref{EqTPS}, which are the free spinor indices of the half-projectors that match the free spinor indices of the quasi-primary operator $\m{O}^{(x)}$.  Meanwhile, the upper set is comprised of the $n_v=\sum_{i=1}^{r-1}iN_i+r\lfloor N_r/2\rfloor$ position space vector indices $\mu_1,\ldots,\mu_{n_v}$ and the position space spinor index $\delta$ (appearing only when $N_r$ is odd), which are dummy indices that must be contracted appropriately.

For the defining irreducible representations, the explicit expressions for the half-projectors, which also appear in \eqref{EqTPS}, are
\eqn{
\begin{gathered}
(\m{T}^{\bs{e}_{i\neq r}})_{\alpha\beta}^{\mu_1\cdots\mu_i}=\frac{1}{\sqrt{2^ri!}}(\gamma^{\mu_1\cdots\mu_i}C^{-1})_{\alpha\beta},\qquad(\m{T}^{\bs{e}_r})_{\alpha}^{\beta}=\delta_\alpha^{\phantom{\alpha}\beta},\\
(\m{T}^{2\bs{e}_r})_{\alpha\beta}^{\mu_1\cdots\mu_r}=\frac{1}{\sqrt{2^rr!}}(\gamma^{\mu_1\cdots\mu_r}C^{-1})_{\alpha\beta},
\end{gathered}
}[EqTPSdef]
where
\eqn{\gamma^{\mu_1\cdots\mu_n}=\gamma^{[\mu_1}\cdots\gamma^{\mu_n]}\equiv\frac{1}{n!}\sum_{\sigma\in S_n}(-1)^\sigma\gamma^{\mu_{\sigma(1)}}\cdots\gamma^{\mu_{\sigma(n)}},}
is the totally antisymmetric product of $\gamma$-matrices.  The corresponding hatted projectors are
\eqn{
\begin{gathered}
(\h{\m{P}}^{\bs{e}_r})_\alpha^{\phantom{\alpha}\beta}=\delta_\alpha^{\phantom{\alpha}\beta},\qquad(\h{\m{P}}^{\bs{e}_{i\neq r}})_{\mu_i\cdots\mu_1}^{\phantom{\mu_i\cdots\mu_1}\nu_1\cdots\nu_i}=\delta_{[\mu_1}^{\phantom{[\mu_1}\nu_1}\cdots\delta_{\mu_i]}^{\phantom{\mu_i]}\nu_i},\\
(\h{\m{P}}^{2\bs{e}_r})_{\mu_r\cdots\mu_1}^{\phantom{\mu_r\cdots\mu_1}\nu_1\cdots\nu_r}=\delta_{[\mu_1}^{\phantom{[\mu_1}\nu_1}\cdots\delta_{\mu_r]}^{\phantom{\mu_r]}\nu_r},
\end{gathered}
}[EqPPSdef]
where $\delta_{[\mu_1}^{\phantom{[\mu_1}\nu_1}\cdots\delta_{\mu_i]}^{\phantom{\mu_i]}\nu_i}$ is the totally antisymmetric normalized product of $\delta_\mu^{\phantom{\mu}\nu}$.  We remark that the half-projectors are essentially ``square roots" of projection operators.  In particular, the hatted projectors can be computed from the half-projectors via the identity $\m{T}_{\bs{N}}*\m{T}^{\bs{N}}=\h{\m{P}}^{\bs{N}}$, where the star product corresponds to a full contraction of the spinor indices.

This observation generalizes to arbitrary irreducible representations.  In fact, the first (second) definition of the half-projectors \eqref{EqTPS} can be seen as the projection of the tensor product $(\bs{e}_1)^{N_1}\otimes(\bs{e}_2)^{N_2}\otimes\cdots$ to the irreducible representation of interest by properly symmetrizing its vector (spinor) indices with the corresponding hatted projectors (projectors).  Hence, for arbitrary irreducible representations we have
\eqn{\m{P}_{\bs{N}}*\m{T}^{\bs{N}}=\m{T}^{\bs{N}}\cdot\h{\m{P}}^{\bs{N}}=\m{T}^{\bs{N}},\qquad\qquad\m{T}_{\bs{N}}*\m{T}^{\bs{N}}=\h{\m{P}}^{\bs{N}},\qquad\qquad\m{T}^{\bs{N}}\cdot\m{T}_{\bs{N}}=\m{P}_{\bs{N}},}[EqTPSId]
where the dot product corresponds to full contraction of the vector indices.  As expected, the first property of \eqref{EqTPSId} generalizes to $\m{P}^{\bs{N'}}*\m{T}^{\bs{N}}=\m{T}^{\bs{N}}\cdot\h{\m{P}}^{\bs{N'}}=\delta_{\bs{N'}\bs{N}}\m{T}^{\bs{N}}$ since the hatted projectors and the projectors satisfy $\h{\m{P}}^{\bs{N}}\cdot\h{\m{P}}^{\bs{N'}}=\delta^{\bs{N'}\bs{N}}\h{\m{P}}^{\bs{N}}$ and $\m{P}_{\bs{N}}*\m{P}_{\bs{N'}}=\delta_{\bs{N'}\bs{N}}\m{P}_{\bs{N}}$, respectively.

We stress that the explicit forms of the half-projectors [see \textit{e.g.}\ \eqref{EqTPSdef}] are not needed for the analysis.  The knowledge of their properties \eqref{EqTPSId} will be sufficient.

%%%%%%%%%%%%%%%%%%%%%%%%%%%%%%%%%%%%%%%%%%%%%%%%%%

\subsection{Irreducible Representations in Embedding Space}

In the embedding space formalism, the above quantities naturally generalize to embedding space quasi-primary operators $\m{O}$, which can be projected back to position space through the use of two supplementary conditions,
\eqn{\eta\cdot\partial\,\m{O}(\eta)=-\tau_{\m{O}}\m{O}(\eta),\qquad\qquad\eta\cdot\Gamma\,\m{O}(\eta)=0.}[EqSupp]
Of these, the first condition is in place to ensure homogeneity of the embedding space quasi-primary operators with twist $\tau_{\m{O}}=\Delta_{\m{O}}-S_{\m{O}}$, where $\Delta_{\m{O}}$ and $S_{\m{O}}$ are the conformal dimension and spin of the quasi-primary operator, respectively.  The second condition, which acts on each embedding space spinor index individually, is present to enforce transversality by halving the number of independent degrees of freedom.  Although the supplementary conditions \eqref{EqSupp} are necessary for projecting embedding space quantities back to position space, we will find that we may bypass position space projection altogether, thanks to techniques we introduce below.

To uplift the machinery of the position-space half-projector operator to embedding space, it is necessary to consider two embedding space coordinates simultaneously.  Indeed, given the two embedding space coordinates present in the definition of the OPE (here taken as $\eta_i$ and $\eta_j$), the embedding space half-projectors \eqref{EqTPS} take the form
\eqna{
(\m{T}_{ij}^\bs{N})&\equiv\left(\left(\frac{\sqrt{2}}{(\ee{i}{j})^\frac{1}{2}}\m{T}^{\bs{e}_2}\eta_i\m{A}_{ij}\right)^{N_1}\cdots\left(\frac{\sqrt{r}}{(\ee{i}{j})^\frac{1}{2}}\m{T}^{\bs{e}_{r_E-1}}\eta_i\m{A}_{ij}\cdots\m{A}_{ij}\right)^{N_{r-1}}\right.\\
&\phantom{=}\qquad\times\left.\left(\frac{\sqrt{r+1}}{(\ee{i}{j})^\frac{1}{2}}\m{T}^{2\bs{e}_{r_E}}\eta_i\m{A}_{ij}\cdots\m{A}_{ij}\right)^{\lfloor N_r/2\rfloor}\left(\m{T}^{\bs{e}_{r_E}}\frac{\eta_i\cdot\Gamma\,\eta_j\cdot\Gamma}{2\ee{i}{j}}\right)^{N_r-2\lfloor N_r/2\rfloor}\right)\cdot\h{\m{P}}_{ij}^\bs{N},
}[EqTES]
which shows that they are built from the embedding space half-projectors for defining irreducible representations \eqref{EqTPSdef},
\eqn{(\m{T}^{\bs{e}_{n+1}}\eta_i\m{A}_{ij}\cdots\m{A}_{ij})_{ab}^{A_1\cdots A_n}\equiv(\m{T}^{\bs{e}_{n+1}})_{ab}^{A_0'\cdots A_n'}\m{A}_{ijA_n'}^{\phantom{ijA_n'}A_n}\cdots\m{A}_{ijA_1'}^{\phantom{ijA_1'}A_1}\eta_{iA_0'},}
and the embedding space metric $\m{A}_{ij}$ discussed below.  Analogously, the embedding space hatted projectors can be constructed from the usual position space hatted projectors \eqref{EqPPSdef} with the help of
\eqn{\h{\m{P}}_{ij}^\bs{N}=\left(\frac{\eta_i\cdot\Gamma\,\eta_j\cdot\Gamma}{2\ee{i}{j}}\right)^{2\xi}\left.\h{\m{P}}^\bs{N}\right|_{\text{PS$\to$ES}}=\left.\h{\m{P}}^\bs{N}\right|_{\text{PS$\to$ES}}\left(\frac{\eta_i\cdot\Gamma\,\eta_j\cdot\Gamma}{2\ee{i}{j}}\right)^{2\xi},}[EqPES]
where $2\xi$ is zero (one) for bosonic (fermionic) irreducible representations.  Here, the position space to embedding space substitution $\text{PS$\to$ES}$ (for the metric, the $\epsilon$-tensor, and the $\gamma$-matrices) is simply
\eqn{
\begin{gathered}
g^{\mu\nu}\to\m{A}_{ij}^{AB}\equiv g^{AB}-\frac{\eta_i^A\eta_j^B}{\ee{i}{j}}-\frac{\eta_i^B\eta_j^A}{\ee{i}{j}},\\
\epsilon^{\mu_1\cdots\mu_d}\to\epsilon_{ij}^{A_1\cdots A_d}\equiv\frac{1}{\ee{i}{j}}\eta_{iA_0'}\epsilon^{A_0'A_1'\cdots A_d'A_{d+1}'}\eta_{jA_{d+1}'}\m{A}_{ijA_d'}^{\phantom{ijA_d'}A_d}\cdots\m{A}_{ijA_1'}^{\phantom{ijA_1'}A_1},\\
\gamma^{\mu_1\cdots\mu_n}\to\Gamma_{ij}^{A_1\cdots A_n}\equiv\Gamma^{A_1'\cdots A_n'}\m{A}_{ijA_n'}^{\phantom{ijA_n'}A_n}\cdots\m{A}_{ijA_1'}^{\phantom{ijA_1'}A_1}\qquad\forall\,n\in\{0,\ldots,r\}.
\end{gathered}
}[EqPStoES]
Hence, both the embedding space half-projectors \eqref{EqTES} and hatted projectors \eqref{EqPES} may be directly obtained from their position space analogs via the substitutions \eqref{EqPStoES}.  Therefore, they are completely fixed by group theory, just as in position space.

The hatted projectors in embedding space satisfy several useful properties.  For example, hatted projectors for contragredient-reflected\footnote{For a given irreducible representation $\bs{N}=\sum_{i=1}^rN_i\bs{e}_i=\{N_1,\ldots,N_r\}$, the associated contragredient-reflected irreducible representation $\bs{N}^{CR}$ is given by
\eqna{
\text{d odd:}&\qquad\bs{N}^{CR}=\{N_1,\ldots,N_r\}=\bs{N},\\
\text{d even:}&\qquad\bs{N}^{CR}=\begin{cases}\{N_1,\ldots,N_r\}=\bs{N}&\text{if $r$ is odd},\\\{N_1,\ldots,N_{r-2},N_r,N_{r-1}\}&\text{if $r$ is even}.\end{cases}
}
Since the associated conjugate irreducible representation $\bs{N}^C$ satisfies
\eqna{
\text{d odd:}&\qquad\bs{N}^C=\{N_1,\ldots,N_r\}=\bs{N},\\
\text{d even:}&\qquad\bs{N}^C=\begin{cases}\{N_1,\ldots,N_r\}=\bs{N}&\text{if $r+q$ is even},\\\{N_1,\ldots,N_{r-2},N_r,N_{r-1}\}&\text{if $r+q$ is odd},\end{cases}
}
in all signatures, in Lorentzian signature ($q=1$) the contragredient-reflected representation $\bs{N}^{CR}$ is equivalent to the conjugate representation $\bs{N}^C$, \textit{i.e.}\ $\bs{N}^{CR}=\bs{N}^C$ for $q=1$.
} irreducible representations $\bs{N}$ and $\bs{N}^{CR}$ are related by
\eqn{(\h{\m{P}}_{ij}^{\bs{N}})_{\{aA\}}^{\phantom{\{aA\}}\{B'b'\}}[(C_\Gamma^{-1})_{b'b}]^{2\xi}(g_{B'B})^{n_v}=[(C_\Gamma^{-1})_{ab'}]^{2\xi}(g_{AB'})^{n_v}(\h{\m{P}}_{ji}^{\bs{N}^{CR}})_{\{bB\}}^{\phantom{\{bB\}}\{B'b'\}}.}[EqPtoPCR]
An equivalent relation between hatted projectors of irreducible representations and their conjugates is $B_\Gamma^{-1}(\h{\m{P}}_{ij}^{\bs{N}})^*B_\Gamma=\h{\m{P}}_{ij}^{\bs{N}^C}$.  As in position space, the embedding space hatted projectors are also related to the embedding space half-projectors through the identity
\eqn{\m{T}_{ij\bs{N}}*\m{T}_{jk}^{\bs{N}}=\left(\frac{\ee{i}{j}}{\ee{j}{k}}\right)^{\frac{1}{2}(S-\xi)}\h{\m{P}}_{ji}^{\bs{N}}\cdot\h{\m{P}}_{jk}^{\bs{N}}=\left(\frac{\ee{i}{j}}{\ee{j}{k}}\right)^{\frac{1}{2}(S-\xi)}\h{\m{P}}_{ji}^{\bs{N}}\left(\frac{\eta_j\cdot\Gamma\,\eta_k\cdot\Gamma}{2\ee{j}{k}}\right)^{2\xi}(\m{A}_{jk})^{n_v},}[EqTESId]
where the equivalence to \eqref{EqTPSId} is exact for $k=i$, or, in other words, for $\m{T}_{ij\bs{N}}*\m{T}_{ji}^{\bs{N}}=\h{\m{P}}_{ji}^{\bs{N}}$.  The last equality is obtained from the identities
\eqn{
\begin{gathered}
\h{\m{P}}_{ij}^{\bs{N}}=(\m{A}_{ij})^{n_v}\left(\frac{\eta_i\cdot\Gamma\,\eta_j\cdot\Gamma}{2\ee{i}{j}}\right)^{2\xi}\h{\m{P}}_{kj}^{\bs{N}}(\m{A}_{ij})^{n_v},\\
\h{\m{P}}_{ji}^{\bs{N}}=(\m{A}_{ji})^{n_v}\h{\m{P}}_{jk}^{\bs{N}}\left(\frac{\eta_j\cdot\Gamma\,\eta_i\cdot\Gamma}{2\ee{i}{j}}\right)^{2\xi}(\m{A}_{ji})^{n_v},
\end{gathered}
}[EqPtoP]
which are valid for any irreducible representation $\bs{N}$ and imply that
\eqn{\m{T}_{ij}^{\bs{N}}=\left(\frac{\ee{i}{k}}{\ee{i}{j}}\right)^{\frac{1}{2}(S-\xi)}\m{T}_{ik}^{\bs{N}}\left(\frac{\eta_i\cdot\Gamma\,\eta_j\cdot\Gamma}{2\ee{i}{j}}\right)^{2\xi}(\m{A}_{ij})^{n_v}.}[EqTtoT]
Another identity useful when computing correlation functions from the embedding space OPE is given by
\eqna{
\m{T}_{ij\bs{N}^{CR}}*\m{T}_{jk}^{\bs{N}^{CR}}\cdot\m{T}_{kj}^{\bs{N}}&=\left(\frac{\ee{i}{j}}{\ee{j}{k}}\right)^{\frac{1}{2}(S-\xi)}\m{T}_{kj}^{\bs{N}}\cdot\h{\m{P}}_{ij}^{\bs{N}}(C_\Gamma^{-T})^{2\xi}(g)^{n_v}\\
&=\left[\frac{(\ee{i}{j})(\ee{i}{k})}{(\ee{j}{k})^2}\right]^{\frac{1}{2}(S-\xi)}\m{T}_{ki}^{\bs{N}}\left(\frac{\eta_k\cdot\Gamma\,\eta_j\cdot\Gamma C_\Gamma^{-T}}{2\ee{j}{k}}\right)^{2\xi}(\m{A}_{jk}\cdot\m{A}_{ij})^{n_v},
}[EqTTT]
obtained with the help of \eqref{EqTESId}, \eqref{EqPtoP} and \eqref{EqTtoT}.  It is understood that the dot product on the left-hand side of \eqref{EqTTT} is realized through \eqref{EqPtoPCR} (since all vector indices are contravariant).  Moreover, we note that all of the above generalizes straightforwardly to even dimensions.

Finally, we reiterate that the explicit (embedding space) half-projectors are not necessary (they are never used); instead we rely on their properties \eqref{EqTESId}, \eqref{EqTtoT}, and \eqref{EqTTT}.  Besides, the transversality and completeness (as a basis) properties of the embedding space half-projectors imply that $\m{O}^{\bs{N}}(\eta_i)\sim\m{T}_{ij}^{\bs{N}}$ for any extra embedding space coordinate $\eta_j$ where $j\neq i$.

%%%%%%%%%%%%%%%%%%%%%%%%%%%%%%%%%%%%%%%%%%%%%%%%%%

\subsection{Operator Product Expansion}

With the formalism in place, we can now consider the embedding space OPE between two embedding space quasi-primary operators in irreducible representations $\bs{N}_i$ and $\bs{N}_j$.  It is given by
\eqn{
\begin{gathered}
\Op{i}{}{1}\Op{j}{}{2}=(\m{T}_{12}^{\bs{N}_i})(\m{T}_{21}^{\bs{N}_j})\cdot\sum_m\sum_{a=1}^{N_{ijm}}\frac{\cOPE{a}{ij}{m}\tOPE{a}{ij}{m}{12}}{(\ee{1}{2})^{p_{ijm}}}\cdot\m{D}_{12}^{(d,h_{ijm}-n_a/2,n_a)}(\m{T}_{12\bs{N}_m})*\Op{m}{}{2},\\
p_{ijm}=\frac{1}{2}(\tau_i+\tau_j-\tau_m),\qquad h_{ijm}=-\frac{1}{2}(\chi_i-\chi_j+\chi_m),\\
\tau_\m{O}=\Delta_\m{O}-S_\m{O},\qquad\chi_\m{O}=\Delta_\m{O}-\xi_\m{O},\qquad\xi_\m{O}=S_\m{O}-\lfloor S_\m{O}\rfloor,
\end{gathered}
}[EqOPE]
where $\cOPE{a}{ij}{m}$ are the $N_{ijm}$ OPE coefficients that depend on the specific CFT under consideration.  In contrast, the remaining quantities are all fixed by group theory and conformal invariance.  Below we describe the two leftover essential ingredients present in the OPE \eqref{EqOPE}, namely the OPE differential operator, which generates conformal descendants, and the OPE tensor structures, which serve to properly contract the embedding space dummy indices.

%%%%%%%%%%%%%%%%%%%%%%%%%%%%%%%%%%%%%%%%%%%%%%%%%%

\subsubsection{OPE Differential Operator}

The OPE differential operator $\m{D}_{ij}^{(d,h,n)}$ is given by
\eqn{
\begin{gathered}
\m{D}_{ij}^{(d,h,n)A_1\cdots A_n}=\m{D}_{ij}^{2(h+n)}\frac{\eta_j^{A_1}}{(\ee{i}{j})^\frac{1}{2}}\cdots\frac{\eta_j^{A_n}}{(\ee{i}{j})^\frac{1}{2}},\\
\m{D}_{ij}^2=(\ee{i}{j})\partial_j^2-(d+2\eta_j\cdot\partial_j)\eta_i\cdot\partial_j,
\end{gathered}
}[EqD]
and its action on arbitrary products of powers of embedding space coordinates was determined in \cite{Fortin:2019dnq}.  Its dummy embedding space vector indices are fully symmetric by construction.  These are contracted with the embedding space half-projectors through the OPE tensor structures.  It has several useful properties including
\eqn{
\begin{gathered}
\m{D}_{ij\{A\}}^{(d,h,n)}(\eta_{jA})^m=(\ee{i}{j})^\frac{m}{2}\m{D}_{ij\{A\}}^{(d,h-m,n+m)},\\
\m{D}_{ij|h+n+m}^{B_m}\cdots\m{D}_{ij|h+n+1}^{B_1}\m{D}_{ij}^{(d,h,n)\{A\}}=\m{D}_{ij}^{(d,h,n+m)\{AB\}}.
\end{gathered}
}[EqId]
where
\eqn{\m{D}_{ij|h}^A=\frac{\eta_j^A}{(\ee{i}{j})^{\frac{1}{2}}}\m{D}_{ij}^2+2h\m{D}_{ij}^A-h(d+2h-2)\frac{\eta_i^A}{(\ee{i}{j})^{\frac{1}{2}}}.}
Moreover, it satisfies contiguous relations of the type
\eqn{
\begin{gathered}
g^{AA}\m{D}_{ij\{A\}}^{(d,h,n)}=0,\\
(\eta_i^A)^m\m{D}_{ij\{A\}}^{(d,h,n)}=(\ee{i}{j})^\frac{m}{2}\m{D}_{ij\{A\}}^{(d,h+m,n-m)},\\
(\eta_j^A)^m\m{D}_{ij\{A\}}^{(d,h,n)}=\rho^{(d,m;-h-n)}(\ee{i}{j})^{m/2}\m{D}_{ij\{A\}}^{(d,h,n-m)},
\end{gathered}
}[EqCR]
where
\eqn{\rho^{(d,h;p)}=(-2)^h(p)_h(p+1-d/2)_h,\qquad\qquad\rho^{(d,h+h';p)}=\rho^{(d,h;p)}\rho^{(d,h';p+h)}.}
The identities \eqref{EqId} follow straightforwardly from the definition \eqref{EqD}.  Meanwhile, the contiguous relations \eqref{EqCR} capture the tracelessness of the OPE differential operator with respect to the $g$-metric and state that it has well-defined contractions with the embedding space coordinates $\eta_i$ and $\eta_j$.

%%%%%%%%%%%%%%%%%%%%%%%%%%%%%%%%%%%%%%%%%%%%%%%%%%

\subsubsection{OPE Tensor Structure Basis}

Since the role of the OPE tensor structures is to intertwine the irreducible representations of the three quasi-primary operators and the OPE differential operator \eqref{EqD} (which can be seen as a symmetric-traceless) to generate singlets, there are as many tensor structures as there are symmetric-traceless irreducible representations in the tensor product decomposition of $\bs{N}_i\otimes\bs{N}_j\otimes\bs{N}_m$, fixing $N_{ijm}$ by group theory arguments.  This observation also implies that there exists an infinite tower of allowed exchanged quasi-primary operators since $N_{ij,m+\ell}\neq0$ for $\bs{N}_m+\ell\bs{e}_1$ if $N_{ijm}\neq0$ for $\bs{N}_m$.

Starting from the OPE building blocks given in \eqref{EqPStoES}, we can construct a convenient basis for the $N_{ijm}$ OPE tensor structures from monomials of the form
\eqn{\tOPE{a}{ij}{m}{12}=(\text{product of $\m{A}_{12}$})\times\epsilon_{12}\times(\Gamma_{12}^{[n]})^{\xi_i+\xi_j+\xi_m}(C_\Gamma^{-1})^{\xi_i+\xi_j-\xi_m}.}[EqTSOPE]
Since products of $\epsilon$-tensors can be re-expressed in terms of metrics times at most one $\epsilon$-tensor, we choose the OPE tensor structures \eqref{EqTSOPE} to have at most one $\epsilon$-tensor.  Similarly, $\Gamma$-matrices, which appear only when there are two fermionic quasi-primary operators present in the OPE [hence the form of its power in \eqref{EqTSOPE}], form a basis $\{\Gamma_{12}^{[n]}\}\,\forall\,n\in\{0,\ldots,r\}$.  Consequently the OPE tensor structures \eqref{EqTSOPE} can be chosen to have at most one $\Gamma$-matrix (with $n$ fixed).

Writing out the embedding space indices explicitly, we have $(\tOPE{a}{ij}{m}{12})_{\{aA\}\{bB\}}^{\phantom{\{aA\}\{bB\}}\{Ee\}\{F\}}$.\footnote{Here the groups of indices $\{aA\}$, $\{bB\}$ and $\{Ee\}$ contract with the dummy indices of the half-projectors for $\m{O}_i$, $\m{O}_j$ and $\m{O}_m$ respectively while the group of indices $\{F\}$ contracts with the dummy indices of the OPE differential operator, see \eqref{EqOPE}.}  This may contain $\m{A}$-metrics with all possible choices of the embedding space indices, with the exception of two indices of the same type, which are disallowed by the tracelessness condition for quasi-primary operators.  The only possibility where we can have a pair of indices of the same type is $\m{A}_{12}^{FF}$ since the OPE differential operator \eqref{EqD} is traceless with respect to the $g$-metric but not the $\m{A}$-metric.\footnote{Although the OPE differential operator could be made symmetric-traceless by simply introducing an extra hatted projector, as in $\m{D}_{ij}^{(d,h,n)}\to\h{\m{P}}_{ij}^{n\bs{e}_1}\cdot\m{D}_{ij}^{(d,h,n)}$, it is more convenient to proceed with \eqref{EqD} instead.}  Hence $\m{A}_{12}^{FF}$ are in principle allowed but redundant due to the contiguous relations \eqref{EqCR}, and as such they may be discarded if desired.  On the contrary, due to their antisymmetry properties, $\epsilon$-tensors and $\Gamma$-matrices can have at most one $F$-index.  However, as argued above, it is always possible to rewrite an $\epsilon$-tensor without an $F$-index as an $\epsilon$-tensor times $\m{A}_{12}^{FF}$ with a modified OPE differential operator; we are therefore free to discard all such $\epsilon$-tensors.  Indeed, using the identity
\eqn{\epsilon_{12}^{X_1\cdots X_d}\m{A}_{12}^{FF}=d\epsilon_{12}^{[X_1\cdots X_{d-1}|F|}\m{A}_{12}^{X_d]F},}[Eqepsilon]
demonstrates that the $\epsilon$-tensor without an $F$-index is linearly dependent and may be re-expressed in terms of $\epsilon$-tensors with a single $F$-index.  Finally, for definiteness, we adopt the convention that the embedding space spinor indices, which appear whenever there are two fermionic quasi-primary operators present in the OPE, are ordered as $(\Gamma_{12}^{[n]}C_\Gamma^{-1})_{ab}$, $(\Gamma_{12}^{[n]})_a^{\phantom{a}e}$ or $(\Gamma_{12}^{[n]})_b^{\phantom{b}e}$, respectively.

%%%%%%%%%%%%%%%%%%%%%%%%%%%%%%%%%%%%%%%%%%%%%%%%%%
%%%%%%%%%%%%%%%%%%%%%%%%%%%%%%%%%%%%%%%%%%%%%%%%%%

\section{Correlation Functions from the OPE}\label{SecCF}

In this section we use the OPE and the identities presented above to compute two-, three-, and four-point correlation functions in embedding space.

%%%%%%%%%%%%%%%%%%%%%%%%%%%%%%%%%%%%%%%%%%%%%%%%%%

\subsection{Two-Point Correlation Functions}

To better understand the OPE \eqref{EqOPE}, it is enlightening to start from the simplest non-trivial correlation functions, \textit{i.e.}\ two-point correlation functions.  We first note that in this case the ``exchanged'' quasi-primary operator is simply the identity operator and that the power of the OPE differential operator must vanish, leading to \cite{Fortin:2019xyr}
\eqn{\Vev{\Op{i}{}{1}\Op{j}{}{2}}=(\m{T}_{12}^{\bs{N}_i})(\m{T}_{21}^{\bs{N}_j})\cdot\frac{\cOPE{}{ij}{\1}\tOPE{}{ij}{\1}{12}}{(\ee{1}{2})^{\tau_i}}=\cOPE{}{ij}{\1}\frac{(\m{T}_{12}^{\bs{N}_i})\cdot(\m{T}_{21}^{\bs{N}_j})}{(\ee{1}{2})^{\tau_i}}.}[Eq2pt]
Here, the two-point correlation function \eqref{Eq2pt} is non-vanishing only if the conformal dimensions of the two quasi-primary operators match, \textit{i.e.}\ $\tau_i=\tau_j$.  Moreover, the irreducible representation of the ``exchanged'' quasi-primary operator (the identity operator) implies that the quasi-primary operators are in irreducible representations that are contragredient-reflected with respect to each other, \textit{i.e.}\ $\bs{N}_j=\bs{N}_i^{CR}$.  In other words, there is a unique tensor structure that intertwines the dummy embedding space indices of the embedding space half-projectors; this natural contraction can occur only if the irreducible representations are in contragredient-reflected representations with respect to each other.  In light of the above, the tensor structure can be chosen such that
\eqn{(\tOPE{}{ij}{\1}{12})_{\{aA\}\{bB\}}=\left[\prod_{r=1}^{n_v^i}\m{A}_{12A_rB_r}\right][(C_\Gamma^{-1})_{ab}]^{2\xi_i},}
with the half-projector dummy indices contracted explicitly as
\eqn{\Vev{\Op{i}{}{1}\Op{j}{}{2}}=\cOPE{}{ij}{\1}\frac{(\m{T}_{12}^{\bs{N}_i})^{\{Aa\}}(\m{T}_{21}^{\bs{N}_i^{CR}})^{\{Bb\}}[(C_\Gamma^{-1})_{ab}]^{2\xi_i}(g_{AB})^{n_v^i}}{(\ee{1}{2})^{\tau_i}},}
\textit{i.e.}\ as in \eqref{EqPtoPCR}.

%%%%%%%%%%%%%%%%%%%%%%%%%%%%%%%%%%%%%%%%%%%%%%%%%%

\subsection{Three-Point Correlation Functions}

Let us next consider the three-point correlation function of three arbitrary quasi-primary operators.  We apply the OPE \eqref{EqOPE} on the first two quasi-primary operators.  This leads to the OPE differential operator \eqref{EqD} acting on the resulting two-point correlation function, for which the explicit solution is \eqref{Eq2pt}.  Using \eqref{EqTTT} then implies that three-point correlation functions can be expressed as \cite{Fortin:2019pep}
\eqn{\Vev{\Op{i}{}{1}\Op{j}{}{2}\Op{m}{}{3}}=\frac{(\m{T}_{12}^{\bs{N}_i})(\m{T}_{21}^{\bs{N}_j})(\m{T}_{31}^{\bs{N}_m})\cdot\sum_{a=1}^{N_{ijm}}\cCF{(a|}{ijm}\mathscr{G}_{(a|}^{ij|m}(\eta_1,\eta_2,\eta_3)}{(\ee{1}{2})^{\frac{1}{2}(\tau_i+\tau_j-\chi_m)}(\ee{1}{3})^{\frac{1}{2}(\chi_i-\chi_j+\tau_m)}(\ee{2}{3})^{\frac{1}{2}(-\chi_i+\chi_j+\chi_m)}}.}[Eq3pt]
Here the OPE coefficients and the OPE tensor structures are given by
\eqn{\cCF{(a|}{ijm}=\sum_n\cOPE{a}{ij}{n}\cOPE{}{nm}{\1},\qquad\qquad\tCF{(a|}{ijm}{12}=\tOPE{a}{ij}{m^{CR}}{12}(C_\Gamma^{-1})^{2\xi_m}(g)^{n_v^m},}[EqCoeff]
while the three-point conformal blocks are defined as
\eqna{
\mathscr{G}_{(a|}^{ij|m}(\eta_1,\eta_2,\eta_3)&=(\h{\m{P}}_{12}^{\bs{N}_i})(\h{\m{P}}_{21}^{\bs{N}_j})(\h{\m{P}}_{31}^{\bs{N}_m})\cdot\tCF{(a|}{ijm}{12}\cdot\frac{(\ee{2}{3})^{\chi_m+h_{ijm}}}{(\ee{1}{3})^{h_{ijm}}}\\
&\phantom{=}\qquad\times\m{D}_{12}^{(d,h_{ijm}-n_a/2,n_a)}\left(\frac{\eta_3\cdot\Gamma\,\eta_2\cdot\Gamma}{2\ee{2}{3}}\right)^{2\xi_m}\frac{(\m{A}_{23}\cdot\m{A}_{12})^{n_v^m}}{(\ee{2}{3})^{\chi_m}}.
}[Eq3ptCB]

In the OPE tensor structure basis \eqref{EqTSOPE}, the tensor structures $(\tCF{(a|}{ijm}{12})_{\{aA\}\{bB\}\{e'E'\}}^{\phantom{\{aA\}\{bB\}\{e'E'\}}\{F\}}$ \eqref{EqCoeff} are monomials of the form
\eqn{\tCF{(a|}{ijm}{12}=(\text{product of $\m{A}_{12}$})\times\epsilon_{12}\times\left(\Gamma_{12}^{[n]}C_\Gamma^{-1}\right)^{\xi_i+\xi_j+\xi_m},}[EqTSOPEbasis]
where the different building blocks are subject to the restrictions that were detailed above.  Hence, in general the three-point conformal blocks $(\mathscr{G}_{(a|}^{ij|m})_{\{aA\}\{bB\}\{eE\}}$ \eqref{Eq3ptCB} are polynomials built from the monomials \eqref{EqTSOPEbasis}, where all $F$-indices are contracted with $\tilde{\eta}_{3F}$, the homogeneized embedding space coordinate of the exchanged quasi-primary operator.  Overall, the three homogeneized three-point coordinates are given by
\eqn{\tilde{\eta}_1^A=\frac{(\ee{2}{3})^{\frac{1}{2}}\eta_1^A}{(\ee{1}{2})^{\frac{1}{2}}(\ee{1}{3})^{\frac{1}{2}}},\qquad\tilde{\eta}_2^A=\frac{(\ee{1}{3})^{\frac{1}{2}}\eta_2^A}{(\ee{1}{2})^{\frac{1}{2}}(\ee{2}{3})^{\frac{1}{2}}},\qquad\tilde{\eta}_3^A=\frac{(\ee{1}{2})^{\frac{1}{2}}\eta_3^A}{(\ee{1}{3})^{\frac{1}{2}}(\ee{2}{3})^{\frac{1}{2}}}.}[Eqetat]

We now consider three-point conformal blocks in \eqref{Eq3pt} corresponding to the exchange of infinite towers of quasi-primary operators $\bs{N}_m+\ell\bs{e}_1$ in the OPE.  We find that the associated OPE tensor structures can be factorized as
\eqn{\tCF{(a|}{ij,m+\ell}{12}=\tCF{(a|}{ij,m+i_a}{12}(\m{A}_{12})^{\ell-i_a},}[EqTSOPEbasisTower]
where the explicit $\m{A}$-metrics contract $\ell-i_a$ of the $\ell\bs{e}_1$ indices of the exchanged quasi-primary operator with $\ell-i_a$ of the OPE differential operator indices.  Meanwhile, the remaining contractions, which are specific to the three irreducible representations, are all included in $\tCF{(a|}{ij,m+i_a}{12}$.

Therefore, upon explicitly exposing the indices, we arrive at the following form for the three-point conformal blocks for $\bs{N}_m+\ell \bs{e}_1$ exchange with the OPE tensor structures \eqref{EqTSOPEbasisTower}
\eqna{
&(\mathscr{G}_{(a|}^{ij|m+\ell})_{\{aA\}\{bB\}\{eE\}}(\eta_1,\eta_2,\eta_3)\\
&\qquad=\left((\h{\m{P}}_{12}^{\bs{N}_i})(\h{\m{P}}_{21}^{\bs{N}_j})\cdot\tCF{(a|}{ij,m+i_a}{12}\left((\ee{1}{2})^{\frac{1}{2}}\frac{\Gamma^T\,\eta_3\cdot\Gamma^T}{2\ee{2}{3}}\right)^{2\xi_m}\right)_{\{aA\}\{bB\}\{e'E'\}}^{\phantom{\{aA\}\{bB\}\{e'E'\}}\{F\}}\\
&\qquad\phantom{=}\times\frac{(\ee{2}{3})^{\chi_{m+\ell}+2\xi_m+h_{ij,m+\ell}}}{(\ee{1}{3})^{h_{ij,m+\ell}}}(\m{A}_{12E'}^{\phantom{12E'}F})^{\ell-i_a}\\
&\qquad\phantom{=}\times\m{D}_{12\{F\}}^{(d,h_{ij,m+\ell}-n_a/2-2\xi_m,n_a+2\xi_m)}\frac{[\m{\h{P}}_{31}^{\bs{N}_m+\ell\bs{e}_1}(\m{A}_{23}\cdot\m{A}_{12})^{n_v^m+\ell}]_{\{eE\}}^{\phantom{\{eE\}}\{E'e'\}}}{(\ee{2}{3})^{\chi_{m+\ell}+2\xi_m}},
}
where the uncontracted $\Gamma^T$ carries an $F$-index.  We may then extract the part that contains special indices, $[(\m{A}_{13}\cdot\m{A}_{23}\cdot\m{A}_{12})^{n_v^m+i_a}]_{\{E\}}^{\phantom{\{E\}}\{E'\}}$, by making the substitution
\eqn{(\m{A}_{13}\cdot\m{A}_{23}\cdot\m{A}_{12})_E^{\phantom{E}E'}\to\m{A}_{13}^{E'E''}\left(g_{E''E}+\frac{(\ee{1}{3})\eta_{2E''}\eta_{2E}}{(\ee{1}{2})(\ee{2}{3})}\right),}
which is valid due to the respective contractions of the $E$- and $E'$-indices.  We further expand the part of interest as
\eqna{
&\left(g_{E''E}+\frac{(\ee{1}{3})\eta_{2E''}\eta_{2E}}{(\ee{1}{2})(\ee{2}{3})}\right)^{n_v^m+i_a}\\
&\qquad=\sum_\sigma\sum_{r_0\geq0}\frac{(-1)^{r_0}(-n_v^m-i_a)_{r_0}}{(n_v^m+i_a)!r_0!}\left(\frac{(\ee{1}{3})}{(\ee{1}{2})(\ee{2}{3})}\right)^{n_v^m+i_a-r_0}\\
&\qquad\phantom{=}\times g_{E''_{\sigma(1)}E_{\sigma(1)}}\cdots g_{E''_{\sigma(r_0)}E_{\sigma(r_0)}}\eta_{2E''_{\sigma(r_0+1)}}\eta_{2E_{\sigma(r_0+1)}}\cdots\eta_{2E''_{\sigma(n_v^m+i_a)}}\eta_{2E_{\sigma(n_v^m+i_a)}},
}
where the $\sigma$-sum is over permutations of the $n_v^m+i_a$ indices and the $r_0$ sum is bounded from above implicitly by the Pochhammer symbol $(-n_v^m-i_a)_{r_0}$.  This allows us to re-express the three-point conformal blocks in the form
\eqna{
&(\mathscr{G}_{(a|}^{ij|m+\ell})_{\{aA\}\{bB\}\{eE\}}(\eta_1,\eta_2,\eta_3)\\
&\qquad=\left((\h{\m{P}}_{12}^{\bs{N}_i})(\h{\m{P}}_{21}^{\bs{N}_j})\cdot\tCF{(a|}{ij,m+i_a}{12}\left(\frac{C_\Gamma\Gamma\,\tilde{\eta}_3\cdot\Gamma C_\Gamma^{-1}}{2}\right)^{2\xi_m}\right)_{\{aA\}\{bB\}\{e'E'''\}}^{\phantom{\{aA\}\{bB\}\{e'E'''\}}\{F\}}(\m{A}_{13}^{E'''E''})^{n_v^m+i_a}\\
&\qquad\phantom{=}\times(\h{\m{P}}_{31}^{\bs{N}_m+\ell\bs{e}_1})_{\{eE\}}^{\phantom{\{eE\}}\{E'e'\}}\sum_\sigma\sum_{r_0\geq0}\frac{(-1)^{r_0}(-n_v^m-i_a)_{r_0}}{(n_v^m+i_a)!r_0!}g_{E'_{\sigma(1)}E''_{\sigma(1)}}\cdots g_{E'_{\sigma(r_0)}E''_{\sigma(r_0)}}\\
&\qquad\phantom{=}\times\frac{\m{G}_{(\chi_{m+\ell}+2\xi_m+n_v^m+i_a-r_0,\ell-i_a)\{E'\}\{FE'_{\sigma(r_0+1)}E''_{\sigma(r_0+1)}\cdots E'_{\sigma(n_v^m+i_a)}E''_{\sigma(n_v^m+i_a)}\}}^{(d,\chi_{m+\ell}/2+h_{ij,m+\ell}-n_v^m/2-i_a/2+r_0/2,n_a-\ell+2\xi_m+2n_v^m+3i_a-2r_0)}(\eta_1,\eta_2,\eta_3)}{R_{(\chi_{m+\ell}+2\xi_m+n_v^m+i_a-r_0,\ell-i_a)}^{(d,\chi_{m+\ell}/2+h_{ij,m+\ell}-n_v^m/2-i_a/2+r_0/2,n_a-\ell+2\xi_m+2n_v^m+3i_a-2r_0)}},
}[Eq3ptCBSoln]
where the uncontracted $\Gamma$-matrix carries an $F$-index, as before.

Here we have written the solution \eqref{Eq3ptCBSoln} in terms of a linear combination of the object
\eqna{
&\m{G}_{(\Delta,\ell)\{E\}\{F\}}^{(d,\Lambda,n)}(\eta_1,\eta_2,\eta_3)\\
&\qquad=R_{(\Delta,\ell)}^{(d,\Lambda,n)}\frac{(\ee{2}{3})^{\Lambda+\Delta/2}}{(\ee{1}{3})^{\Lambda-\Delta/2}}(\m{A}_{12E'}^{\phantom{12E'}F})^\ell\m{D}_{12\{F\}}^{(d,\Lambda-\Delta/2-\ell/2-n/2,\ell+n)}\frac{[\h{\m{P}}_{31}^{\ell\bs{e}_1}(\m{A}_{23}\cdot\m{A}_{12})^\ell]_{\{E\}}^{\phantom{\{E\}}\{E'\}}}{(\ee{2}{3})^\Delta}\\
&\qquad=R_{(\Delta,\ell)}^{(d,\Lambda,n)}\frac{(\ee{2}{3})^{\Lambda+\Delta/2}}{(\ee{1}{3})^{\Lambda-\Delta/2}}\left[-\frac{\eta_2^{E'}}{(\ee{1}{2})^{\frac{1}{2}}}\right]^\ell\m{D}_{12\{F\}}^{(d,\Lambda-\Delta/2+\ell/2-n/2,n)}\frac{[\h{\m{P}}_{31}^{\ell\bs{e}_1}(\m{A}_{23}\cdot\m{A}_{12})^\ell]_{\{E\}\{E'\}}}{(\ee{2}{3})^\Delta}.
}[EqG3p]
It turns out that this object is in fact a tensorial generalization of the scalar-scalar-(spin-$\ell$) three-point conformal blocks, as we describe in the next section.  This tensorial generalization of the three-point scalar block has vanishing degrees of homogeneity in all embedding space coordinates.  Although we find it more convenient to proceed as detailed above, it is important to note here that the explicit hatted projector $\h{\m{P}}_{31}^{\ell\bs{e}_1}$ can ultimately be absorbed by the hatted projector $\h{\m{P}}_{31}^{\bs{N}_m+\ell\bs{e}_1}$ in \eqref{Eq3ptCBSoln} since it can be commuted through the OPE differential operator.

%%%%%%%%%%%%%%%%%%%%%%%%%%%%%%%%%%%%%%%%%%%%%%%%%%

\subsection{Four-Point Correlation Functions}

Let us now turn to the four-point correlation functions.  This time, we apply the OPE \eqref{EqOPE} twice to arrive at the following general form for the four-point object:
\eqna{
&\Vev{\Op{i}{}{1}\Op{j}{}{2}\Op{l}{}{4}\Op{k}{}{3}}\\
&\qquad=\frac{(\m{T}_{12}^{\bs{N}_i})(\m{T}_{21}^{\bs{N}_j})(\m{T}_{43}^{\bs{N}_l})(\m{T}_{34}^{\bs{N}_k})\cdot\sum_{m,n}\sum_{a=1}^{N_{ijm}}\sum_{b=1}^{N_{klm}}\cCF{(a|}{ijm}\cCF{(b|}{lkn}G^{nm}\mathscr{G}_{(a|b)}^{ij|m|lk}(\eta_1,\eta_2,\eta_4,\eta_3)}{(\ee{1}{2})^{\frac{1}{2}(\tau_i+\tau_j)}(\ee{3}{4})^{\frac{1}{2}(\tau_k+\tau_l)}(\ee{1}{3})^{\frac{1}{2}(\chi_k-\chi_l)}(\ee{2}{4})^{\frac{1}{2}(-\chi_i+\chi_j)}(\ee{1}{4})^{\frac{1}{2}(\chi_i-\chi_j-\chi_k+\chi_l)}},
}[Eq4pt]
where $G^{ij}$ is such that $\sum_j\cOPE{}{ij}{\1}G^{jk}=\delta_i^{\phantom{i}k}$ and $\sum_jG^{ij}\cOPE{}{jk}{\1}=\delta_{\phantom{i}k}^i$.  Here the $\mathscr{G}_{(a|b)}^{ij|m|lk}(\eta_1,\eta_2,\eta_4,\eta_3)$ are the four-point conformal blocks.  These are given explicitly by
\eqna{
&\mathscr{G}_{(a|b)}^{ij|m|lk}(\eta_1,\eta_2,\eta_4,\eta_3)\\
&\qquad=(\h{\m{P}}_{12}^{\bs{N}_i})(\h{\m{P}}_{21}^{\bs{N}_j})(\h{\m{P}}_{43}^{\bs{N}_l})(\h{\m{P}}_{34}^{\bs{N}_k})\cdot\tCF{(a|}{ijm^{CR}}{12}\,\tCF{(b|}{lkm}{43}\cdot\frac{(\ee{1}{4})^{\frac{1}{2}(\chi_i-\chi_j-\chi_k+\chi_l)}}{(\ee{1}{3})^{\frac{1}{2}(-\chi_k+\chi_l)}(\ee{2}{4})^{\frac{1}{2}(\chi_i-\chi_j)}}\\
&\qquad\phantom{=}\times(\ee{1}{2})^{\frac{1}{2}\chi_m}(\ee{3}{4})^{\frac{1}{2}\chi_m}\m{D}_{12}^{(d,h_{ijm}-n_a/2,n_a)}\m{D}_{43}^{(d,h_{lkm}-n_b/2,n_b)}\frac{(\m{A}_{12})^{n_v^m}(C_\Gamma^T)^{2\xi_m}\h{\m{P}}_{23}^{\bs{N}_m}(\m{A}_{34})^{n_v^m}}{(\ee{2}{3})^{\chi_m}},
}[Eq4ptCB]
and are homogeneous of vanishing degrees in all four embedding space coordinates.  Consequently, they are functions of
\eqn{
\begin{gathered}
u=\frac{(\ee{1}{2})(\ee{3}{4})}{(\ee{1}{3})(\ee{2}{4})},\qquad\qquad v=\frac{(\ee{1}{4})(\ee{2}{3})}{(\ee{1}{3})(\ee{2}{4})},\\
\bar{\eta}_1^A=\frac{(\ee{2}{4})^{\frac{1}{2}}\eta_1^A}{(\ee{1}{2})^{\frac{1}{2}}(\ee{1}{4})^{\frac{1}{2}}},\qquad\qquad\bar{\eta}_2^A=\frac{(\ee{1}{4})^{\frac{1}{2}}\eta_2^A}{(\ee{1}{2})^{\frac{1}{2}}(\ee{2}{4})^{\frac{1}{2}}},\\
\bar{\eta}_3^A=\frac{(\ee{1}{4})^{\frac{1}{2}}\eta_3^A}{(\ee{1}{3})^{\frac{1}{2}}(\ee{3}{4})^{\frac{1}{2}}},\qquad\qquad\bar{\eta}_4^A=\frac{(\ee{1}{3})^{\frac{1}{2}}\eta_4^A}{(\ee{1}{4})^{\frac{1}{2}}(\ee{3}{4})^{\frac{1}{2}}}.
\end{gathered}
}[Eqetab]
The form of \eqref{Eq4ptCB} stems from the simultaneous action of two OPEs \eqref{EqOPE} on the resulting two-point correlation function \eqref{Eq2pt}.

We now again consider the conformal block corresponding to the exchange of an arbitrary quasi-primary operator in an infinite tower in the irreducible representations $\bs{N}_m\to\bs{N}_m+\ell\bs{e}_1$ with the tensor structures \eqref{EqTSOPEbasis}.  Upon decomposing the hatted projectors as
\eqn{\h{\m{P}}_{23}^{\bs{N}_m+\ell\bs{e}_1}=\sum_t\mathscr{A}_t(d,\ell)\left(\frac{\eta_2\cdot\Gamma\,\eta_3\cdot\Gamma}{2\ee{2}{3}}\right)^{2\xi_m}\h{\m{Q}}_{23|t}^{\bs{N}_m+\ell_t\bs{e}_1}\h{\m{P}}_{23}^{(\ell-\ell_t)\bs{e}_1},}[EqPExp]
we find that the conformal blocks are given by
\eqna{
&\mathscr{G}_{(a|b)}^{ij|m+\ell|lk}(\eta_1,\eta_2,\eta_4,\eta_3)\\
&\qquad=(\h{\m{P}}_{12}^{\bs{N}_i})(\h{\m{P}}_{21}^{\bs{N}_j})(\h{\m{P}}_{43}^{\bs{N}_l})(\h{\m{P}}_{34}^{\bs{N}_k})\cdot\tCF{(a|}{ij,m^{CR}+i_a}{12}\,\tCF{(b|}{lk,m+i_b}{43}\cdot\frac{(\ee{1}{4})^{\frac{1}{2}(\chi_i-\chi_j-\chi_k+\chi_l)}}{(\ee{1}{3})^{\frac{1}{2}(-\chi_k+\chi_l)}(\ee{2}{4})^{\frac{1}{2}(\chi_i-\chi_j)}}\\
&\qquad\phantom{=}\times(\ee{1}{2})^{\frac{1}{2}\chi_{m+\ell}}(\ee{3}{4})^{\frac{1}{2}\chi_{m+\ell}}\sum_t\mathscr{A}_t(d,\ell)(\m{A}_{12})^{\ell-i_a}(\m{A}_{34})^{\ell-i_b}\\
&\qquad\phantom{=}\times\m{D}_{12}^{(d,h_{ij,m+\ell}-n_a/2,n_a)}\m{D}_{43}^{(d,h_{lk,m+\ell}-n_b/2,n_b)}\left(\frac{C_\Gamma^T\eta_2\cdot\Gamma\,\eta_3\cdot\Gamma}{2\ee{2}{3}}\right)^{2\xi_m}\\
&\qquad\phantom{=}\times(\m{A}_{12})^{n_v^m+\ell_t}\h{\m{Q}}_{23|t}^{\bs{N}_m+\ell_t\bs{e}_1}(\m{A}_{34})^{n_v^m+\ell_t}\frac{(\m{A}_{12})^{\ell-\ell_t}\h{\m{P}}_{23}^{(\ell-\ell_t)\bs{e}_1}(\m{A}_{34})^{\ell-\ell_t}}{(\ee{2}{3})^{\chi_{m+\ell}}},
}
or
\eqna{
&(\mathscr{G}_{(a|b)}^{ij|m+\ell|lk})_{\{aA\}\{bB\}\{dD\}\{cC\}}(\eta_1,\eta_2,\eta_4,\eta_3)\\
&\qquad=\left((\h{\m{P}}_{12}^{\bs{N}_i})(\h{\m{P}}_{21}^{\bs{N}_j})\cdot\tCF{(a|}{ij,m^{CR}+i_a}{12}\left(\frac{C_\Gamma^T\Gamma\,\Gamma'}{2}\right)^{2\xi_m}(g)^{n_v^m+i_a}\right)_{\{aA\}\{bB\}}^{\phantom{\{aA\}\{bB\}}\{Ee\}\{F\}}\\
&\qquad\phantom{=}\times\left((\h{\m{P}}_{43}^{\bs{N}_l})(\h{\m{P}}_{34}^{\bs{N}_k})\cdot\tCF{(b|}{lk,m+i_b}{43}\right)_{\{dD\}\{cC\}\{e'E'\}}^{\phantom{\{dD\}\{cC\}\{e'E'\}}\{F'\}}\frac{(\ee{1}{4})^{\frac{1}{2}(\chi_i-\chi_j-\chi_k+\chi_l)}}{(\ee{1}{3})^{\frac{1}{2}(-\chi_k+\chi_l)}(\ee{2}{4})^{\frac{1}{2}(\chi_i-\chi_j)}}\\
&\qquad\phantom{=}\times(\ee{1}{2})^{\frac{1}{2}\chi_{m+\ell}+\xi_m}(\ee{3}{4})^{\frac{1}{2}\chi_{m+\ell}+\xi_m}\sum_t\mathscr{A}_t(d,\ell)(\m{A}_{12}^{EF})^{\ell-i_a}(\m{A}_{34E'}^{\phantom{34E'}F'})^{\ell-i_b}\\
&\qquad\phantom{=}\times\m{D}_{12\{F\}}^{(d,h_{ij,m+\ell}-n_a/2-2\xi_m,n_a+2\xi_m)}\m{D}_{43\{F'\}}^{(d,h_{lk,m+\ell}-n_b/2-2\xi_m,n_b+2\xi_m)}\\
&\qquad\phantom{=}\times[(\m{A}_{12})^{n_v^m+\ell_t}\h{\m{Q}}_{23|t}^{\bs{N}_m+\ell_t\bs{e}_1}(\m{A}_{34})^{n_v^m+\ell_t}]_{\{eE\}}^{\phantom{\{eE\}}\{E'e'\}}\frac{[(\m{A}_{12})^{\ell-\ell_t}\h{\m{P}}_{23}^{(\ell-\ell_t)\bs{e}_1}(\m{A}_{34})^{\ell-\ell_t}]_{\{E\}}^{\phantom{\{E\}}\{E'\}}}{(\ee{2}{3})^{\chi_{m+\ell}+2\xi_m}},
}
where the uncontracted $\Gamma$-matrix carries an $F$-index while the uncontracted $\Gamma'$-matrix carries an $F'$-index.

For convenience, let us denote the $\bs{N}_m$ indices as $E_s$ ($E'_s$) and separate the $\ell\bs{e}_1$ $E$-indices ($E'$-indices) into two individually symmetrized groups $\{E^{i_a}\}$ ($\{E'^{i_b}\}$) and $\{E_c^{\ell-i_a}\}$ ($\{E'_c{}^{\ell-i_b}\}$), where the second group is directly contracted with the $(\m{A}_{12}^{E_cF})^{\ell-i_a}$ ($(\m{A}_{34}{}_{E'_c}{}^{F'})^{\ell-i_a}$).  This leads to
\eqna{
&(\mathscr{G}_{(a|b)}^{ij|m+\ell|lk})_{\{aA\}\{bB\}\{dD\}\{cC\}}(\eta_1,\eta_2,\eta_4,\eta_3)\\
&\qquad=\left((\h{\m{P}}_{12}^{\bs{N}_i})(\h{\m{P}}_{21}^{\bs{N}_j})\cdot\tCF{(a|}{ij,m^{CR}+i_a}{12}\left(\frac{C_\Gamma^T\Gamma\,\Gamma'}{2}\right)^{2\xi_m}(g)^{n_v^m+i_a}\right)_{\{aA\}\{bB\}}^{\phantom{\{aA\}\{bB\}}\{E^{i_a}E_s^{n_v^m}e\}\{F\}}\\
&\qquad\phantom{=}\times\left((\h{\m{P}}_{43}^{\bs{N}_l})(\h{\m{P}}_{34}^{\bs{N}_k})\cdot\tCF{(b|}{lk,m+i_b}{43}(C_\Gamma^T)^{2\xi_m}(g)^{n_v^m+i_b}\right)_{\{dD\}\{cC\}}^{\phantom{\{dD\}\{cC\}}\{E'^{i_b}E_s'^{n_v^m}e'\}\{F'\}}\\
&\qquad\phantom{=}\times\frac{(\ee{1}{4})^{\frac{1}{2}(\chi_i-\chi_j-\chi_k+\chi_l)}}{(\ee{1}{3})^{\frac{1}{2}(-\chi_k+\chi_l)}(\ee{2}{4})^{\frac{1}{2}(\chi_i-\chi_j)}}(\ee{1}{2})^{\frac{1}{2}\chi_{m+\ell}+\xi_m}(\ee{3}{4})^{\frac{1}{2}\chi_{m+\ell}+\xi_m}\\
&\qquad\phantom{=}\times\sum_t\sum_{\substack{j_a,j_b\\k_a,k_b}}\mathscr{A}_t(d,\ell)\left[-\frac{\eta_2^{E_c}}{(\ee{1}{2})^{\frac{1}{2}}}\right]^{\ell-i_a}\left[-\frac{\eta_3^{E_c'}}{(\ee{3}{4})^{\frac{1}{2}}}\right]^{\ell-i_b}\Sym_{\substack{\{E_c\},\{E'_c\}\\\{E\},\{E'\}}}\\
&\qquad\phantom{=}\times\genfrac{(}{)}{0pt}{0}{i_a}{j_a+k_a}\frac{(-\ell_t-p_t-q'_t+r_t-r'_t)_{i_a-j_a-k_a}(-\ell+\ell_t+p_t+q'_t+r'_t)_{j_a}(-r_t)_{k_a}}{(-\ell)_{i_a}}\\
&\qquad\phantom{=}\times\genfrac{(}{)}{0pt}{0}{i_b}{j_b+k_b}\frac{(-\ell_t-p'_t-q_t+r'_t-r_t)_{i_b-j_b-k_b}(-\ell+\ell_t+p'_t+q_t+r_t)_{j_b}(-r'_t)_{k_b}}{(-\ell)_{i_b}}\\
&\qquad\phantom{=}\times\m{D}_{12\{F\}}^{(d,h_{ij,m+\ell}-n_a/2-2\xi_m+\ell-i_a,n_a+2\xi_m-\ell+i_a)}\m{D}_{43\{F'\}}^{(d,h_{lk,m+\ell}-n_b/2-2\xi_m+\ell-i_b,n_b+2\xi_m-\ell+i_b)}\\
&\qquad\phantom{=}\times[(\m{A}_{12})^{n_v^m+\ell_t}\h{\m{Q}}_{23|t}^{\bs{N}_m+\ell_t\bs{e}_1}(C_\Gamma^{-1})^{2\xi_m}(\m{A}_{34})^{n_v^m+\ell_t}]_{\substack{\{eE_s^{n_v^m-p_t-q_t}(E_c^{\ell_t+p_t+q'_t-r_t+r'_t-i_a+j_a+k_a}E^{i_a-j_a-k_a})\}\\\{(E'^{i_b-j_b-k_b}E_c'^{\ell_t+p'_t+q_t-r'_t+r_t-i_b+j_b+k_b})E_s'^{n_v^m-p'_t-q'_t}e'\}}}\\
&\qquad\phantom{=}\times\frac{[(\m{A}_{12})^{\ell-\ell_t}\h{\m{P}}_{23}^{(\ell-\ell_t)\bs{e}_1}(\m{A}_{34})^{\ell-\ell_t}]_{\substack{\{E^{j_a}E_c^{\ell-\ell_t-p_t-q'_t-r'_t-j_a}E'^{k_b}E_c'^{r'_t-k_b}E_s^{p_t}E_s'^{q'_t}\}\\\{E_s^{q_t}E_s'^{p'_t}E_c^{r_t-k_a}E^{k_a}E_c'^{\ell-\ell_t-p'_t-q_t-r_t-j_b}E'^{j_b}\}}}}{(\ee{2}{3})^{\chi_{m+\ell}+2\xi_m}},
}
where $p_t$ ($p'_t$) is the number of ``special'' $\bs{N}_m$ $E_s$-indices ($E'_s$-indices) in the unprime (prime) slot of $\h{\m{P}}_{23}^{(\ell-\ell_t)\bs{e}_1}$, $q_t$ ($q'_t$) is the number of ``special'' $\bs{N}_m$ $E_s$-indices ($E'_s$-indices) in the prime (unprime) slot of $\h{\m{P}}_{23}^{(\ell-\ell_t)\bs{e}_1}$ and $r_t$ ($r'_t$) is the number of $\ell\bs{e}_1$ $E$-indices ($E'$-indices) in the wrong slot of $\h{\m{P}}_{23}^{(\ell-\ell_t)\bs{e}_1}$.  This can now be re-formulated in terms of well-defined objects as
\eqna{
&(\mathscr{G}_{(a|b)}^{ij|m+\ell|lk})_{\{aA\}\{bB\}\{dD\}\{cC\}}(\eta_1,\eta_2,\eta_4,\eta_3)\\
&\qquad=\sum_t\sum_{\substack{j_a,j_b\\k_a,k_b}}\mathscr{A}_t(d,\ell)\genfrac{(}{)}{0pt}{0}{i_a}{j_a+k_a}\frac{(-\ell_t-p_t-q'_t+r_t-r'_t)_{i_a-j_a-k_a}(-\ell+\ell_t+p_t+q'_t+r'_t)_{j_a}(-r_t)_{k_a}}{(-\ell)_{i_a}}\\
&\qquad\phantom{=}\times\genfrac{(}{)}{0pt}{0}{i_b}{j_b+k_b}\frac{(-\ell_t-p'_t-q_t+r'_t-r_t)_{i_b-j_b-k_b}(-\ell+\ell_t+p'_t+q_t+r_t)_{j_b}(-r'_t)_{k_b}}{(-\ell)_{i_b}}\\
&\qquad\phantom{=}\times\left((\h{\m{P}}_{12}^{\bs{N}_i})(\h{\m{P}}_{21}^{\bs{N}_j})\cdot\tCF{(a|}{ij,m^{CR}+i_a}{12}\left(\frac{C_\Gamma^T\Gamma\,\Gamma'}{2}\right)^{2\xi_m}(g)^{n_v^m+i_a}\right)_{\{aA\}\{bB\}}^{\phantom{\{aA\}\{bB\}}\{(E^{i_a})E_s^{n_v^m}e\}\{F\}}\\
&\qquad\phantom{=}\times\left((\h{\m{P}}_{43}^{\bs{N}_l})(\h{\m{P}}_{34}^{\bs{N}_k})\cdot\tCF{(b|}{lk,m+i_b}{43}(C_\Gamma^T)^{2\xi_m}(g)^{n_v^m+i_b}\right)_{\{dD\}\{cC\}}^{\phantom{\{dD\}\{cC\}}\{(E'^{i_b})E_s'^{n_v^m}e'\}\{F'\}}\\
&\qquad\phantom{=}\times(-\bar{\eta}_2^{E_c})^{\ell_t+p_t+q'_t+r'_t-i_a+j_a}(-\bar{\eta}_3^{E_c'})^{\ell_t+p'_t+q_t+r_t-i_b+j_b}\\
&\qquad\phantom{=}\times[(\m{A}_{12})^{n_v^m+\ell_t}\h{\m{Q}}_{23|t}^{\bs{N}_m+\ell_t\bs{e}_1}(C_\Gamma^{-1})^{2\xi_m}(\m{A}_{34})^{n_v^m+\ell_t}]_{\substack{\{eE_s^{n_v^m-p_t-q_t}(E_c^{\ell_t+p_t+q'_t-r_t+r'_t-i_a+j_a+k_a}E^{i_a-j_a-k_a})\}\\\{(E'^{i_b-j_b-k_b}E_c'^{\ell_t+p'_t+q_t-r'_t+r_t-i_b+j_b+k_b})E_s'^{n_v^m-p'_t-q'_t}e'\}}}\\
&\qquad\phantom{=}\circ\frac{\m{G}_{(\chi_{m+\ell}+2\xi_m,\ell-\ell_t)\{F\}\{E^{j_a}E'^{k_b}E_c'^{r'_t-k_b}E_s^{p_t}E_s'^{q'_t}\}\{F'\}\{E'^{j_b}E^{k_a}E_c^{r_t-k_a}E_s'^{p'_t}E_s^{q_t}\}}^{(d,\chi_{m+\ell}/2+h_{ij,m+\ell}+\ell_t/2-i_a/2,n_a+2\xi_m-\ell+i_a,j_a+p_t+q'_t+r'_t,\chi_{m+\ell}/2+h_{lk,m+\ell}+\ell_t/2-i_b/2,n_b+2\xi_m-\ell+i_b,j_b+p'_t+q_t+r_t)}}{c_{(d,\ell-\ell_t)}R_{(\chi_{m+\ell}+2\xi_m,\ell-\ell_t)}^{(d,\chi_{m+\ell}/2+h_{ij,m+\ell}+\ell_t/2-i_a/2,n_a+2\xi_m-\ell+i_a)}R_{(\chi_{m+\ell}+2\xi_m,\ell-\ell_t)}^{(d,\chi_{m+\ell}/2+h_{lk,m+\ell}+\ell_t/2-i_b/2,n_b+2\xi_m-\ell+i_b)}},
}[Eq4ptCBSoln]
where the $\circ$-product between $(\m{A}_{12})\h{\m{Q}}_{23}(\m{A}_{34})$ and $\m{G}$ is a special product whose action is described below.  Here, the object
\eqna{
&\m{G}_{(\Delta,\ell)\{F\}\{E\}\{F'\}\{E'\}}^{(d,\Lambda,n,m,\Lambda',n',m')}(\eta_1,\eta_2,\eta_4,\eta_3)\\
&\qquad=c_{(d,\ell)}R_{(\Delta,\ell)}^{(d,\Lambda,n)}R_{(\Delta,\ell)}^{(d,\Lambda',n')}\frac{(\ee{2}{3})^{\Lambda+m/2}}{(\ee{1}{3})^{\Lambda-\Delta/2+m/2}}\frac{(\ee{2}{3})^{\Lambda'+m'/2}}{(\ee{2}{4})^{\Lambda'-\Delta/2+m'/2}}\\
&\qquad\phantom{=}\times u^{\Delta/2}v^{-\Lambda-m/2-\Lambda'-m'/2}\left[-\frac{\eta_2^E}{(\ee{1}{2})^{\frac{1}{2}}}\right]^{\ell-m}\left[-\frac{\eta_3^{E'}}{(\ee{3}{4})^{\frac{1}{2}}}\right]^{\ell-m'}\\
&\qquad\phantom{=}\times\m{D}_{12\{F\}}^{(d,\Lambda-\Delta/2+\ell/2-n/2,n)}\m{D}_{43\{F'\}}^{(d,\Lambda'-\Delta/2+\ell/2-n'/2,n')}\frac{[(\m{A}_{12})^\ell\h{\m{P}}_{23}^{\ell\bs{e}_1}(\m{A}_{34})^\ell]_{\{E\}\{E'\}}}{(\ee{2}{3})^\Delta},
}[EqG4p]
can be interpreted as the tensorial generalization of the four-point scalar block with spin-$\ell$ for symmetric-traceless exchange, since for $n=m=n'=m'=0$ \eqref{EqG4p} corresponds to the standard four-point scalar block for spin-$\ell$ exchange.  We remark that the explicit contractions between $(-\bar{\eta}_2^{E_c})^{q_t-i_a+j_a}$ or $(-\bar{\eta}_3^{E_c'})^{q'_t-i_b+j_b}$ and the special part of the hatted projection operator in \eqref{Eq4ptCBSoln} cannot be performed due to the $\circ$-product between the last two lines (in fact, if they could, the blocks would vanish identically).

The $\circ$-product between the last two lines of \eqref{Eq4ptCBSoln}, which includes products with $\m{A}_{12}\cdot\m{A}_{23}\cdot\m{A}_{34}$ for non-trace terms and $\m{A}_{12}\m{A}_{34}$ for trace terms from the projection operator, acts repeatedly as in
\eqna{
&(\m{A}_{12}\cdot\m{A}_{23}\cdot\m{A}_{34})_{XX'}\circ\m{G}_{(\Delta,\ell)\{F\}\{E\}\{F'\}\{E'\}}^{(d,\Lambda,n,m,\Lambda',n',m')}\\
&\qquad=g_{XX'}\m{G}_{(\Delta,\ell)\{F\}\{E\}\{F'\}\{E'\}}^{(d,\Lambda,n,m,\Lambda',n',m')}-\bar{\eta}_{1X'}\frac{\omega_{(\Delta,\ell)}^{(d,\Lambda,n,\Lambda',n')}}{\omega_{(\Delta,\ell)}^{(d,\Lambda-1/2,n+1,\Lambda',n')}}\m{G}_{(\Delta,\ell)\{FX\}\{E\}\{F'\}\{E'\}}^{(d,\Lambda-1/2,n+1,m,\Lambda',n',m')}\\
&\qquad\phantom{=}-\bar{\eta}_{4X}\frac{\omega_{(\Delta,\ell)}^{(d,\Lambda,n,\Lambda',n')}}{\omega_{(\Delta,\ell)}^{(d,\Lambda,n,\Lambda'-1/2,n'+1)}}\m{G}_{(\Delta,\ell)\{F\}\{E\}\{F'X'\}\{E'\}}^{(d,\Lambda,n,m,\Lambda'-1/2,n'+1,m')}\\
&\qquad\phantom{=}+\bar{\eta}_1^{F'}\frac{\omega_{(\Delta,\ell)}^{(d,\Lambda,n,\Lambda',n')}}{\omega_{(\Delta+1,\ell)}^{(d,\Lambda-1/2,n+2,\Lambda',n'+1)}}\m{G}_{(\Delta+1,\ell)\{FXX'\}\{E\}\{F'\}\{E'\}}^{(d,\Lambda-1/2,n+2,m,\Lambda',n'+1,m')}\\
&\qquad\phantom{=}+\bar{\eta}_4^F\frac{\omega_{(\Delta,\ell)}^{(d,\Lambda,n,\Lambda',n')}}{\omega_{(\Delta+1,\ell)}^{(d,\Lambda,n+1,\Lambda'-1/2,n'+2)}}\m{G}_{(\Delta+1,\ell)\{F\}\{E\}\{F'XX'\}\{E'\}}^{(d,\Lambda,n+1,m,\Lambda'-1/2,n'+2,m')}\\
&\qquad\phantom{=}-\frac{\omega_{(\Delta,\ell)}^{(d,\Lambda,n,\Lambda',n')}}{\omega_{(\Delta+1,\ell)}^{(d,\Lambda,n+1,\Lambda',n'+1)}}\m{G}_{(\Delta+1,\ell)\{FX'\}\{E\}\{F'X\}\{E'\}}^{(d,\Lambda,n+1,m,\Lambda',n'+1,m')}\\
&\qquad\phantom{=}+\frac{1}{u^{\frac{1}{2}}}\frac{\omega_{(\Delta,\ell)}^{(d,\Lambda,n,\Lambda',n')}}{\omega_{(\Delta,\ell)}^{(d,\Lambda-1/2,n+1,\Lambda'-1/2,n'+1)}}\m{G}_{(\Delta,\ell)\{FX\}\{E\}\{F'X'\}\{E'\}}^{(d,\Lambda-1/2,n+1,m,\Lambda'-1/2,n'+1,m')}\\
&\qquad\phantom{=}-\bar{\eta}_1^{F'}\bar{\eta}_4^F\frac{\omega_{(\Delta,\ell)}^{(d,\Lambda,n,\Lambda',n')}}{\omega_{(\Delta+1,\ell)}^{(d,\Lambda-1/2,n+2,\Lambda'-1/2,n'+2)}}\m{G}_{(\Delta+1,\ell)\{FX\}\{E\}\{F'X'\}\{E'\}}^{(d,\Lambda-1/2,n+2,m,\Lambda'-1/2,n'+2,m')},
}[EqCircProd]
\eqna{
&\m{A}_{12XY}\m{A}_{34X'Y'}\circ\m{G}_{(\Delta,\ell)\{F\}\{E\}\{F'\}\{E'\}}^{(d,\Lambda,n,m,\Lambda',n',m')}\\
&\qquad=g_{XY}g_{X'Y'}\m{G}_{(\Delta,\ell)\{F\}\{E\}\{F'\}\{E'\}}^{(d,\Lambda,n,m,\Lambda',n',m')}-\bar{\eta}_{1X}g_{X'Y'}\frac{\omega_{(\Delta,\ell)}^{(d,\Lambda,n,\Lambda',n')}}{\omega_{(\Delta,\ell)}^{(d,\Lambda-1/2,n+1,\Lambda',n')}}\m{G}_{(\Delta,\ell)\{FY\}\{E\}\{F'\}\{E'\}}^{(d,\Lambda-1/2,n+1,m,\Lambda',n',m')}\\
&\qquad\phantom{=}-\bar{\eta}_{1Y}g_{X'Y'}\frac{\omega_{(\Delta,\ell)}^{(d,\Lambda,n,\Lambda',n')}}{\omega_{(\Delta,\ell)}^{(d,\Lambda-1/2,n+1,\Lambda',n')}}\m{G}_{(\Delta,\ell)\{FX\}\{E\}\{F'\}\{E'\}}^{(d,\Lambda-1/2,n+1,m,\Lambda',n',m')}\\
&\qquad\phantom{=}-g_{XY}\bar{\eta}_{4X'}\frac{\omega_{(\Delta,\ell)}^{(d,\Lambda,n,\Lambda',n')}}{\omega_{(\Delta,\ell)}^{(d,\Lambda,n,\Lambda'-1/2,n'+1)}}\m{G}_{(\Delta,\ell)\{F\}\{E\}\{F'Y'\}\{E'\}}^{(d,\Lambda,n,m,\Lambda'-1/2,n'+1,m')}\\
&\qquad\phantom{=}-g_{XY}\bar{\eta}_{4Y'}\frac{\omega_{(\Delta,\ell)}^{(d,\Lambda,n,\Lambda',n')}}{\omega_{(\Delta,\ell)}^{(d,\Lambda,n,\Lambda'-1/2,n'+1)}}\m{G}_{(\Delta,\ell)\{F\}\{E\}\{F'X'\}\{E'\}}^{(d,\Lambda,n,m,\Lambda'-1/2,n'+1,m')}\\
&\qquad\phantom{=}+\bar{\eta}_{1X}\bar{\eta}_{4X'}\frac{\omega_{(\Delta,\ell)}^{(d,\Lambda,n,\Lambda',n')}}{\omega_{(\Delta,\ell)}^{(d,\Lambda-1/2,n+1,\Lambda'-1/2,n'+1)}}\m{G}_{(\Delta,\ell)\{FY\}\{E\}\{F'Y'\}\{E'\}}^{(d,\Lambda-1/2,n+1,m,\Lambda'-1/2,n'+1,m')}\\
&\qquad\phantom{=}+\bar{\eta}_{1X}\bar{\eta}_{4Y'}\frac{\omega_{(\Delta,\ell)}^{(d,\Lambda,n,\Lambda',n')}}{\omega_{(\Delta,\ell)}^{(d,\Lambda-1/2,n+1,\Lambda'-1/2,n'+1)}}\m{G}_{(\Delta,\ell)\{FY\}\{E\}\{F'X'\}\{E'\}}^{(d,\Lambda-1/2,n+1,m,\Lambda'-1/2,n'+1,m')}\\
&\qquad\phantom{=}+\bar{\eta}_{1Y}\bar{\eta}_{4X'}\frac{\omega_{(\Delta,\ell)}^{(d,\Lambda,n,\Lambda',n')}}{\omega_{(\Delta,\ell)}^{(d,\Lambda-1/2,n+1,\Lambda'-1/2,n'+1)}}\m{G}_{(\Delta,\ell)\{FX\}\{E\}\{F'Y'\}\{E'\}}^{(d,\Lambda-1/2,n+1,m,\Lambda'-1/2,n'+1,m')}\\
&\qquad\phantom{=}+\bar{\eta}_{1Y}\bar{\eta}_{4Y'}\frac{\omega_{(\Delta,\ell)}^{(d,\Lambda,n,\Lambda',n')}}{\omega_{(\Delta,\ell)}^{(d,\Lambda-1/2,n+1,\Lambda'-1/2,n'+1)}}\m{G}_{(\Delta,\ell)\{FX\}\{E\}\{F'X'\}\{E'\}}^{(d,\Lambda-1/2,n+1,m,\Lambda'-1/2,n'+1,m')},
}
where we introduced
\eqn{\omega_{(\Delta,\ell)}^{(d,\Lambda,n,\Lambda',n')}=c_{(d,\ell)}R_{(\Delta,\ell)}^{(d,\Lambda,n)}R_{(\Delta,\ell)}^{(d,\Lambda',n')},}[Eqomega]
to simplify the notation.

The action of the product in these equations is derived directly from the expansion of the $\m{A}$-metrics in terms of the relevant embedding space coordinates, while taking into account that the latter are inside the OPE differential operators.

%%%%%%%%%%%%%%%%%%%%%%%%%%%%%%%%%%%%%%%%%%%%%%%%%%
%%%%%%%%%%%%%%%%%%%%%%%%%%%%%%%%%%%%%%%%%%%%%%%%%%

\section{Tensorial Generalizations of Scalar Blocks with Spins}\label{SecTensorGeneralizations}

In this section, we study the three- and four-point tensorial generalizations of the scalar conformal blocks for symmetric-traceless exchange in embedding space introduced in the previous section.  It emerges that several important properties of three- and four-point scalar blocks for spin-$\ell$ exchange conveniently carry over to their tensorial generalizations.  We consider the three-point and four-tensorial blocks in turn.

%%%%%%%%%%%%%%%%%%%%%%%%%%%%%%%%%%%%%%%%%%%%%%%%%%

\subsection{Three-Point Tensorial Blocks}

As discussed above, the tensorial generalization of the scalar-scalar-(spin-$\ell$) three-point conformal blocks \eqref{EqG3p} is given by
\eqna{
&\m{G}_{(\Delta,\ell)\{E\}\{F\}}^{(d,\Lambda,n)}(\eta_1,\eta_2,\eta_3)\\
&\qquad=R_{(\Delta,\ell)}^{(d,\Lambda,n)}\frac{(\ee{2}{3})^{\Lambda+\Delta/2}}{(\ee{1}{3})^{\Lambda-\Delta/2}}\left[-\frac{\eta_2^{E'}}{(\ee{1}{2})^{\frac{1}{2}}}\right]^\ell\m{D}_{12\{F\}}^{(d,\Lambda-\Delta/2+\ell/2-n/2,n)}\frac{[\h{\m{P}}_{31}^{\ell\bs{e}_1}(\m{A}_{23}\cdot\m{A}_{12})^\ell]_{\{E\}\{E'\}}}{(\ee{2}{3})^\Delta}.
}[EqG3]
This object has $\ell$ $E$-indices and $n$ $F$-indices, where each set is separately fully symmetrized.  The $E$-indices denote the usual set of indices carried by the spin-$\ell$ quasi-primary operator in the $\Vev{SS\m{O}^{(\ell)}}$ block, while the $F$-indices are the extra indices that lead to the tensorial generalization.  In this language, the tensorial generalization corresponding to the conformal block for spin-$\ell$ exchange has no $F$-indices, while the associated quantity corresponding to some nontrivial spinning exchange would necessarily feature $F$-indices.  Here $R_{(\Delta,\ell)}^{(d,\Lambda,n)}$ is a normalization factor that represents the three-point analog of the rotation matrix between the OPE basis and the three-point basis, as discussed in the next section.

It emerges that the three-point tensorial block satisfies a host of interesting properties.  To begin with, we find that it verifies the following contiguous relations directly from its definition:
\eqn{
\begin{gathered}
g\cdot\m{G}_{(\Delta,\ell)}^{(d,\Lambda,n)}=\m{G}_{(\Delta,\ell)}^{(d,\Lambda,n)}\cdot g=0,\qquad\tilde{\eta}_1\cdot\m{G}_{(\Delta,\ell)}^{(d,\Lambda,n)}=\tilde{\eta}_3\cdot\m{G}_{(\Delta,\ell)}^{(d,\Lambda,n)}=0,\\
\m{G}_{(\Delta,\ell)}^{(d,\Lambda,n)}\cdot\tilde{\eta}_1=\frac{R_{(\Delta,\ell)}^{(d,\Lambda,n)}}{R_{(\Delta,\ell)}^{(d,\Lambda+1/2,n-1)}}\m{G}_{(\Delta,\ell)}^{(d,\Lambda+1/2,n-1)},\\
\m{G}_{(\Delta,\ell)}^{(d,\Lambda,n)}\cdot\tilde{\eta}_2=\rho^{(d,1;-\Lambda+\Delta/2-\ell/2-n/2)}\frac{R_{(\Delta,\ell)}^{(d,\Lambda,n)}}{R_{(\Delta,\ell)}^{(d,\Lambda-1/2,n-1)}}\m{G}_{(\Delta,\ell)}^{(d,\Lambda-1/2,n-1)},\\
\m{G}_{(\Delta,\ell)}^{(d,\Lambda,n)}\cdot\tilde{\eta}_3=\frac{R_{(\Delta,\ell)}^{(d,\Lambda,n)}}{R_{(\Delta-1,\ell)}^{(d,\Lambda,n-1)}}\m{G}_{(\Delta-1,\ell)}^{(d,\Lambda,n-1)},
\end{gathered}
}[EqG3CR]
where a left (right) contraction is a full contraction with $E$-indices ($F$-indices) and the homogeneized three-point embedding space coordinates are given by \eqref{Eqetat}.  We find that the remaining two contractions (in particular, with $\tilde{\eta}_2^E$ and $g^{EF}$) are more complicated to derive; however, they are not necessary, since the $E$-indices always contract with the half-projector $\m{T}_{31}$.

Following \cite{Fortin:2020ncr}, \eqref{EqG3} can be computed explicitly and gives
\eqna{
&\m{G}_{(\Delta,\ell)\{E\}\{F\}}^{(d,\Lambda,n)}(\eta_1,\eta_2,\eta_3)\\
&\qquad=\sum_{s_0,s_3,t\geq0}\frac{(-1)^{s_0}(-2)^{s_0-t}(-\ell)_{s_0+s_3}(-n)_{s_0+s_3}(-s_0)_t}{s_0!s_3!t!}\\
&\qquad\phantom{=}\times\frac{(-\Lambda+\Delta/2+\ell/2-n/2)_{s_0-t}(-\Lambda+\Delta/2-\ell/2-n/2+1-d/2)_{s_0-t}}{(\Lambda-\ell-n+2-d)_{s_0+s_3}}\\
&\qquad\phantom{=}\times[\m{\h{P}}_{31}^{\ell\bs{e}_1}(\m{A}_{12}\cdot\tilde{\eta}_3)^{\ell-s_0}(g)^{s_0}]_{\{E\}\{F\}}(\tilde{\eta}_{3F})^{s_3}\frac{\bar{I}_{12\{F\}}^{(d+2\ell,\Lambda-\Delta/2-\ell/2-n/2+s_3+t,n-s_0-s_3;\Delta+\ell)}}{\rho^{(d+2\ell,\Lambda-\Delta/2-\ell/2-n/2;\Delta+\ell)}},
}[EqG3Soln]
where the overall coefficient is chosen as
\eqn{\frac{1}{R_{(\Delta,\ell)}^{(d,\Lambda,n)}}=(-2)^\ell(-\Lambda+\Delta/2-\ell/2-n/2)_\ell(\Delta-\ell-n+2-d)_\ell\rho^{(d+2\ell,\Lambda-\Delta/2-\ell/2-n/2;\Delta+\ell)}.}[EqG3R]
The choice \eqref{EqG3R} leads to the usual three-point scalar block with unit coefficient.  Indeed, setting $n=0$ implies $\m{G}_{(\Delta,\ell)\{E\}}^{(d,\Lambda,0)}=[\m{\h{P}}_{31}^{\ell\bs{e}_1}(\m{A}_{12}\cdot\tilde{\eta}_3)^\ell]_{\{E\}}$ as expected.

Another identity for the three-point tensorial block \eqref{EqG3} comes from the property $\frac{\bar{I}_{21}^{(d,h,n;p)}}{\rho^{(d,h;p)}}=\frac{\bar{I}_{12}^{(d,-h-n-p,n;p)}}{\rho^{(d,-h-n-p;p)}}$ and the solution \eqref{EqG3Soln} which imply that it satisfies
\eqn{\m{G}_{(\Delta,\ell)\{E\}\{F\}}^{(d,-\Lambda,n)}(\eta_2,\eta_1,\eta_3)=(-1)^\ell(\m{A}_{23E}^{\phantom{23E}E'})^\ell\m{G}_{(\Delta,\ell)\{E'\}\{F\}}^{(d,\Lambda,n)}(\eta_1,\eta_2,\eta_3).}[EqG3Sym]
This is in agreement with the expected result for $\Vev{SS\m{O}^{(\ell)}}$.

For future convenience, it is advantageous to rewrite the three-point tensorial blocks as
\eqn{
\begin{gathered}
\m{G}_{(\Delta,\ell)\{E\}\{F\}}^{(d,\Lambda,n)}(\eta_1,\eta_2,\eta_3)=\sum_{\substack{s_0,\bs{q}\geq0\\\bar{q}=n-s_0}}\Omega_{(\Delta,\ell)}^{(d,\Lambda,n)}(s_0,\bs{q})[\m{\h{P}}_{31}^{\ell\bs{e}_1}(\m{A}_{12}\cdot\tilde{\eta}_3)^{\ell-s_0}(g)^{s_0}]_{\{E\}\{F\}}S_{(\bs{q})\{F\}},\\
S_{(\bs{q})\{F\}}=(g_{FF})^{q_0}(\tilde{\eta}_{1F})^{q_1}(\tilde{\eta}_{2F})^{q_2}(\tilde{\eta}_{3F})^{q_3},
\end{gathered}
}[EqG3Omega]
where the embedding space coordinates with vanishing degree of homogeneity in $S_{(\bs{q})}$ are built as expected.  Here the coefficients in \eqref{EqG3Omega} are given by
\eqna{
\Omega_{(\Delta,\ell)}^{(d,\Lambda,n)}(s_0,\bs{q})&=\frac{(-1)^{n-s_0}(-2)^{n-q_0}n!}{s_0!q_0!q_1!q_2!q_3!}(-\ell)_{s_0}F_{1,1,1}^{1,2,2}\left[\left.\begin{array}{c}a_1;c_1,c_2;f_1,f_2\\b_1;d_1;g_1\end{array}\right|1,1\right]\\
&\phantom{=}\qquad\times(\Lambda+\Delta/2+\ell/2-n/2)_{n-s_0-q_0-q_1}(-\Lambda+\Delta/2+\ell/2-n/2)_{n-q_0-q_2}\\
&\phantom{=}\qquad\times\frac{(-\Lambda+\Delta/2-\ell/2-n/2+1-d/2)_{s_0}(-\Delta+d/2)_{q_0+q_1+q_2}}{(\Lambda-\ell-n+2-d)_{s_0}},
}[EqOmega]
where the arguments of the Kamp\'e de F\'eriet function are
\eqn{
\begin{gathered}
a_1=\Lambda+\Delta/2-\ell/2-n/2+1-d/2,\qquad\qquad b_1=\Lambda-\Delta/2-\ell/2-n/2+q_0+q_2+1,\\
c_1=-\ell+s_0,\qquad\qquad c_2=-q_3,\qquad\qquad d_1=\Delta-\ell-n+s_0+2-d,\\
f_1=-s_0,\qquad\qquad f_2=\Lambda+\Delta/2+\ell/2+n/2-s_0-q_0-q_1,\\
g_1=\Lambda-\Delta/2+\ell/2+n/2-s_0+d/2.
\end{gathered}
}
We may then invoke \eqref{EqG3Omega} to determine an expansion that is useful for deriving contiguous relations of four-point tensorial blocks.  This is given by
\eqn{\m{G}_{(\Delta,\ell)\{E\}\{F\}}^{(d,\Lambda,n)}(\eta_1,\eta_2,\eta_3)=\sum_{\substack{\bs{q}\geq0\\\bar{q}\leq n}}c_{(\Delta,\ell)(\bs{q})}^{(d,\Lambda,n;n_1,n_2)}\m{G}_{(\Delta-n_1-n_2+q_3,\ell)\{E\}\{F\}}^{(d,\Lambda+n_1/2-n_2/2-q_1/2+q_2/2,n-\bar{q})}(\eta_1,\eta_2,\eta_3)S_{(\bs{q})\{F\}},}[EqG3Exp]
which leads to the relations
\eqn{\sum_{\substack{\bs{p}\geq0\\\bs{p}\leq\bs{q}\\\bar{q}=n-s_0}}c_{(\Delta,\ell)(\bs{p})}^{(d,\Lambda,n;n_1,n_2)}\Omega_{(\Delta-n_1-n_2+p_3,\ell)}^{(d,\Lambda+n_1/2-n_2/2-p_1/2+p_2/2,n-\bar{p})}(s_0,\bs{p})=\Omega_{(\Delta,\ell)}^{(d,\Lambda,n)}(s_0,\bs{q}),}[EqRRc]
for the coefficients in \eqref{EqG3Exp}.  We may accordingly apply these relations to recursively determine the coefficients by starting from $s_0=n$ and decreasing $s_0$ at each step.

%%%%%%%%%%%%%%%%%%%%%%%%%%%%%%%%%%%%%%%%%%%%%%%%%%

\subsection{Four-Point Tensorial Blocks}

We now follow the above analysis of the three-point case to write down analogous results for the four-point tensorial blocks \eqref{EqG4p}.  These are given by
\eqna{
&\m{G}_{(\Delta,\ell)\{F\}\{E\}\{F'\}\{E'\}}^{(d,\Lambda,n,m,\Lambda',n',m')}(\eta_1,\eta_2,\eta_4,\eta_3)\\
&\qquad=c_{(d,\ell)}R_{(\Delta,\ell)}^{(d,\Lambda,n)}R_{(\Delta,\ell)}^{(d,\Lambda',n')}\frac{(\ee{2}{3})^{\Lambda+m/2}}{(\ee{1}{3})^{\Lambda-\Delta/2+m/2}}\frac{(\ee{2}{3})^{\Lambda'+m'/2}}{(\ee{2}{4})^{\Lambda'-\Delta/2+m'/2}}\\
&\qquad\phantom{=}\times u^{\Delta/2}v^{-\Lambda-m/2-\Lambda'-m'/2}\left[-\frac{\eta_2^E}{(\ee{1}{2})^{\frac{1}{2}}}\right]^{\ell-m}\left[-\frac{\eta_3^{E'}}{(\ee{3}{4})^{\frac{1}{2}}}\right]^{\ell-m'}\\
&\qquad\phantom{=}\times\m{D}_{12\{F\}}^{(d,\Lambda-\Delta/2+\ell/2-n/2,n)}\m{D}_{43\{F'\}}^{(d,\Lambda'-\Delta/2+\ell/2-n'/2,n')}\frac{[(\m{A}_{12})^\ell\h{\m{P}}_{23}^{\ell\bs{e}_1}(\m{A}_{34})^\ell]_{\{E\}\{E'\}}}{(\ee{2}{3})^\Delta},
}[EqG4]
where the various sets of indices are symmetrized separately in groups.  As mentioned above, the four-point tensorial blocks can be seen as tensorial generalizations of the scalar blocks for spin-$\ell$ exchange, since for $n=m=n'=m'=0$ we find that they are directly related to the $\Vev{SSSS}_{\m{O}^{(\ell)}}$ blocks.  Here the $R$-normalization factors come from the three-point correlation functions (or the OPE-to-correlator computation) while $c_{(d,\ell)}$ is an extra normalization factor that is chosen to match with \cite{Dolan:2011dv} (the same is true for the choice of OPE and the powers of $u$ and $v$).

Directly from the definition \eqref{EqG4}, it is straightforward to verify the following contiguous relations for the four-point tensorial blocks with the $g$-metric
\eqn{
\begin{gathered}
g^{FF}\m{G}_{(\Delta,\ell)\{F\}\{E\}\{F'\}\{E'\}}^{(d,\Lambda,n,m,\Lambda',n',m')}=g^{EE}\m{G}_{(\Delta,\ell)\{F\}\{E\}\{F'\}\{E'\}}^{(d,\Lambda,n,m,\Lambda',n',m')}=0,\\
g^{F'F'}\m{G}_{(\Delta,\ell)\{F\}\{E\}\{F'\}\{E'\}}^{(d,\Lambda,n,m,\Lambda',n',m')}=g^{E'E'}\m{G}_{(\Delta,\ell)\{F\}\{E\}\{F'\}\{E'\}}^{(d,\Lambda,n,m,\Lambda',n',m')}=0,\\
g^{FE}\m{G}_{(\Delta,\ell)\{F\}\{E\}\{F'\}\{E'\}}^{(d,\Lambda,n,m,\Lambda',n',m')}=g^{F'E'}\m{G}_{(\Delta,\ell)\{F\}\{E\}\{F'\}\{E'\}}^{(d,\Lambda,n,m,\Lambda',n',m')}=0,\\
g^{FF'}\m{G}_{(\Delta,\ell)\{F\}\{E\}\{F'\}\{E'\}}^{(d,\Lambda,n,m,\Lambda',n',m')}=\frac{\omega_{(\Delta,\ell)}^{(d,\Lambda,n,\Lambda',n')}}{\omega_{(\Delta-1,\ell)}^{(d,\Lambda,n-1,\Lambda',n'-1)}}\m{G}_{(\Delta-1,\ell)\{F\}\{E\}\{F'\}\{E'\}}^{(d,\Lambda,n-1,m,\Lambda',n'-1,m')},
\end{gathered}
}[EqG4CRg1]
as well as
\eqna{
g^{FE'}\m{G}_{(\Delta,\ell)\{F\}\{E\}\{F'\}\{E'\}}^{(d,\Lambda,n,m,\Lambda',n',m')}&=\frac{1}{2(\Lambda'-\Delta/2+\ell/2+n'/2+1)(d+\ell-\Delta-1)}\\
&\phantom{=}\qquad\times\frac{\omega_{(\Delta,\ell)}^{(d,\Lambda,n,\Lambda',n')}}{\omega_{(\Delta-1,\ell)}^{(d,\Lambda,n-1,\Lambda'+1/2,n')}}\m{G}_{(\Delta-1,\ell)\{F\}\{E\}\{F'\}\{E'\}}^{(d,\Lambda,n-1,m,\Lambda'+1/2,n',m'-1)}\\
&\phantom{=}\qquad-\frac{n'}{d+\ell-\Delta-1}\frac{\omega_{(\Delta,\ell)}^{(d,\Lambda,n,\Lambda',n')}}{\omega_{(\Delta-1,\ell)}^{(d,\Lambda,n-1,\Lambda',n'-1)}}\m{G}_{(\Delta-1,\ell)\{F\}\{E\}\{(F'\}\{F')E'\}}^{(d,\Lambda,n-1,m,\Lambda',n'-1,m')},\\
g^{EF'}\m{G}_{(\Delta,\ell)\{F\}\{E\}\{F'\}\{E'\}}^{(d,\Lambda,n,m,\Lambda',n',m')}&=\frac{1}{2(\Lambda-\Delta/2+\ell/2+n/2+1)(d+\ell-\Delta-1)}\\
&\phantom{=}\qquad\times\frac{\omega_{(\Delta,\ell)}^{(d,\Lambda,n,\Lambda',n')}}{\omega_{(\Delta-1,\ell)}^{(d,\Lambda+1/2,n,\Lambda',n'-1)}}\m{G}_{(\Delta-1,\ell)\{F\}\{E\}\{F'\}\{E'\}}^{(d,\Lambda+1/2,n,m-1,\Lambda',n'-1,m')}\\
&\phantom{=}\qquad-\frac{n}{d+\ell-\Delta-1}\frac{\omega_{(\Delta,\ell)}^{(d,\Lambda,n,\Lambda',n')}}{\omega_{(\Delta-1,\ell)}^{(d,\Lambda,n-1,\Lambda',n'-1)}}\m{G}_{(\Delta-1,\ell)\{(F\}\{F)E\}\{F'\}\{E'\}}^{(d,\Lambda,n-1,m,\Lambda',n'-1,m')},
}
and
\eqna{
&g^{EE'}\mathscr{G}_{(\Delta,\ell)\{F\}\{E\}\{F'\}\{E'\}}^{(d,\Lambda,n,m,\Lambda',n',m')}\\
&\qquad=\frac{(d+\ell-\Delta-2)(d+\ell-3)(d+2\ell-2)}{\ell(d+\ell-\Delta-1)(d+2\ell-4)}\frac{\omega_{(\Delta,\ell)}^{(d,\Lambda,n,\Lambda',n')}}{\omega_{(\Delta,\ell-1)}^{(d,\Lambda+1/2,n,\Lambda'+1/2,n')}}\mathscr{G}_{(\Delta,\ell-1)\{F\}\{E\}\{F'\}\{E'\}}^{(d,\Lambda+1/2,n,m-1,\Lambda'+1/2,n',m'-1)}\\
&\qquad\phantom{=}+\frac{1}{4(d+\ell-\Delta-1)^2(\Lambda-\Delta/2+\ell/2+n/2+1)(\Lambda'-\Delta/2+\ell/2+n'/2+1)}\\
&\qquad\phantom{=}\times\frac{\omega_{(\Delta,\ell)}^{(d,\Lambda,n,\Lambda',n')}}{\omega_{(\Delta-1,\ell)}^{(d,\Lambda+1/2,n,\Lambda'+1/2,n')}}\mathscr{G}_{(\Delta-1,\ell)\{F\}\{E\}\{F'\}\{E'\}}^{(d,\Lambda+1/2,n,m-1,\Lambda'+1/2,n',m'-1)}\\
&\qquad\phantom{=}-\frac{n'}{2(d+\ell-\Delta-1)^2(\Lambda-\Delta/2+\ell/2+n/2+1)}\frac{\omega_{(\Delta,\ell)}^{(d,\Lambda,n,\Lambda',n')}}{\omega_{(\Delta-1,\ell)}^{(d,\Lambda+1/2,n,\Lambda',n'-1)}}\mathscr{G}_{(\Delta-1,\ell)\{F\}\{E\}\{(F'\}\{F')E'\}}^{(d,\Lambda+1/2,n,m-1,\Lambda',n'-1,m')}\\
&\qquad\phantom{=}-\frac{n}{2(d+\ell-\Delta-1)^2(\Lambda'-\Delta/2+\ell/2+n'/2+1)}\frac{\omega_{(\Delta,\ell)}^{(d,\Lambda,n,\Lambda',n')}}{\omega_{(\Delta-1,\ell)}^{(d,\Lambda,n-1,\Lambda'+1/2,n')}}\mathscr{G}_{(\Delta-1,\ell)\{(F\}\{F)E\}\{F'\}\{E'\}}^{(d,\Lambda,n-1,m,\Lambda'+1/2,n',m'-1)}\\
&\qquad\phantom{=}+\frac{nn'}{(d+\ell-\Delta-1)^2}\frac{\omega_{(\Delta,\ell)}^{(d,\Lambda,n,\Lambda',n')}}{\omega_{(\Delta-1,\ell)}^{(d,\Lambda,n-1,\Lambda',n'-1)}}\mathscr{G}_{(\Delta-1,\ell)\{(F\}\{F)E\}\{(F'\}\{F')E'\}}^{(d,\Lambda,n-1,m,\Lambda',n'-1,m')}.
}[EqG4CRg2]
We may accordingly obtain a variety of contiguous relations that arise from the contraction of this object with the embedding space coordinates.  These are
\eqn{
\begin{gathered}
\bar{\eta}_1^F\m{G}_{(\Delta,\ell)\{F\}\{E\}\{F'\}\{E'\}}^{(d,\Lambda,n,m,\Lambda',n',m')}=\frac{R_{(\Delta,\ell)}^{(d,\Lambda,n)}}{R_{(\Delta,\ell)}^{(d,\Lambda+1/2,n-1)}}\m{G}_{(\Delta,\ell)\{F\}\{E\}\{F'\}\{E'\}}^{(d,\Lambda+1/2,n-1,m,\Lambda',n',m')},\\
\bar{\eta}_2^F\m{G}_{(\Delta,\ell)\{F\}\{E\}\{F'\}\{E'\}}^{(d,\Lambda,n,m,\Lambda',n',m')}=\rho^{(d,1;-\Lambda+\Delta/2-\ell/2-n/2)}\frac{R_{(\Delta,\ell)}^{(d,\Lambda,n)}}{R_{(\Delta,\ell)}^{(d,\Lambda-1/2,n-1)}}\m{G}_{(\Delta,\ell)\{F\}\{E\}\{F'\}\{E'\}}^{(d,\Lambda-1/2,n-1,m,\Lambda',n',m')},\\
\bar{\eta}_1^E\m{G}_{(\Delta,\ell)\{F\}\{E\}\{F'\}\{E'\}}^{(d,\Lambda,n,m,\Lambda',n',m')}=0,\\
\bar{\eta}_2^E\m{G}_{(\Delta,\ell)\{F\}\{E\}\{F'\}\{E'\}}^{(d,\Lambda,n,m,\Lambda',n',m')}=-\m{G}_{(\Delta,\ell)\{F\}\{E\}\{F'\}\{E'\}}^{(d,\Lambda,n,m-1,\Lambda',n',m')},
\end{gathered}
}[EqG4CReta1]
as well as
\eqna{
&\bar{\eta}_3^F\m{G}_{(\Delta,\ell)\{F\}\{E\}\{F'\}\{E'\}}^{(d,\Lambda,n,m,\Lambda',n',m')}\\
&\qquad=\frac{R_{(\Delta,\ell)}^{(d,\Lambda,n)}}{R_{(\Delta,\ell)}^{(d,\Lambda+1/2,n-1)}}\frac{v}{u^{\frac{1}{2}}}\m{G}_{(\Delta,\ell)\{F\}\{E\}\{F'\}\{E'\}}^{(d,\Lambda+1/2,n-1,m,\Lambda',n',m')}+2(\Lambda-\Delta/2+\ell/2+n/2)\\
&\qquad\phantom{=}\times\left[-\frac{(\Lambda-\Delta/2-\ell/2+n/2+m-1+d/2)+(\Lambda+m/2-1/2)(u-v-1)}{u^{\frac{1}{2}}}+\bar{\eta}_3\cdot\bar{\m{D}}_{12}\right]\\
&\qquad\phantom{=}\times\frac{R_{(\Delta,\ell)}^{(d,\Lambda,n)}}{R_{(\Delta,\ell)}^{(d,\Lambda-1/2,n-1)}}\m{G}_{(\Delta,\ell)\{F\}\{E\}\{F'\}\{E'\}}^{(d,\Lambda-1/2,n-1,m,\Lambda',n',m')}\\
&\qquad\phantom{=}+2(\ell-m)(\Lambda-\Delta/2+\ell/2+n/2)\frac{R_{(\Delta,\ell)}^{(d,\Lambda,n)}}{R_{(\Delta,\ell)}^{(d,\Lambda-1/2,n-1)}}\bar{\eta}_3^E\m{G}_{(\Delta,\ell)\{F\}\{E\}\{F'\}\{E'\}}^{(d,\Lambda-1/2,n-1,m+1,\Lambda',n',m')},\\
&\bar{\eta}_4^F\m{G}_{(\Delta,\ell)\{F\}\{E\}\{F'\}\{E'\}}^{(d,\Lambda,n,m,\Lambda',n',m')}\\
&\qquad=\frac{R_{(\Delta,\ell)}^{(d,\Lambda,n)}}{R_{(\Delta,\ell)}^{(d,\Lambda+1/2,n-1)}}\frac{1}{u^{\frac{1}{2}}}\m{G}_{(\Delta,\ell)\{F\}\{E\}\{F'\}\{E'\}}^{(d,\Lambda+1/2,n-1,m,\Lambda',n',m')}+2(\Lambda-\Delta/2+\ell/2+n/2)\\
&\qquad\phantom{=}\times\left[\frac{(\Lambda+\Delta/2+\ell/2-n/2-d/2)}{u^{\frac{1}{2}}}+\bar{\eta}_4\cdot\bar{\m{D}}_{12}\right]\frac{R_{(\Delta,\ell)}^{(d,\Lambda,n)}}{R_{(\Delta,\ell)}^{(d,\Lambda-1/2,n-1)}}\m{G}_{(\Delta,\ell)\{F\}\{E\}\{F'\}\{E'\}}^{(d,\Lambda-1/2,n-1,m,\Lambda',n',m')}\\
&\qquad\phantom{=}+2(\ell-m)(\Lambda-\Delta/2+\ell/2+n/2)\frac{R_{(\Delta,\ell)}^{(d,\Lambda,n)}}{R_{(\Delta,\ell)}^{(d,\Lambda-1/2,n-1)}}\bar{\eta}_4^E\m{G}_{(\Delta,\ell)\{F\}\{E\}\{F'\}\{E'\}}^{(d,\Lambda-1/2,n-1,m+1,\Lambda',n',m')},
}[EqG4CReta2]
with
\eqna{
&\bar{\eta}_3^E\m{G}_{(\Delta,\ell)\{F\}\{E\}\{F'\}\{E'\}}^{(d,\Lambda,n,m,\Lambda',n',m')}\\
&\qquad=-\frac{1}{u^{\frac{1}{2}}}\m{G}_{(\Delta,\ell)\{F\}\{E\}\{F'\}\{E'\}}^{(d,\Lambda,n,m-1,\Lambda',n',m')}\\
&\qquad\phantom{=}+\sum_{i\geq0}^{m-1}\frac{(-1)^i(-m+1)_iR_{(\Delta,\ell)}^{(d,\Lambda,n)}}{(-2)^{i+1}(\ell-m+1)_{i+1}(\Lambda-\Delta/2+\ell/2+n/2+1)_{i+1}(-\Delta+n-1+d)_{i+1}}\\
&\qquad\phantom{=}\times\left\{\left[\frac{2(\Lambda'+m'/2)v+(\Lambda+m/2+\Lambda'+m'/2+1/2)(u-v-1)}{u^{\frac{1}{2}}}\right.\right.\\
&\qquad\phantom{=}\left.+\bar{\eta}_3\cdot\bar{\m{D}}_{21}\right]\frac{\m{G}_{(\Delta,\ell)\{F(E\}\{E)\}\{F'\}\{E'\}}^{(d,\Lambda+1+i/2,n+i,m-1-i,\Lambda',n',m')}}{R_{(\Delta,\ell)}^{(d,\Lambda+1+i/2,n+i)}}\\
&\qquad\phantom{=}-2(\Lambda-\Delta/2+\ell/2+n/2+i+1)(\Lambda+\Delta/2+\ell/2-n/2+1-d/2)\\
&\qquad\phantom{=}\left.\times\left[\frac{(\Lambda+m/2-1/2)(u-v-1)}{u^{\frac{1}{2}}}+\bar{\eta}_3\cdot\bar{\m{D}}_{12}\right]\frac{\m{G}_{(\Delta,\ell)\{F(E\}\{E)\}\{F'\}\{E'\}}^{(d,\Lambda+i/2,n+i,m-1-i,\Lambda',n',m')}}{R_{(\Delta,\ell)}^{(d,\Lambda+i/2,n+i)}}\right\},
}[EqG4CReta3]
and finally
\eqna{
&\bar{\eta}_4^E\m{G}_{(\Delta,\ell)\{F\}\{E\}\{F'\}\{E'\}}^{(d,\Lambda,n,m,\Lambda',n',m')}\\
&\qquad=-\frac{1}{u^{\frac{1}{2}}}\m{G}_{(\Delta,\ell)\{F\}\{E\}\{F'\}\{E'\}}^{(d,\Lambda,n,m-1,\Lambda',n',m')}\\
&\qquad\phantom{=}+\sum_{i\geq0}^{m-1}\frac{(-1)^i(-m+1)_iR_{(\Delta,\ell)}^{(d,\Lambda,n)}}{(-2)^{i+1}(\ell-m+1)_{i+1}(\Lambda-\Delta/2+\ell/2+n/2+1)_{i+1}(-\Delta+n-1+d)_{i+1}}\\
&\qquad\phantom{=}\times\left\{\left[-\frac{2(\Lambda+m/2+\Lambda'+m'/2+1/2)+(\Lambda'+m'/2)(u-v-1)}{u^{\frac{1}{2}}}\right.\right.\\
&\qquad\phantom{=}\left.+\bar{\eta}_4\cdot\bar{\m{D}}_{21}\right]\frac{\m{G}_{(\Delta,\ell)\{F(E\}\{E)\}\{F'\}\{E'\}}^{(d,\Lambda+1+i/2,n+i,m-1-i,\Lambda',n',m')}}{R_{(\Delta,\ell)}^{(d,\Lambda+1+i/2,n+i)}}\\
&\qquad\phantom{=}-2(\Lambda-\Delta/2+\ell/2+n/2+i+1)(\Lambda+\Delta/2+\ell/2-n/2+1-d/2)\\
&\qquad\phantom{=}\left.\times\left[-\frac{2(\Lambda+m/2-1/2)}{u^{\frac{1}{2}}}+\bar{\eta}_4\cdot\bar{\m{D}}_{12}\right]\frac{\m{G}_{(\Delta,\ell)\{F(E\}\{E)\}\{F'\}\{E'\}}^{(d,\Lambda+i/2,n+i,m-1-i,\Lambda',n',m')}}{R_{(\Delta,\ell)}^{(d,\Lambda+i/2,n+i)}}\right\},
}[EqG4CReta4]
where the four-point homogeneous embedding space coordinates are given in \eqref{Eqetab}.  Due to the symmetries of the four-point tensorial blocks, the remaining contiguous relations corresponding to contractions with the embedding space coordinates with $F'$- or $E'$-indices are easily generated from \eqref{EqG4CReta1}, \eqref{EqG4CReta2}, \eqref{EqG4CReta3} and \eqref{EqG4CReta4}.  A sketch of the proofs for the contiguous relations above is presented in Appendix \ref{SAppProofs}.

Once the four-point tensorial blocks have been fully contracted by applying the above contiguous relations, we are in general left with finite $\ell$-independent linear combinations of four-point scalar blocks and first-order derivatives acting on these objects.  Note that the action of the derivatives can be further simplified using \cite{Dolan:2011dv} (see Appendix \ref{SAppProofs}).

We remark that although the contiguous relations are sufficient for computing the four-point blocks, the explicit sums in \eqref{EqG4CReta3} and \eqref{EqG4CReta4} can be cumbersome.  We therefore consider another way to proceed.  In particular, we first set $m=m'=0$ using the contiguous relations [see \eqref{EqG4CReta2}]
\eqna{
&\bar{\eta}_3^E\m{G}_{(\Delta,\ell)\{F\}\{E\}\{F'\}\{E'\}}^{(d,\Lambda,n,m,\Lambda',n',m')}\\
&\qquad=-\frac{R_{(\Delta,\ell)}^{(d,\Lambda,n)}}{R_{(\Delta,\ell)}^{(d,\Lambda+1,n)}}\frac{v}{u^{\frac{1}{2}}}\frac{\m{G}_{(\Delta,\ell)\{F\}\{E\}\{F'\}\{E'\}}^{(d,\Lambda+1,n,m-1,\Lambda',n',m')}}{2(\ell-m+1)(\Lambda-\Delta/2+\ell/2+n/2+1)}\\
&\qquad\phantom{=}-\left[-\frac{(\Lambda-\Delta/2-\ell/2+n/2+m-1+d/2)+(\Lambda+m/2-1/2)(u-v-1)}{u^{\frac{1}{2}}}+\bar{\eta}_3\cdot\bar{\m{D}}_{12}\right]\\
&\qquad\phantom{=}\times\frac{\m{G}_{(\Delta,\ell)\{F\}\{E\}\{F'\}\{E'\}}^{(d,\Lambda,n,m-1,\Lambda',n',m')}}{(\ell-m+1)}+\frac{R_{(\Delta,\ell)}^{(d,\Lambda,n)}}{R_{(\Delta,\ell)}^{(d,\Lambda+1/2,n+1)}}\frac{\bar{\eta}_3^F\m{G}_{(\Delta,\ell)\{F\}\{E\}\{F'\}\{E'\}}^{(d,\Lambda+1/2,n+1,m-1,\Lambda',n',m')}}{2(\ell-m+1)(\Lambda-\Delta/2+\ell/2+n/2+1)},\\
&\bar{\eta}_4^E\m{G}_{(\Delta,\ell)\{F\}\{E\}\{F'\}\{E'\}}^{(d,\Lambda,n,m,\Lambda',n',m')}\\
&\qquad=-\frac{R_{(\Delta,\ell)}^{(d,\Lambda,n)}}{R_{(\Delta,\ell)}^{(d,\Lambda+1,n)}}\frac{1}{u^{\frac{1}{2}}}\frac{\m{G}_{(\Delta,\ell)\{F\}\{E\}\{F'\}\{E'\}}^{(d,\Lambda+1,n,m-1,\Lambda',n',m')}}{2(\ell-m+1)(\Lambda-\Delta/2+\ell/2+n/2+1)}\\
&\qquad\phantom{=}-\left[\frac{(\Lambda+\Delta/2+\ell/2-n/2-d/2)}{u^{\frac{1}{2}}}+\bar{\eta}_4\cdot\bar{\m{D}}_{12}\right]\frac{\m{G}_{(\Delta,\ell)\{F\}\{E\}\{F'\}\{E'\}}^{(d,\Lambda,n,m-1,\Lambda',n',m')}}{(\ell-m+1)}\\
&\qquad\phantom{=}+\frac{R_{(\Delta,\ell)}^{(d,\Lambda,n)}}{R_{(\Delta,\ell)}^{(d,\Lambda+1/2,n+1)}}\frac{\bar{\eta}_4^F\m{G}_{(\Delta,\ell)\{F\}\{E\}\{F'\}\{E'\}}^{(d,\Lambda+1/2,n+1,m-1,\Lambda',n',m')}}{2(\ell-m+1)(\Lambda-\Delta/2+\ell/2+n/2+1)},
}[EqG4CRmmp0]
as well as the analogous ones for the $F'$ and $E'$ contractions.  At that point, the four-point tensorial blocks can be rewritten in terms of the three-point tensorial blocks as
\eqna{
&\m{G}_{(\Delta,\ell)\{F\}\{F'\}}^{(d,\Lambda,n,\Lambda',n')}(\eta_1,\eta_2,\eta_4,\eta_3)\\
&\qquad=c_{(d,\ell)}R_{(\Delta,\ell)}^{(d,\Lambda,n)}\frac{(\ee{2}{3})^\Lambda}{(\ee{1}{3})^{\Lambda-\Delta/2}}\frac{(\ee{2}{3})^{\Lambda'}}{(\ee{2}{4})^{\Lambda'-\Delta/2}}u^{\Delta/2}v^{-\Lambda-\Lambda'}\\
&\qquad\phantom{=}\times\frac{(-\eta_2^E)^\ell}{(\ee{1}{2})^{\frac{\ell}{2}}}\m{D}_{12\{F\}}^{(d,\Lambda-\Delta/2+\ell/2-n/2,n)}\frac{(\ee{2}{4})^{\Lambda'-\Delta/2}}{(\ee{2}{3})^{\Lambda'+\Delta/2}}(\m{A}_{12E}^{\phantom{12E}E''})^\ell\m{G}_{(\Delta,\ell)\{E''\}\{F'\}}^{(d,\Lambda',n')}(\eta_4,\eta_3,\eta_2)\\
&\qquad=c_{(d,\ell)}R_{(\Delta,\ell)}^{(d,\Lambda',n')}\frac{(\ee{2}{3})^\Lambda}{(\ee{1}{3})^{\Lambda-\Delta/2}}\frac{(\ee{2}{3})^{\Lambda'}}{(\ee{2}{4})^{\Lambda'-\Delta/2}}u^{\Delta/2}v^{-\Lambda-\Lambda'}\\
&\qquad\phantom{=}\times\frac{(-\eta_3^{E'})^\ell}{(\ee{3}{4})^{\frac{\ell}{2}}}\m{D}_{43\{F'\}}^{(d,\Lambda'-\Delta/2+\ell/2-n'/2,n')}\frac{(\ee{1}{3})^{\Lambda-\Delta/2}}{(\ee{2}{3})^{\Lambda+\Delta/2}}(\m{A}_{34E'}^{\phantom{34E'}E''})^\ell\m{G}_{(\Delta,\ell)\{E''\}\{F\}}^{(d,\Lambda,n)}(\eta_1,\eta_2,\eta_3),
}[EqG4asG3]
where $\m{G}_{(\Delta,\ell)\{F\}\{F'\}}^{(d,\Lambda,n,\Lambda',n')}(\eta_1,\eta_2,\eta_4,\eta_3)\equiv\m{G}_{(\Delta,\ell)\{F\}\{F'\}}^{(d,\Lambda,n,0,\Lambda',n',0)}(\eta_1,\eta_2,\eta_4,\eta_3)$.  As for the symmetry properties of the four-point tensorial block, it is straightforward to see that from \eqref{EqG4}, \eqref{EqG4asG3} and \eqref{EqG3Sym}, we may obtain
\eqn{
\begin{gathered}
\m{G}_{(\Delta,\ell)\{F'\}\{F\}}^{(d,\Lambda',n',\Lambda,n)}(\eta_4,\eta_3,\eta_1,\eta_2)=\m{G}_{(\Delta,\ell)\{F\}\{F'\}}^{(d,\Lambda,n,\Lambda',n')}(\eta_1,\eta_2,\eta_4,\eta_3),\\
\m{G}_{(\Delta,\ell)\{F\}\{F'\}}^{(d,\Lambda,n,-\Lambda',n')}(\eta_1,\eta_2,\eta_3,\eta_4)=(-1)^\ell v^\Lambda\m{G}_{(\Delta,\ell)\{F\}\{F'\}}^{(d,\Lambda,n,\Lambda',n')}(\eta_1,\eta_2,\eta_4,\eta_3),\\
\m{G}_{(\Delta,\ell)\{F\}\{F'\}}^{(d,-\Lambda,n,\Lambda',n')}(\eta_2,\eta_1,\eta_4,\eta_3)=(-1)^\ell v^{\Lambda'}\m{G}_{(\Delta,\ell)\{F\}\{F'\}}^{(d,\Lambda,n,\Lambda',n')}(\eta_1,\eta_2,\eta_4,\eta_3).
\end{gathered}
}[EqG4Sym]
Note that these symmetry properties are obvious generalizations of the ones for $\Vev{SSSS}_{\m{O}^{(\ell)}}$.

The main advantage of \eqref{EqG4asG3} is that it allows for simpler contiguous relations with $\bar{\eta}_3^F$ and $\bar{\eta}_4^F$ (and thus $\bar{\eta}_1^{F'}$ and $\bar{\eta}_2^{F'}$) with the help of \eqref{EqG3Exp}.  These correspond to
\eqna{
\bar{\eta}_{\frac{7+\sigma}{2}}^F\m{G}_{(\Delta,\ell)\{F\}\{F'\}}^{(d,\Lambda,n,\Lambda',n')}&=\sum_{\substack{\bs{q}'\geq0\\\bar{q}'\leq n'}}\frac{R_{(\Delta,\ell)}^{(d,\Lambda,n)}}{R_{(\Delta-1+q'_2,\ell)}^{(d,\Lambda,n-1+q'_2)}}c_{(\Delta,\ell)(\bs{q}')}^{(d,\Lambda',n',1/2+\sigma/2,1/2-\sigma/2)}\\
&\phantom{=}\qquad\times(g_{F'F'})^{q'_0}(\bar{\eta}_{3F'})^{q'_3}(\bar{\eta}_{4F'})^{q'_4}\m{G}_{(\Delta-1+q'_2,\ell)\{FF'\}\{F'\}}^{(d,\Lambda,n-1+q'_2,\Lambda'+\sigma/2-q'_4/2+q'_3/2,n'-\bar{q}')},\\
\m{G}_{(\Delta,\ell)\{F\}\{F'\}}^{(d,\Lambda,n,\Lambda',n')}\bar{\eta}_{\frac{3-\sigma}{2}}^{F'}&=\sum_{\substack{\bs{q}\geq0\\\bar{q}\leq n}}\frac{R_{(\Delta,\ell)}^{(d,\Lambda',n')}}{R_{(\Delta-1+q_3,\ell)}^{(d,\Lambda',n'-1+q_3)}}c_{(\Delta,\ell)(\bs{q})}^{(d,\Lambda,n;1/2+\sigma/2,1/2-\sigma/2)}\\
&\phantom{=}\qquad\times(g_{FF})^{q_0}(\bar{\eta}_{1F})^{q_1}(\bar{\eta}_{2F})^{q_2}\m{G}_{(\Delta-1+q_3,\ell)\{F\}\{F'F\}}^{(d,\Lambda+\sigma/2-q_1/2+q_2/2,n-\bar{q},\Lambda',n'-1+q_3)},
}[EqG4CRetap]
for $\sigma=\pm1$.  In the same spirit, complete contractions can also be derived from \eqref{EqG3Exp} for
\eqn{\m{G}_{(\Delta,\ell)}^{(d,\Lambda,\Lambda';\bs{n})}(u,v)=(\bar{\eta}_3^F)^{n_3}(\bar{\eta}_4^F)^{n_4}(\bar{\eta}_1^{F'})^{n_1}(\bar{\eta}_2^{F'})^{n_2}\m{G}_{(\Delta,\ell)\{F\}\{F'\}}^{(d,\Lambda,n_3+n_4,\Lambda',n_1+n_2)}(\eta_1,\eta_2,\eta_4,\eta_3).}[EqG4Scalar]
Using \eqref{EqG3Exp} leads to
\eqna{
\m{G}_{(\Delta,\ell)}^{(d,\Lambda,\Lambda';\bs{n})}(u,v)&=(\bar{\eta}_1^{F'})^{n_1}(\bar{\eta}_2^{F'})^{n_2}\sum_{\substack{\bs{q}'\geq0\\\bar{q}'\leq n_1+n_2}}\frac{R_{(\Delta,\ell)}^{(d,\Lambda,n_3+n_4)}}{R_{(\Delta-n_3-n_4+q'_2,\ell)}^{(d,\Lambda,q'_2)}}c_{(\Delta,\ell)(\bs{q}')}^{(d,\Lambda',n_1+n_2;n_4,n_3)}\\
&\phantom{=}\qquad\times(g_{F'F'})^{q'_0}(\bar{\eta}_{4F'})^{q'_4}(\bar{\eta}_{3F'})^{q'_3}\m{G}_{(\Delta-n_3-n_4+q'_2,\ell)\{F'\}\{F'\}}^{(d,\Lambda,q'_2,\Lambda'+n_4/2-n_3/2-q'_4/2+q'_3/2,n_1+n_2-\bar{q}')}\\
&=(\bar{\eta}_3^F)^{n_3}(\bar{\eta}_4^F)^{n_4}\sum_{\substack{\bs{q}\geq0\\\bar{q}\leq n_3+n_4}}\frac{R_{(\Delta,\ell)}^{(d,\Lambda',n_1+n_2)}}{R_{(\Delta-n_1-n_2+q_3,\ell)}^{(d,\Lambda',q_3)}}c_{(\Delta,\ell)(\bs{q})}^{(d,\Lambda,n_3+n_4;n_1,n_2)}\\
&\phantom{=}\qquad\times(g_{FF})^{q_0}(\bar{\eta}_{1F})^{q_1}(\bar{\eta}_{2F})^{q_2}\m{G}_{(\Delta-n_1-n_2+q_3,\ell)\{F\}\{F\}}^{(d,\Lambda+n_1/2-n_2/2-q_1/2+q_2/2,n_3+n_4-\bar{q},\Lambda',q_3)},
}
depending on the order of the contractions.\footnote{By contracting randomly, this most likely generates several identities for usual scalar conformal blocks.}  Performing all the contractions results in
\eqna{
\m{G}_{(\Delta,\ell)}^{(d,\Lambda,\Lambda';\bs{n})}(u,v)&=\sum_{\substack{\bs{q}'\geq0\\\bar{q}'\leq n_1+n_2\\\bs{p}'\geq0}}\frac{R_{(\Delta,\ell)}^{(d,\Lambda,n_3+n_4)}}{R_{(\Delta-n_3-n_4+q'_2,\ell)}^{(d,\Lambda,q'_2)}}c_{(\Delta,\ell)(\bs{q}')}^{(d,\Lambda',n_1+n_2;n_4,n_3)}\frac{(-1)^{\bar{p}'}(-q'_4)_{p'_4}(-q'_3)_{p'_3}(-q'_2)_{p'_2}}{p'_2!p'_3!p'_4!}\\
&\phantom{=}\qquad\times\frac{2^{q'_0}(-n_2)_{q'_0+\bar{p}'}(-n_1)_{\bar{q}'-q'_0-\bar{p}'}}{(-n_1-n_2)_{\bar{q}'}}\frac{v^{p'_3}}{u^{\frac{q'_4}{2}+\frac{q'_3}{2}}}(\bar{\eta}_2)^{n_2-q'_0-\bar{p}'}(\bar{\eta}_1)^{n_1-\bar{q}'+q'_0+\bar{p}'}\\
&\phantom{=}\qquad\cdot\m{G}_{(\Delta-n_3-n_4+q'_2,\ell)}^{(d,\Lambda,q'_2,\Lambda'+n_4/2-n_3/2-q'_4/2+q'_3/2,n_1+n_2-\bar{q}')}\cdot(\bar{\eta}_2)^{p'_2}(\bar{\eta}_1)^{q'_2-p'_2}\\
&=\sum_{\substack{\bs{q}\geq0\\\bar{q}\leq n_3+n_4\\\bs{p}\geq0}}\frac{R_{(\Delta,\ell)}^{(d,\Lambda',n_1+n_2)}}{R_{(\Delta-n_1-n_2+q_3,\ell)}^{(d,\Lambda',q_3)}}c_{(\Delta,\ell)(\bs{q})}^{(d,\Lambda,n_3+n_4;n_1,n_2)}\frac{(-1)^{\bar{p}}(-q_1)_{p_1}(-q_2)_{p_2}(-q_3)_{p_3}}{p_1!p_2!p_3!}\\
&\phantom{=}\qquad\times\frac{2^{q_0}(-n_3)_{q_0+\bar{p}}(-n_4)_{\bar{q}-q_0-\bar{p}}}{(-n_3-n_4)_{\bar{q}}}\frac{v^{p_2}}{u^{\frac{q_1}{2}+\frac{q_2}{2}}}(\bar{\eta}_3)^{n_3-q_0-\bar{p}}(\bar{\eta}_4)^{n_4-\bar{q}+q_0+\bar{p}}\\
&\phantom{=}\qquad\cdot\m{G}_{(\Delta-n_1-n_2+q_3,\ell)}^{(d,\Lambda+n_1/2-n_2/2-q_1/2+q_2/2,n_3+n_4-\bar{q},\Lambda',q_3)}\cdot(\bar{\eta}_3)^{p_3}(\bar{\eta}_4)^{q_3-p_3},
}
with $\bar{p}=p_1+p_2+p_3$ and $\bar{p}'=p'_2+p'_3+p'_4$.  It is now straightforward to rewrite the fully-contracted four-point tensorial blocks in terms of the usual scalar blocks by using first the contiguous relations \eqref{EqG4CReta1} and \eqref{EqG4CReta2} followed by the then trivial contiguous relations \eqref{EqG4CRetap}.  The final answer is thus
\eqna{
&\m{G}_{(\Delta,\ell)}^{(d,\Lambda,\Lambda';\bs{n})}(u,v)\\
&\qquad=\sum_{\substack{\bs{q}'\geq0\\\bar{q}'\leq n_1+n_2\\\bs{p}'\geq0}}\frac{R_{(\Delta-n_3-n_4+q'_2,\ell)}^{(d,\Lambda'+n_4/2-n_3/2-q'_4/2+q'_3/2,n_1+n_2-\bar{q}')}}{R_{(\Delta-\bar{n}+\bar{q}'+q'_2,\ell)}^{(d,\Lambda'+n_4/2-n_3/2-q'_4/2+q'_3/2,0)}}\frac{R_{(\Delta,\ell)}^{(d,\Lambda,n_3+n_4)}}{R_{(\Delta-n_3-n_4+q'_2,\ell)}^{(d,\Lambda+q'_2/2-p'_2,0)}}c_{(\Delta,\ell)(\bs{q}')}^{(d,\Lambda',n_1+n_2;n_4,n_3)}\\
&\qquad\phantom{=}\times\frac{(-1)^{\bar{p}'}(-q'_4)_{p'_4}(-q'_3)_{p'_3}(-q'_2)_{p'_2}}{p'_2!p'_3!p'_4!}\frac{2^{q'_0}(-n_2)_{q'_0+\bar{p}'}(-n_1)_{\bar{q}'-q'_0-\bar{p}'}}{(-n_1-n_2)_{\bar{q}'}}\rho^{(d,p'_2;-\Lambda+\Delta/2-\ell/2-n_3/2-n_4/2)}\\
&\qquad\phantom{=}\times\frac{v^{p'_3}}{u^{\frac{q'_3}{2}+\frac{q'_4}{2}}}\m{G}_{(\Delta-\bar{n}+\bar{q}'+q'_2,\ell)}^{(d,\Lambda+n_1/2-n_2/2-q'_4/2-q'_3/2+p'_4+p'_3,0,\Lambda'+n_4/2-n_3/2-q'_4/2+q'_3/2,0)}\\
&\qquad=\sum_{\substack{\bs{q}\geq0\\\bar{q}\leq n_3+n_4\\\bs{p}\geq0}}\frac{R_{(\Delta-n_1-n_2+q_3,\ell)}^{(d,\Lambda+n_1/2-n_2/2-q_1/2+q_2/2,n_3+n_4-\bar{q})}}{R_{(\Delta-\bar{n}+\bar{q}+q_3,\ell)}^{(d,\Lambda+n_1/2-n_2/2-q_1/2+q_2/2,0)}}\frac{R_{(\Delta,\ell)}^{(d,\Lambda',n_1+n_2)}}{R_{(\Delta-n_1-n_2+q_3,\ell)}^{(d,\Lambda'+q_3/2-p_3,0)}}c_{(\Delta,\ell)(\bs{q})}^{(d,\Lambda,n_3+n_4;n_1,n_2)}\\
&\qquad\phantom{=}\times\frac{(-1)^{\bar{p}}(-q_1)_{p_1}(-q_2)_{p_2}(-q_3)_{p_3}}{p_1!p_2!p_3!}\frac{2^{q_0}(-n_3)_{q_0+\bar{p}}(-n_4)_{\bar{q}-q_0-\bar{p}}}{(-n_3-n_4)_{\bar{q}}}\rho^{(d,p_3;-\Lambda'+\Delta/2-\ell/2-n_1/2-n_2/2)}\\
&\qquad\phantom{=}\times\frac{v^{p_2}}{u^{\frac{q_1}{2}+\frac{q_2}{2}}}\m{G}_{(\Delta-\bar{n}+\bar{q}+q_3,\ell)}^{(d,\Lambda+n_1/2-n_2/2-q_1/2+q_2/2,0,\Lambda'+n_4/2-n_3/2-q_1/2-q_2/2+p_1+p_2,0)},
}[EqG4ScalarSoln]
where $\bar{n}=n_1+n_2+n_3+n_4$.

%%%%%%%%%%%%%%%%%%%%%%%%%%%%%%%%%%%%%%%%%%%%%%%%%%
%%%%%%%%%%%%%%%%%%%%%%%%%%%%%%%%%%%%%%%%%%%%%%%%%%

\section{Conformal Bootstrap Equations}\label{SecBootstrap}

The conformal bootstrap technique rests on the principle that correlation functions are independent of the choice of OPE.  By choosing to perform the OPE between different pairs of external quasi-primary operators, one obtains different yet equivalent expansions of the correlation functions in terms of conformal blocks.  The equivalence of these different expansions accordingly leads to constraints on the OPE coefficients.  In this section, we describe how to set up the conformal bootstrap equations directly in the embedding space, without projecting back to position space.  We begin by describing the general strategy for generating the conformal bootstrap equations in our framework.   We stress that by $M$-point conformal bootstrap equations, we mean the use of associativity of the $M$-point correlation functions to constrain the OPE coefficients, with the standard conformal bootstrap equations stemming from $M=4$.  Therefore, by two- and three-point conformal bootstrap equations, we mean the associativity of the two- and three-point correlation functions, respectively.  Although non-standard, it leads to a consistent nomenclature.

%%%%%%%%%%%%%%%%%%%%%%%%%%%%%%%%%%%%%%%%%%%%%%%%%%

\subsection{General Strategy}

It was shown in \cite{Fortin:2020yjz} that for $M$-point correlation functions, the number of independent bootstrap equations at the level of the topologies of the correlation functions is given by $N_B=\text{max}\{1,T_0(M)-1\}$, where $T_0(M)$ is the number of unrooted binary trees with $M$ unlabeled leaves.  Since $M$-point correlation functions decompose into a finite set of independent tensor structures, the actual number of independent bootstrap equations is larger.

Let us present an algorithm for counting the number of independent $M$-point tensor structures in embedding space.  Starting from the $M$-point correlation function $\Vev{\m{O}_{i_1}(\eta_1)\cdots\m{O}_{i_M}(\eta_M)}$, it is convenient to contract the correlation function by $(\m{T}_{a_11\bs{N}_{i_1}})\cdots(\m{T}_{a_MM\bs{N}_{i_M}})*$, generating hatted projectors $(\h{\m{P}}_{1a_1}^{\bs{N}_{i_1}})\cdots(\h{\m{P}}_{Ma_M}^{\bs{N}_{i_M}})$ with the help of \eqref{EqTESId}.  By repeated use of \eqref{EqPtoP}, it is possible to re-express all hatted projectors at the same two (arbitrarily chosen) embedding coordinates, \textit{e.g.} $(\h{\m{P}}_{12}^{\bs{N}_{i_1}})\cdots(\h{\m{P}}_{12}^{\bs{N}_{i_M}})$.  Hence, by decomposing the tensor product of these irreducible representations in the space of invariant tensors \eqref{EqPStoES} with $i=1$ and $j=2$, we see that a $M$-point correlation function can be expanded in a basis of conformal blocks encoding the exchange of the irreducible representations
\eqn{\bs{N}_{i_1}\otimes\cdots\otimes\bs{N}_{i_M}=\bigoplus_{\bs{R}}m_{\bs{i}}^{\bs{R}}\bs{R},}
which must be contracted with objects made out of the remaining embedding space coordinates $\{\eta_3,\ldots,\eta_M\}$ to form singlets.\footnote{Contractions with the two chosen (but arbitrary) embedding space coordinates vanish automatically.}  Here $m_{\bs{i}}^{\bs{R}}$ is the multiplicity of the irreducible representation $\bs{R}$ that appears in the tensor product decomposition of the $M$ irreducible representations of the quasi-primary operators.  Hence, for a specific irreducible representation $\bs{R}$, the irreducible representations that can occur upon contracting with the remaining embedding space coordinates are given by
\eqn{\bs{R}\otimes\ell_1\bs{e}_1\otimes\cdots\otimes\ell_{N_\ell}\bs{e}_1=\bigoplus_{\bs{S}}m_{\bs{R}\bs{\ell}}^{\bs{S}}\bs{S},}
where $m_{\bs{R}\bs{\ell}}^{\bs{S}}$ is again the multiplicity of the irreducible representation $\bs{S}$ appearing in the tensor product decomposition and
\eqn{N_\ell=\text{min}\{d,M-2\}=\begin{cases}M-2\quad\text{for}\quad M-3<d\\d\quad\text{for}\quad M-3\geq d\end{cases}.}[EqNl]
To understand the bound on $N_\ell$ \eqref{EqNl}, \textit{i.e.}\ the number of independent $\ell\bs{e}_1$ that can be built from the remaining embedding space coordinates, we need to first understand the constraints on embedding space coordinates when $d$ is small compared to $M$.  In contrast to the situation in position space, the constraints on embedding space coordinates are subtle since linear combinations of embedding space coordinates may not necessarily be on the light-cone.\footnote{The cases $d=1$ and $d=2$ are not considered here since the $2d$ (global) conformal algebra factorizes into two copies of the $1d$ conformal algebra, and the $1d$ conformal algebra only has trivial irreducible representations.}

For the case that $M-3<d$, all the embedding space coordinates $\{\eta_3,\ldots,\eta_M\}$ are linearly independent and thus $N_\ell=M-2$ where each $\ell_m\bs{e}_1$ is built from the embedding space coordinates $\eta_{m+2}$.  Meanwhile, for the case $M-3\geq d$, some constraints arise that are better understood from the point of view of the appropriate metric.  Starting from the metric of our choice (\textit{e.g.}\ $\m{A}_{12}$), it is possible to build additional metrics recursively with the help of
\eqn{\m{A}_{12\cdots m+1}^{AB}=\m{A}_{12\cdots m}^{AB}-\frac{(\m{A}_{12\cdots m}\cdot\eta_{m+1})^A(\m{A}_{12\cdots m}\cdot\eta_{m+1})^B}{(\eta_{m+1}\cdot\m{A}_{12\cdots m}\cdot\eta_{m+1})},}[EqA]
for $2\leq m<d+2$.  Then $\m{A}_{12\cdots m}$ is symmetric under permutations of its embedding space coordinates\footnote{The proof proceeds by recurrence.  We first assume that $\m{A}_{12\cdots m}^{AB}$ is fully symmetric under permutations.  Using the definition \eqref{EqA} twice for $\m{A}_{12\cdots m+1}^{AB}$ leads to an expression written in terms of $\m{A}_{12\cdots m-1}^{AB}$.  By recombining the terms in a different order, it is straightforward to show that $\m{A}_{12\cdots m+1}^{AB}=\m{A}_{12\cdots m-1,m+1,m}^{AB}$.  Since $\m{A}_{12\cdots m}^{AB}$ is fully symmetric, then from \eqref{EqA} and the identity previously obtained, $\m{A}_{12\cdots m+1}$ is also fully symmetric.  This thus completes the proof since the base case $\m{A}_{12}^{AB}$ is fully symmetric $\m{A}_{12}^{AB}=\m{A}_{21}^{AB}$.} and is transverse with respect to $\{\eta_1,\eta_2,\ldots,\eta_m\}$, as can be seen by recursion.  Thus, $\m{A}_{12\cdots m}$ has rank $d+2-m$, which implies that $\m{A}_{12\cdots d+2}^{AB}=0$ identically.  Inverting \eqref{EqA} gives
\eqn{\m{A}_{12}^{AB}=\m{A}_{12\cdots m}^{AB}+\sum_{k\geq3}^m\frac{(\m{A}_{12\cdots k-1}\cdot\eta_k)^A(\m{A}_{12\cdots k-1}\cdot\eta_k)^B}{(\eta_k\cdot\m{A}_{12\cdots k-1}\cdot\eta_k)}.}[EqAA]
Thereupon, we observe that since embedding space coordinates are implicitly contracted with the correlation functions through the chosen metric, we can therefore rewrite every possible contraction in the above tensor product as a contraction of the form
\eqn{(\m{A}_{12}\cdot\eta_i)^A=\sum_{k\geq3}^{d+2}\frac{(\eta_i\cdot\m{A}_{12\cdots k-1}\cdot\eta_k)}{(\eta_k\cdot\m{A}_{12\cdots k-1}\cdot\eta_k)}(\m{A}_{12\cdots k-1}\cdot\eta_k)^A,}[EqAetaA]
where \eqref{EqAA} with $m=d+2$ and $\m{A}_{12\cdots d+2}=0$ were used.  We conclude that all contractions are linear combinations of $\m{A}_{12\cdots k-1}\cdot\eta_k$ for $k\in\{3,\ldots,d+2\}$ from which the $\ell\bs{e}_1$'s can be built, implying $N_\ell=d$.  In other words, of the $M-2$ remaining embedding space coordinates, there are $M-2-d$ constraints leading to $N_\ell=d$, as stated in \eqref{EqNl}.\footnote{A similar analysis can be performed for the number of independent cross-ratios $N_\text{cr}$,
\eqn{N_\text{cr}=\begin{cases}\frac{M(M-3)}{2}\quad\text{for}\quad M-3<d\\d(M-3)-\frac{(d-1)(d-2)}{2}\quad\text{for}\quad M-3\geq d\end{cases},}
when $M-3\geq d$.  Since $A_{12\cdots d+2}$ is built from the embedding space coordinates $\{\eta_1,\eta_2,\ldots,\eta_{d+2}\}$, \eqref{EqAetaA} contracted with $\eta_j$ is identically satisfied if $i$ or $j$ is in $\{1,2,\ldots,d+2\}$.  It is however not the case when both $i$ and $j$ are in $\{d+3,\ldots,M\}$, where \eqref{EqAetaA} contracted with $\eta_j$ gives non-trivial relations
\eqn{1=\sum_{k\geq3}^{d+2}\frac{(\eta_i\cdot\m{A}_{12\cdots k-1}\cdot\eta_k)}{(\eta_k\cdot\m{A}_{12\cdots k-1}\cdot\eta_k)}\frac{(\eta_j\cdot\m{A}_{12\cdots k-1}\cdot\eta_k)}{(\eta_j\cdot\m{A}_{12}\cdot\eta_i)},}
between the cross-ratios for every $d+3\leq i\leq j\leq M$.  Considering that there are $(M-d-1)(M-d-2)/2$ non-trivial independent relations, there are $N_\text{cr}=M(M-3)/2-(M-d-1)(M-d-2)/2$ independent cross-ratios when $M-3\geq d$, as expected.  In addition to giving the proper counting for the number of independent cross-ratios, the previous analysis gives all the explicit identities between the cross-ratios.}

It follows that the number of independent tensor structures for a $M$-point correlation function is
\eqn{N_S=\sum_{\substack{\bs{R},\bs{\ell}\\|\bs{\ell}|=n_v^{\bs{R}}}}m_{\bs{i}}^{\bs{R}}m_{\bs{R}\bs{\ell}}^{\bs{0}},}[EqNumberTS]
where $|\bs{\ell}|=\sum_m\ell_m$ is fixed to avoid trace contractions between the different embedding space coordinates and $\bs{S}$ is set to $\bs{0}$ to only extract singlets.\footnote{To clear up potential confusion, $\bs{i}=\{i_1,\ldots,i_M\}$ is a vector of numbers indexing the $M$ quasi-primary operators, $\bs{\ell}=\{\ell_1,\ldots,\}$ is a vector of non-negative integers denoting the symmetric-traceless irreducible representations constructed from the remaining embedding space coordinates, while $\bs{R}$ and $\bs{S}$ are vectors of Dynkin indices denoting irreducible representations appearing in the corresponding tensor product decompositions.}  The above argument implies that the total number of independent $M$-point conformal bootstrap equations is $N_BN_S$.\footnote{Since the number of independent tensor structures is smaller when correlation functions include conserved quasi-primary operators, the result \eqref{EqNumberTS} must be modified accordingly when one of more of the quasi-primary operators are conserved.}

\begin{table}[t]
\centering
\resizebox{11cm}{!}{%
\begin{tabular}{|c|c|c|c|}\hline
 & $M=2$ & $M=3$ & $M=4$\\\hline\hline
%$d=2$ & $1$ per $\bs{0}$ & $1$ per $\bs{0}$ & $1$ per $\bs{0}$\\\hline
$d=3$ & $1$ per $\bs{0}$ & $1$ per $2\ell\bs{e}_1$ & $\ell+1$ per $2\ell\bs{e}_1$\\\hline
$d=4$ & $1$ per $\bs{0}$ & $1$ per $\ell\bs{e}_1+\ell\bs{e}_2$ & \begin{tabular}{c}$\ell+1$ per $(\ell+2m)\bs{e}_1+\ell\bs{e}_2$\\$\ell+1$ per $\ell\bs{e}_1+(\ell+2m)\bs{e}_2$\end{tabular}\\\hline
$d=5$ & $1$ per $\bs{0}$ & $1$ per $\ell\bs{e}_1$ & $\ell+1$ per $\ell\bs{e}_1+2m\bs{e}_2$\\\hline
$d=6$ & $1$ per $\bs{0}$ & $1$ per $\ell\bs{e}_1$ & $\ell+1$ per $\ell\bs{e}_1+m\bs{e}_2+m\bs{e}_3$\\\hline
$d\geq7$ & $1$ per $\bs{0}$ & $1$ per $\ell\bs{e}_1$ & $\ell+1$ per $\ell\bs{e}_1+m\bs{e}_2$\\\hline
\end{tabular}
}
\caption{Number of independent $M$-point conformal bootstrap equations per irreducible representations appearing in the tensor product decomposition of the irreducible representations of the $M$ quasi-primary operators.  For smaller spacetime dimensions, the irreducible representations are explicitly expressed in terms of their appropriate Dynkin indices.}
\label{TabBoot}
\end{table}

Let us consider a few examples.  For two-point correlation functions, there is only one topology ($N_B=1$) and no remaining embedding space coordinates ($\bs{\ell}=\bs{0}$).  Hence, the number of independent tensor structures is $N_BN_S=\sum_{\bs{R}}m_{i_1i_2}^{\bs{R}}m_{\bs{R}\bs{0}}^{\bs{0}}=m_{i_1i_2}^{\bs{0}}=\delta_{i_1,i_2^{CR}}$ since $\bs{R}=\bs{0}$ otherwise $m_{\bs{R}\bs{0}}^{\bs{0}}=0$.  Next, $m_{i_1i_2}^{\bs{0}}=1$ if the two irreducible representations are contragredient-reflected with respect to each other and vanishes otherwise.  Meanwhile, for three-point correlation functions, there is also one topology ($N_B=1$) and one remaining embedding space coordinate (for $d>0$).  This implies that the number of independent tensor structures is $N_BN_S=\sum_{\substack{\bs{R},\bs{\ell}\\|\bs{\ell}|=n_v^{\bs{R}}}}m_{\bs{i}}^{\bs{R}}m_{\bs{R}\bs{\ell}}^{\bs{0}}=\sum_{\ell\geq0}m_{i_1i_2i_3}^{\ell\bs{e}_1}$ since $\bs{R}=\ell\bs{e}_1$ otherwise $m_{\bs{R}\bs{\ell}}^{\bs{0}}=0$.  Thus, as expected, there is only one tensor structure per symmetric-traceless representation appearing in the tensor product decomposition of the irreducible representations of the three quasi-primary operators.  The case of $M=4$ is the last simple example where multiplicities do not lead to complications.  For the four-point correlation functions, the number of topologies is again one ($N_B=1$), but there are now two remaining embedding space coordinates (for $d>1$) left to contract.  Thus, the number of independent four-point tensor structures is $N_BN_S=\sum_{\substack{\bs{R},\bs{\ell}\\|\bs{\ell}|=n_v^{\bs{R}}}}m_{\bs{i}}^{\bs{R}}m_{\bs{R}\bs{\ell}}^{\bs{0}}=\sum_{m,\ell\geq0}(\ell+1)m_{i_1i_2i_3i_4}^{\ell\bs{e}_1+m\bs{e}_2}$ where $\bs{R}=\ell\bs{e}_1+m\bs{e}_2$ with $m_{\ell{e}_1+m\bs{e}_2,\ell_1\ell_2}^{\bs{0}}=\ell+1$.  Indeed, there are $\ell+1$ different ways of constructing the irreducible representation $\ell\bs{e}_1+m\bs{e}_2$ from products of $(\eta_3)^{\ell_1}$ and $(\eta_4)^{\ell_2}$, by taking $\ell_1=m+i$ and $\ell_2=\ell+m-i$ such that $\ell_1+\ell_2=\ell+2m$ with $i\in\{0,1,\ldots,\ell\}$.  For $M\geq5$, the computation of multiplicities $m_{\bs{R}\bs{\ell}}^{\bs{0}}$ is much more elaborate due to the Fock conditions and do not seem to lead to such simple formulas.  Moreover, the constraints for small spacetime dimensions start becoming relevant at $M=6$ where there is a distinction between $d>3$ and $d=3$.  The cases $2\leq M\leq4$ are summarized in Table~\ref{TabBoot}.

The aforementioned counting of independent tensor structures suggests a simple path for generating the $M$-point conformal bootstrap equations directly in embedding space.  Indeed, upon multiplying by the appropriate half-projectors and subsequently contracting with the embedding space invariant tensors as well as the remaining available embedding space coordinates, we arrive at bootstrap equations expressed purely in terms of embedding space scalar objects that project trivially to position space.  In what follows, we elucidate this strategy for two-, three-, and four-point conformal bootstrap equations in turn.

%%%%%%%%%%%%%%%%%%%%%%%%%%%%%%%%%%%%%%%%%%%%%%%%%%

\subsection{Two-Point Bootstrap Equations}

Although we do not expect to generate any nontrivial bootstrap constraints on the OPE coefficients from the associativity of one- and two-point correlation functions, we can nevertheless hope to extract a true symmetry property of the two-point OPE coefficients from the two-point bootstrap equations.  Starting from the two-point correlation function and using the OPE in the two possible orders, the two-point conformal bootstrap equations are given by
\eqn{\Vev{\Op{i}{}{1}\Op{j}{}{2}}=(-1)^{2\xi_i}\Vev{\Op{j}{}{2}\Op{i}{}{1}}.}
Upon contracting with $(\m{T}_{21\bs{N}_i})(\m{T}_{12\bs{N}_j})*$, the last equation assumes the form
\eqn{\Vev{\m{T}_{21\bs{N}_i}*\Op{i}{}{1}\m{T}_{12\bs{N}_j}*\Op{j}{}{2}}=(-1)^{2\xi_i}\Vev{\m{T}_{12\bs{N}_j}*\Op{j}{}{2}\m{T}_{21\bs{N}_i}*\Op{i}{}{1}},}
which implies that
\eqna{
&\cOPE{}{ij}{\1}(\h{\m{P}}_{12}^{\bs{N}_i})_{\{aA\}}^{\phantom{\{aA\}}\{A'a'\}}(\h{\m{P}}_{21}^{\bs{N}_i^{CR}})_{\{bB\}}^{\phantom{\{bB\}}\{B'b'\}}[(C_\Gamma^{-1})_{a'b'}]^{2\xi_i}(g_{A'B'})^{n_v^i}\\
&\qquad=(-1)^{2\xi_i}\cOPE{}{ji}{\1}(\h{\m{P}}_{21}^{\bs{N}_i^{CR}})_{\{bB\}}^{\phantom{\{bB\}}\{B'b'\}}(\h{\m{P}}_{12}^{\bs{N}_i})_{\{aA\}}^{\phantom{\{aA\}}\{A'a'\}}[(C_\Gamma^{-1})_{b'a'}]^{2\xi_i}(g_{B'A'})^{n_v^i},
}
where we used \eqref{Eq2pt}.  Using \eqref{EqPtoPCR} and the symmetry property of $C_\Gamma$ gives
\eqna{
&\cOPE{}{ij}{\1}(\h{\m{P}}_{12}^{\bs{N}_i})_{\{aA\}}^{\phantom{\{aA\}}\{A'a'\}}(\h{\m{P}}_{12}^{\bs{N}_i})_{\{a'A'\}}^{\phantom{\{a'A'\}}\{B'b'\}}[(C_\Gamma^{-1})_{bb'}]^{2\xi_i}(g_{BB'})^{n_v^i}\\
&\qquad=(-1)^{2\xi_i[1+(r+1)(r+2)/2]}\cOPE{}{ji}{\1}(\h{\m{P}}_{12}^{\bs{N}_i})_{\{a'A'\}}^{\phantom{\{a'A'\}}\{B'b'\}}(\h{\m{P}}_{12}^{\bs{N}_i})_{\{aA\}}^{\phantom{\{aA\}}\{A'a'\}}[(C_\Gamma^{-1})_{bb'}]^{2\xi_i}(g_{BB'})^{n_v^i},
}
or
\eqna{
&\cOPE{}{ij}{\1}(\h{\m{P}}_{12}^{\bs{N}_i})_{\{aA\}}^{\phantom{\{aA\}}\{B'b'\}}[(C_\Gamma^{-1})_{bb'}]^{2\xi_i}(g_{BB'})^{n_v^i}\\
&\qquad=(-1)^{2\xi_i[1+(r+1)(r+2)/2]}\cOPE{}{ji}{\1}(\h{\m{P}}_{12}^{\bs{N}_i})_{\{aA\}}^{\phantom{\{aA\}}\{B'b'\}}[(C_\Gamma^{-1})_{bb'}]^{2\xi_i}(g_{BB'})^{n_v^i}.
}
At this point, determining the symmetry property of $\cOPE{}{ij}{\1}$ is straightforward.  However, in the interest of illustrating the overall technique that generalizes well to higher-point conformal bootstrap equations, we may proceed further by contracting the remaining free embedding space indices.  In this case, the only nontrivial contraction is with $(\m{A}_{12}^{AB})^{n_v^i}[(C_\Gamma)^{ab}]^{2\xi_i}$, giving
\eqn{\cOPE{}{ij}{\1}(\h{\m{P}}_{12}^{\bs{N}_i})_{\{aA\}}^{\phantom{\{aA\}}\{Aa\}}=(-1)^{\xi_ir(r+3)}\cOPE{}{ji}{\1}(\h{\m{P}}_{12}^{\bs{N}_i})_{\{aA\}}^{\phantom{\{aA\}}\{Aa\}}.}
Since $(\h{\m{P}}_{12}^{\bs{N}_i})_{\{aA\}}^{\phantom{\{aA\}}\{Aa\}}=\text{dim}(\bs{N}_i)$ is simply a nonzero number corresponding to the dimension of the irreducible representation $\bs{N}_i$, it drops out, leading to the following reduced form of the two-point conformal bootsrap equation
\eqn{\cOPE{}{ij}{\1}=(-1)^{\xi_ir(r+3)}\cOPE{}{ji}{\1}.}[EqBoot2]
This is none other than the symmetry property of the two-point OPE coefficients mentioned above.

%%%%%%%%%%%%%%%%%%%%%%%%%%%%%%%%%%%%%%%%%%%%%%%%%%

\subsection{Three-Point Bootstrap Equations}

Turning to the three-point case, we remark that just like for two-point correlation functions, associativity of the three-point correlation functions \eqref{Eq3pt} implies symmetry properties of the three-point OPE coefficients.  These symmetry properties take the simplest possible form in a particular basis of tensor structures,
\eqn{\Vev{\Op{i}{}{1}\Op{j}{}{2}\Op{m}{}{3}}=\frac{(\m{T}_{12}^{\bs{N}_i})(\m{T}_{21}^{\bs{N}_j})(\m{T}_{31}^{\bs{N}_m})\cdot\sum_{a=1}^{N_{ijm}}\cCF{[a|}{ijm}\mathscr{G}_{[a|}^{ij|m}}{(\ee{1}{2})^{\frac{1}{2}(\tau_i+\tau_j-\chi_m)}(\ee{1}{3})^{\frac{1}{2}(\chi_i-\chi_j+\tau_m)}(\ee{2}{3})^{\frac{1}{2}(-\chi_i+\chi_j+\chi_m)}},}[Eq3ptCF]
where the change of basis is implemented by rotation matrices $R_{ijm}$ such that
\eqn{\cCF{(a|}{ijm}=\sum_{a'=1}^{N_{ijm}}\cCF{[a'|}{ijm}(R_{ijm})_{a'a},\qquad\qquad\mathscr{G}_{(a|}^{ij|m}=\sum_{a'=1}^{N_{ijm}}(R_{ijm}^{-1})_{aa'}\mathscr{G}_{[a'|}^{ij|m}.}[EqRM]
In this new basis, the three-point conformal blocks are monomials of the form
\eqna{
\mathscr{G}_{[a|}^{ij|m}&=(\h{\m{P}}_{12}^{\bs{N}_i})(\h{\m{P}}_{21}^{\bs{N}_j})(\h{\m{P}}_{31}^{\bs{N}_m})\cdot(\text{product of $\m{A}_{123}$})\times[(\m{A}_{12}\cdot\tilde{\eta}_3)_A]^{\delta_i}[(\m{A}_{12}\cdot\tilde{\eta}_3)_B]^{\delta_j}\\
&\phantom{=}\qquad\times[(\m{A}_{12}\cdot\tilde{\eta}_3)_E]^{\delta_m}(\epsilon_{123})^{\delta_\epsilon}\times\left(\Gamma_{123}^{[n-\delta_\Gamma]}\,(\tilde{\eta}_3\cdot\Gamma_{12})^{\delta_\Gamma}\frac{\tilde{\eta}_3\cdot\Gamma_{12}\,\tilde{\eta}_3\cdot\Gamma C_\Gamma^{-1}}{2}\right)^{\xi_i+\xi_j+\xi_m},
}[EqTS3pt]
where
\eqn{
\begin{gathered}
\m{A}_{123}^{AB}=\m{A}_{12}^{AB}-\frac{(\m{A}_{12}\cdot\eta_3)^A(\m{A}_{12}\cdot\eta_3)^B}{\eta_3\cdot\m{A}_{12}\cdot\eta_3},\\
\epsilon_{123}^{A_1\cdots A_{d-1}}=\epsilon_{12}^{A'_1\cdots A'_{d-1}A_d}\tilde{\eta}_{3A_d}\m{A}_{123A'_{d-1}}^{\phantom{123A'_{d-1}}A_{d-1}}\cdots\m{A}_{123A'_1}^{\phantom{123A'_1}A_1},\\
\Gamma_{123}^{A_1\cdots A_n}=\Gamma_{12}^{A'_1\cdots A'_n}\m{A}_{123A'_n}^{\phantom{123A'_n}A_n}\cdots\m{A}_{123A'_1}^{\phantom{123A'_1}A_1},
\end{gathered}
}[EqA123]
and the homogeneized three-point embedding space coordinates are defined in \eqref{Eqetat}.  Note that due to the constraints on the building blocks described below [see \eqref{EqTSOPE}], the powers $\delta_\epsilon$ and $\delta_\Gamma$ in \eqref{EqTS3pt} can only be zero or one.\footnote{For three-point correlation functions involving one or more conserved currents, there are relationships between the tensor structures due to the conservation condition that lead to linear combinations of OPE coefficients that vanish.}

In \eqref{EqA123}, it is readily apparent that $\m{A}_{123}$ is the new metric defined in \eqref{EqA} with three eigenvectors $\{\eta_1,\eta_2,\eta_3\}$ with vanishing eigenvalues (and thus with rank $d-1$) that is fully symmetric under its embedding space coordinates, \textit{i.e.}\ $\m{A}_{ijk}=\m{A}_{jik}=\m{A}_{ikj}=\ldots$.  Similarly, we introduce a new $\epsilon$-tensor $\epsilon_{123}$ with $d-1$ indices [as expected from \eqref{Eqepsilon}] since a full contraction of $\epsilon_{12}$ with the new metric $\m{A}_{123}$ is identically zero from the rank of $\m{A}_{123}$.  Finally, we introduce another $\Gamma$-matrix $\Gamma_{123}$, with a possible factor $\tilde{\eta}_3\cdot\Gamma_{12}$ to allow for contractions with $\m{A}_{12}\cdot\tilde{\eta}_3$, whose role is similar to that of the OPE differential operator (it carries missing free indices) at the level of the three-point correlation functions.

To implement the three-point conformal bootstrap, we must consider the associativity of the three-point correlation functions.  Although this can be done directly from the point of view of three-point correlation function \eqref{Eq3ptCF}, for the sake of higher-point generalizability, it is useful to obtain the bootstrap equations by following the strategy outlined above.

Upon interchanging $\m{O}_i\leftrightarrow\m{O}_j$, we find that the resulting three-point conformal bootstrap equation is
\eqn{\Vev{\Op{i}{}{1}\Op{j}{}{2}\Op{m}{}{3}}=(-1)^{\xi_i+\xi_j-\xi_m}\Vev{\Op{j}{}{2}\Op{i}{}{1}\Op{m}{}{3}}.}
Writing it out explicitly, we obtain
\eqna{
&\frac{(\m{T}_{12}^{\bs{N}_i})(\m{T}_{21}^{\bs{N}_j})(\m{T}_{31}^{\bs{N}_m})\cdot\sum_{a=1}^{N_{ijm}}\cCF{[a|}{ijm}\mathscr{G}_{[a|}^{ij|m}(\eta_1,\eta_2,\eta_3)}{(\ee{1}{2})^{\frac{1}{2}(\tau_i+\tau_j-\chi_m)}(\ee{1}{3})^{\frac{1}{2}(\chi_i-\chi_j+\tau_m)}(\ee{2}{3})^{\frac{1}{2}(-\chi_i+\chi_j+\chi_m)}}\\
&\qquad=\frac{(-1)^{\xi_i+\xi_j-\xi_m}(\m{T}_{21}^{\bs{N}_j})(\m{T}_{12}^{\bs{N}_i})(\m{T}_{32}^{\bs{N}_m})\cdot\sum_{a=1}^{N_{ijm}}\cCF{[a|}{jim}\mathscr{G}_{[a|}^{ji|m}(\eta_2,\eta_1,\eta_3)}{(\ee{1}{2})^{\frac{1}{2}(\tau_j+\tau_i-\chi_m)}(\ee{2}{3})^{\frac{1}{2}(\chi_j-\chi_i+\tau_m)}(\ee{1}{3})^{\frac{1}{2}(-\chi_j+\chi_i+\chi_m)}},
}
which can be simplified using properties of the half-projectors \eqref{EqTtoT}.  As described above, the half-projectors can be also discarded via appropriate contractions with $(\m{T}_{21\bs{N}_i})(\m{T}_{12\bs{N}_j})(\m{T}_{13\bs{N}_m})*$, which instead leaves us with bootstrap equations involving hatted projectors.  Applying the identities \eqref{EqPtoP} ultimately yields the following bootstrap equation
\eqna{
&\sum_{a=1}^{N_{ijm}}\cCF{[a|}{ijm}(\mathscr{G}_{[a|}^{ij|m})_{\{aA\}\{bB\}\{eE\}}(\eta_1,\eta_2,\eta_3)\\
&\qquad=\sum_{a=1}^{N_{ijm}}(-1)^{\xi_i+\xi_j-\xi_m}\cCF{[a|}{jim}\left((\m{A}_{13})^{n_v^m}\mathscr{G}_{[a|}^{ji|m}(\eta_2,\eta_1,\eta_3)\right)_{\{bB\}\{aA\}\{eE\}}.
}
We next use the explicit basis of tensor structures \eqref{EqTS3pt} to rewrite the three-point conformal bootstrap equation as
\eqna{
&(\h{\m{P}}_{12}^{\bs{N}_i})(\h{\m{P}}_{21}^{\bs{N}_j})(\h{\m{P}}_{31}^{\bs{N}_m})\cdot\sum_{a=1}^{N_{ijm}}\cCF{[a|}{ijm}\left(\prod\m{A}_{123}\right)(\m{A}_{12}\cdot\tilde{\eta}_3)^{\delta_i}(\m{A}_{12}\cdot\tilde{\eta}_3)^{\delta_j}(\m{A}_{12}\cdot\tilde{\eta}_3)^{\delta_m}\\
&\qquad\phantom{=}\times(\epsilon_{123})^{\delta_\epsilon}\left(\Gamma_{123}^{[n-\delta_\Gamma]}\,(\tilde{\eta}_3\cdot\Gamma_{12})^{\delta_\Gamma}\frac{\tilde{\eta}_3\cdot\Gamma_{12}\,\tilde{\eta}_3\cdot\Gamma C_\Gamma^{-1}}{2}\right)^{\xi_i+\xi_j+\xi_m}\\
&\qquad=(\h{\m{P}}_{12}^{\bs{N}_i})(\h{\m{P}}_{21}^{\bs{N}_j})[(\m{A}_{13})^{n_v^m}\h{\m{P}}_{32}^{\bs{N}_m}]\cdot\sum_{a=1}^{N_{ijm}}(-1)^{\xi_i+\xi_j-\xi_m}\cCF{[a|}{jim}\left(\prod\m{A}_{123}\right)(\m{A}_{12}\cdot\tilde{\eta}_3)^{\delta_i}\\
&\qquad\phantom{=}\times(\m{A}_{12}\cdot\tilde{\eta}_3)^{\delta_j}(\m{A}_{12}\cdot\tilde{\eta}_3)^{\delta_m}(\epsilon_{213})^{\delta_\epsilon}\left(\Gamma_{123}^{[n-\delta_\Gamma]}\,(\tilde{\eta}_3\cdot\Gamma_{12})^{\delta_\Gamma}\frac{\tilde{\eta}_3\cdot\Gamma_{12}\,\tilde{\eta}_3\cdot\Gamma C_\Gamma^{-1}}{2}\right)^{\xi_i+\xi_j+\xi_m},
}[EqBoot3p]
where it is important to keep in mind that the powers implicitly depend on the tensor structure through the $a$-index, which specifies the tensor structure; meanwhile, the spinor indices on the $\Gamma$'s  are ordered as $a'b'$, $a'e'$ or $b'e'$ ($b'a'$, $a'e'$ or $b'e'$) on the LHS (RHS).  Invoking the identities \eqref{EqPtoP}, we may re-express the RHS of \eqref{EqBoot3p} as
\eqna{
&(\h{\m{P}}_{12}^{\bs{N}_i})(\h{\m{P}}_{21}^{\bs{N}_j})\left[\h{\m{P}}_{31}^{\bs{N}_m}(\m{A}_{23})^{n_v^m}\left(\frac{\tilde{\eta}_3\cdot\Gamma\,\tilde{\eta}_2\cdot\Gamma}{2}\right)^{2\xi_m}\right]\cdot\sum_{a=1}^{N_{ijm}}(-1)^{\xi_i+\xi_j-\xi_m+\delta_\epsilon}\cCF{[a|}{jim}\left(\prod\m{A}_{123}\right)\\
&\qquad\phantom{=}\times(\m{A}_{12}\cdot\tilde{\eta}_3)^{\delta_i}(\m{A}_{12}\cdot\tilde{\eta}_3)^{\delta_j}(\m{A}_{12}\cdot\tilde{\eta}_3)^{\delta_m}(\epsilon_{123})^{\delta_\epsilon}\left(\Gamma_{123}^{[n-\delta_\Gamma]}\,(\tilde{\eta}_3\cdot\Gamma_{12})^{\delta_\Gamma}\frac{\tilde{\eta}_3\cdot\Gamma_{12}\,\tilde{\eta}_3\cdot\Gamma C_\Gamma^{-1}}{2}\right)^{\xi_i+\xi_j+\xi_m}\\
&\qquad=(\h{\m{P}}_{12}^{\bs{N}_i})(\h{\m{P}}_{21}^{\bs{N}_j})(\h{\m{P}}_{31}^{\bs{N}_m})\cdot\sum_{a=1}^{N_{ijm}}(-1)^{\xi_i+\xi_j-\xi_m+\delta_\epsilon}\cCF{[a|}{jim}\left(\prod\m{A}_{123}\right)(\m{A}_{12}\cdot\tilde{\eta}_3)^{\delta_i}(\m{A}_{12}\cdot\tilde{\eta}_3)^{\delta_j}\\
&\qquad\phantom{=}\times(\m{A}_{23}\cdot\m{A}_{12}\cdot\tilde{\eta}_3)^{\delta_m}(\epsilon_{123})^{\delta_\epsilon}\left(\Gamma_{123}^{[n-\delta_\Gamma]}\,(\tilde{\eta}_3\cdot\Gamma_{12})^{\delta_\Gamma}\frac{\tilde{\eta}_3\cdot\Gamma_{12}\,\tilde{\eta}_3\cdot\Gamma C_\Gamma^{-1}}{2}\right)^{\xi_i+\xi_j+\xi_m},
}
where we have noted that $\m{A}_{23}\cdot\m{A}_{123}=\m{A}_{123}$ and that the factor $\left(\frac{\tilde{\eta}_3\cdot\Gamma\,\tilde{\eta}_2\cdot\Gamma}{2}\right)^{2\xi_m}$ disappears through its contraction with the $\Gamma$'s.

We next observe that we may make the replacement $\m{A}_{23}\cdot\m{A}_{12}\cdot\tilde{\eta}_3\to-\m{A}_{12}\cdot\tilde{\eta}_3$ due to the contraction of $\m{A}_{23}\cdot\m{A}_{12}\cdot\tilde{\eta}_3$ with $\h{\m{P}}_{31}^{\bs{N}_m}$.  Moreover, an extra factor of $(-1)^{(\xi_i+\xi_j-\xi_m)T_n}$, where
\eqn{T_n=\frac{n(n-1)}{2}+n(r+1)+\frac{(r+1)(r+2)}{2},}[EqT]
arises upon properly accounting for the symmetry properties of the $\Gamma$'s and the contraction with the hatted projectors such that the spinor indices are ordered as per our convention (here $a'b'$ instead of $b'a'$).\footnote{The factor $\xi_i+\xi_j-\xi_m$ in the exponent appears since the possible sign difference occurs only when the fermionic quasi-primary operators are $\m{O}_i$ and $\m{O}_j$.  In that case, the relevant factor on the RHS is
\eqn{\left(\frac{\tilde{\eta}_1\cdot\Gamma\,\tilde{\eta}_2\cdot\Gamma}{2}\right)_a^{\phantom{a}a'}\left(\frac{\tilde{\eta}_2\cdot\Gamma\,\tilde{\eta}_1\cdot\Gamma}{2}\right)_b^{\phantom{b}b'}\left(\Gamma_{123}^{[n-\delta_\Gamma]}\,(\tilde{\eta}_3\cdot\Gamma_{12})^{\delta_\Gamma}\frac{\tilde{\eta}_3\cdot\Gamma_{12}\,\tilde{\eta}_3\cdot\Gamma C_\Gamma^{-1}}{2}\right)_{b'a'},}
where the $\Gamma$'s inside the hatted projectors have been extracted [see \eqref{EqPES}].  Re-ordering the terms leads to
\eqna{
&\left(\frac{\tilde{\eta}_1\cdot\Gamma\,\tilde{\eta}_2\cdot\Gamma}{2}\frac{\tilde{\eta}_3\cdot\Gamma\,\tilde{\eta}_3\cdot\Gamma_{12}}{2}\right)_a^{\phantom{a}a'}\left(\frac{\tilde{\eta}_2\cdot\Gamma\,\tilde{\eta}_1\cdot\Gamma}{2}\right)_b^{\phantom{b}b'}\left(\Gamma_{123}^{[n-\delta_\Gamma]}\,(\tilde{\eta}_3\cdot\Gamma_{12})^{\delta_\Gamma}C_\Gamma^{-1}\right)_{b'a'}\\
&\qquad=(-1)^{1+T_n}\left(\frac{\tilde{\eta}_1\cdot\Gamma\,\tilde{\eta}_2\cdot\Gamma}{2}\frac{\tilde{\eta}_3\cdot\Gamma\,\tilde{\eta}_2\cdot\Gamma}{2}\right)_a^{\phantom{a}a'}\left(\frac{\tilde{\eta}_2\cdot\Gamma\,\tilde{\eta}_1\cdot\Gamma}{2}\right)_b^{\phantom{b}b'}\left(\Gamma_{123}^{[n-\delta_\Gamma]}\,(\tilde{\eta}_3\cdot\Gamma_{12})^{\delta_\Gamma}C_\Gamma^{-1}\right)_{a'b'}\\
&\qquad=(-1)^{1+T_n}\left(\frac{\tilde{\eta}_1\cdot\Gamma\,\tilde{\eta}_2\cdot\Gamma}{2}\right)_a^{\phantom{a}a'}\left(\frac{\tilde{\eta}_2\cdot\Gamma\,\tilde{\eta}_1\cdot\Gamma}{2}\frac{\tilde{\eta}_3\cdot\Gamma\,\tilde{\eta}_1\cdot\Gamma}{2}\right)_b^{\phantom{b}b'}\left(\Gamma_{123}^{[n-\delta_\Gamma]}\,(\tilde{\eta}_3\cdot\Gamma_{12})^{\delta_\Gamma}C_\Gamma^{-1}\right)_{a'b'}\\
&\qquad=(-1)^{T_n}\left(\frac{\tilde{\eta}_1\cdot\Gamma\,\tilde{\eta}_2\cdot\Gamma}{2}\right)_a^{\phantom{a}a'}\left(\frac{\tilde{\eta}_2\cdot\Gamma\,\tilde{\eta}_1\cdot\Gamma}{2}\right)_b^{\phantom{b}b'}\left(\Gamma_{123}^{[n-\delta_\Gamma]}\,(\tilde{\eta}_3\cdot\Gamma_{12})^{\delta_\Gamma}\frac{\tilde{\eta}_3\cdot\Gamma_{12}\,\tilde{\eta}_3\cdot\Gamma C_\Gamma^{-1}}{2}\right)_{a'b'},
}
where the extracted $\Gamma$'s can be reabsorbed in the corresponding hatted projectors following \eqref{EqPES}, leading to the sign stated above.
}

With this, the RHS of \eqref{EqBoot3p} can be recast in the same form as the LHS such that
\eqna{
&(\h{\m{P}}_{12}^{\bs{N}_i})(\h{\m{P}}_{21}^{\bs{N}_j})(\h{\m{P}}_{31}^{\bs{N}_m})\cdot\sum_{a=1}^{N_{ijm}}(-1)^{(\xi_i+\xi_j-\xi_m)(1+T_n)+\delta_\epsilon+\delta_m}\cCF{[a|}{jim}\left(\prod\m{A}_{123}\right)(\m{A}_{12}\cdot\tilde{\eta}_3)^{\delta_i}(\m{A}_{12}\cdot\tilde{\eta}_3)^{\delta_j}\\
&\qquad\phantom{=}\times(\m{A}_{12}\cdot\tilde{\eta}_3)^{\delta_m}(\epsilon_{123})^{\delta_\epsilon}\left(\Gamma_{123}^{[n-\delta_\Gamma]}\,(\tilde{\eta}_3\cdot\Gamma_{12})^{\delta_\Gamma}\frac{\tilde{\eta}_3\cdot\Gamma_{12}\,\tilde{\eta}_3\cdot\Gamma C_\Gamma^{-1}}{2}\right)^{\xi_i+\xi_j+\xi_m}.
}
Hence, we find that under the interchange $\m{O}_i\leftrightarrow\m{O}_j$, the three-point conformal blocks in the basis \eqref{EqTS3pt} do not mix, with
\eqn{\cCF{[a|}{ijm}=(-1)^{(\xi_i+\xi_j-\xi_m)(1+T_n)+\delta_\epsilon+\delta_m}\cCF{[a|}{jim}.}

At this point, we may carry out the same analysis for all three-point conformal bootstrap equations (\textit{i.e.}\ any ordering of the three quasi-primary operators), giving
\eqna{
\cCF{[a|}{ijm}&=(-1)^{(\xi_i+\xi_j-\xi_m)(1+T_n)+\delta_\epsilon+\delta_m}\cCF{[a|}{jim}\\
&=(-1)^{(\xi_i+\xi_j+\xi_m)(1+T_n)+2\xi_j\delta_\Gamma+\delta_\epsilon+\delta_j}\cCF{[a|}{mji}\\
&=(-1)^{(-\xi_i+\xi_j+\xi_m)(1+T_n)+2\xi_i\delta_\Gamma+\delta_\epsilon+\delta_i+\delta_j+\delta_m}\cCF{[a|}{imj}\\
&=(-1)^{2\xi_i(1+T_n)+2\xi_j\delta_\Gamma+\delta_i+\delta_j}\cCF{[a|}{jmi}\\
&=(-1)^{2\xi_m(1+T_n)+2\xi_i\delta_\Gamma+\delta_i+\delta_m}\cCF{[a|}{mij},
}[EqBoot3]
from which it is evident that there is no mixing of the tensor structures \eqref{EqTS3pt}.  Given that the permutation group of three elements $\{\m{O}_i,\m{O}_j,\m{O}_m\}$ is generated by $\m{O}_i\leftrightarrow\m{O}_j$ and $\m{O}_j\leftrightarrow\m{O}_m$, it is easy to check that the three-point conformal bootstrap equations \eqref{EqBoot3} are consistent under $S_3$.  The bootstrap equations \eqref{EqBoot3} manifestly demonstrate the superiority of the three-point tensor structure basis \eqref{EqTS3pt} since in this basis the associativity of the three-point correlation functions directly translates into simple symmetry properties of the OPE coefficients themselves, as opposed to linear combinations of different OPE coefficients.

Finally, we note that we did not exploit the general strategy of contracting with the remaining available embedding space coordinates to obtain the three-point conformal bootstrap equations since they were not necessary, just like in the two-point case.  We can however employ this strategy to determine the rotation matrices that transform three-point conformal blocks in the OPE basis \eqref{Eq3ptCB} to three-point conformal blocks in the three-point basis \eqref{Eq3ptCF}.

Indeed, from \eqref{EqRM} we have
\eqn{\mathscr{G}_{(a|}^{ij|m}=\sum_{a'=1}^{N_{ijm}}(R_{ijm}^{-1})_{aa'}\mathscr{G}_{[a'|}^{ij|m},}
which can be converted into a scalar relation by contracting with the invariant tensors $\m{A}_{12}$, $\epsilon_{12}$ and $\Gamma_{12}$ as well as the extra embedding space coordinate $\m{A}_{12}\cdot\tilde{\eta}_3$.  Considering that there is one tensor structure per three-point conformal block, this procedure generates $N_{ijm}$ independent equations\footnote{We note that a different choice of coordinates on the invariant tensors would lead to different linear combinations, albeit with the same solutions.} that relate contracted three-point conformal blocks in the OPE basis to linear combinations of contracted three-point conformal blocks in the three-point basis, effectively determining the rotation matrix $R_{ijm}^{-1}$.  Obviously, it is also possible to determine the rotation matrix by simply rewriting the three-point conformal blocks in the OPE basis in terms of the quantities \eqref{EqA123} and directly comparing with \eqref{EqTS3pt} (see Appendix \ref{SAppRM}).  However, in contrast to the strategy outlined above, a direct comparison necessitates the use of the Fock conditions on the tensor structures to ensure that the final answer is not expressed as an overcomplete basis.

%%%%%%%%%%%%%%%%%%%%%%%%%%%%%%%%%%%%%%%%%%%%%%%%%%

\subsection{Four-Point Bootstrap Equations}

It is well-known that the full set of conformal bootstrap equations can be extracted from the four-point correlation functions of all quasi-primary operators \cite{Ferrara:1973yt,Polyakov:1974gs}.  In this context, there are three different channels:
\eqn{
\begin{gathered}
\text{$s$-channel}:\qquad\Vev{\Op{i}{}{1}\Op{j}{}{2}\Op{l}{}{4}\Op{k}{}{3}}=\Vev{\wick{\c1{\m{O}_i}(\eta_1)\c1{\m{O}_j}(\eta_2)\c1{\m{O}_l}(\eta_4)\c1{\m{O}_k}(\eta_3)}},\\
\text{$t$-channel}:\qquad\Vev{\Op{i}{}{1}\Op{j}{}{2}\Op{l}{}{4}\Op{k}{}{3}}=\Vev{\wick{\c2{\m{O}_i}(\eta_1)\c1{\m{O}_j}(\eta_2)\c2{\m{O}_l}(\eta_4)\c1{\m{O}_k}(\eta_3)}},\\
\text{$u$-channel}:\qquad\Vev{\Op{i}{}{1}\Op{j}{}{2}\Op{l}{}{4}\Op{k}{}{3}}=\Vev{\wick{\c2{\m{O}_i}(\eta_1)\c1{\m{O}_j}(\eta_2)\c1{\m{O}_l}(\eta_4)\c2{\m{O}_k}(\eta_3)}}.
\end{gathered}
}[EqChannel]
The idea of the conformal bootstrap is to demand that the correlation functions in the three channels agree.  In particular, this is the notion that the four-point functions must satisfy consistency conditions due to crossing symmetry, which accordingly lead to constraints on the OPE coefficients.

We note that the symmetry group of the diagram associated with a four-point correlation function (the comb, see \cite{Fortin:2020yjz}) is $H_{4|\text{comb}}=(\mathbb{Z}_2)^2\rtimes\mathbb{Z}_2$.  Since the permutation group of four points is $S_4$, there are $|S_4/H_{4|\text{comb}}|=24/8=3$ different channels associated with a four-point correlation function \cite{Fortin:2020yjz}.  Hence the four-point conformal blocks have nice transformation properties under $H_{4|\text{comb}}$ and the three different channels (which lead to constraints on the OPE coefficients) correspond to the $s$-, $t$-, and $u$-channels described in \eqref{EqChannel}.  Moreover, if we write down an equation between the first two channels and then apply the symmetry properties of the four-point blocks, it is easy to argue that the second equation with the third channel is redundant.  Hence, for reasons that will become clear later, we henceforth consider only on the conformal bootstrap equation between the $s$-channel and the $t$-channel, \textit{i.e.}\
\eqn{\Vev{\wick{\c1{\m{O}_i}(\eta_1)\c1{\m{O}_j}(\eta_2)\c1{\m{O}_l}(\eta_4)\c1{\m{O}_k}(\eta_3)}}=\Vev{\wick{\c2{\m{O}_i}(\eta_1)\c1{\m{O}_j}(\eta_2)\c2{\m{O}_l}(\eta_4)\c1{\m{O}_k}(\eta_3)}}.}[EqBootstrap]

It is simple to re-express the $t$-channel four-point correlation functions as $s$-channel four-point correlation functions with substitutions.  For example, we have
\eqn{\Vev{\wick{\c2{\m{O}_i}(\eta_1)\c1{\m{O}_j}(\eta_2)\c2{\m{O}_l}(\eta_4)\c1{\m{O}_k}(\eta_3)}}=(-1)^{4\xi_j\xi_l}\Vev{\wick{\c1{\m{O}_i}(\eta_1)\c1{\m{O}_l}(\eta_4)\c1{\m{O}_l}(\eta_2)\c1{\m{O}_k}(\eta_3)}},}
which implies that the formula for four-point conformal blocks conformal blocks presented above [see \eqref{Eq4ptCB}] can be used for the $t$-channel with the substitutions $\eta_2\leftrightarrow\eta_4$ and $j\leftrightarrow l$.

It is apparent from the above discussion of the symmetry properties of the OPE coefficients in the three-point basis that the four-point conformal bootstrap equations must also be simpler in the same basis.  Starting from \eqref{Eq4pt}, we thus have
\eqna{
&\Vev{\Op{i}{}{1}\Op{j}{}{2}\Op{l}{}{4}\Op{k}{}{3}}\\
&\qquad=\frac{(\m{T}_{12}^{\bs{N}_i})(\m{T}_{21}^{\bs{N}_j})(\m{T}_{43}^{\bs{N}_l})(\m{T}_{34}^{\bs{N}_k})\cdot\sum_{m,n}\sum_{a=1}^{N_{ijm}}\sum_{b=1}^{N_{klm}}\cCF{[a|}{ijm}\cCF{[b|}{lkn}G^{nm}\mathscr{G}_{[a|b]}^{ij|m|lk}(\eta_1,\eta_2,\eta_4,\eta_3)}{(\ee{1}{2})^{\frac{1}{2}(\tau_i+\tau_j)}(\ee{3}{4})^{\frac{1}{2}(\tau_k+\tau_l)}(\ee{1}{3})^{\frac{1}{2}(\chi_k-\chi_l)}(\ee{2}{4})^{\frac{1}{2}(-\chi_i+\chi_j)}(\ee{1}{4})^{\frac{1}{2}(\chi_i-\chi_j-\chi_k+\chi_l)}}\\
&\qquad=\frac{(-1)^{4\xi_j\xi_l}(\m{T}_{14}^{\bs{N}_i})(\m{T}_{41}^{\bs{N}_l})(\m{T}_{23}^{\bs{N}_j})(\m{T}_{32}^{\bs{N}_k})\cdot\sum_{m,n}\sum_{a=1}^{N_{ilm}}\sum_{b=1}^{N_{kjm}}\cCF{[a|}{ilm}\cCF{[b|}{jkn}G^{nm}\mathscr{G}_{[a|b]}^{il|m|jk}(\eta_1,\eta_4,\eta_2,\eta_3)}{(\ee{1}{4})^{\frac{1}{2}(\tau_i+\tau_l)}(\ee{2}{3})^{\frac{1}{2}(\tau_k+\tau_j)}(\ee{1}{3})^{\frac{1}{2}(\chi_k-\chi_j)}(\ee{2}{4})^{\frac{1}{2}(-\chi_i+\chi_l)}(\ee{1}{2})^{\frac{1}{2}(\chi_i-\chi_l-\chi_k+\chi_j)}},
}
which transforms into
\eqna{
&\sum_{m,n}\sum_{a,a'=1}^{N_{ijm}}\sum_{b,b'=1}^{N_{klm}}\cCF{[a|}{ijm}(R_{ijm})_{aa'}\cCF{[b|}{lkn}(R_{lkm})_{bb'}G^{nm}\\
&\qquad\times v^{\frac{1}{2}(\chi_j+\chi_k)}\left(\mathscr{G}_{(a'|b')}^{ij|m|lk}(\eta_1,\eta_2,\eta_4,\eta_3)\right)_{\{aA\}\{bB\}\{dD\}\{cC\}}\\
&\qquad=(-1)^{4\xi_j\xi_l}\sum_{m,n}\sum_{a,a'=1}^{N_{ilm}}\sum_{b,b'=1}^{N_{kjm}}\cCF{[a|}{ilm}(R_{ilm})_{aa'}\cCF{[b|}{jkn}(R_{jkm})_{bb'}G^{nm}u^{\frac{1}{2}(\chi_k+\chi_l)}\\
&\qquad\phantom{=}\times\left((\m{A}_{12})^{n_v^i}(\m{A}_{34})^{n_v^l}(\m{A}_{12})^{n_v^j}(\m{A}_{34})^{n_v^k}\mathscr{G}_{(a'|b')}^{il|m|jk}(\eta_1,\eta_4,\eta_2,\eta_3)\right)_{\{aA\}\{dD\}\{bB\}\{cC\}},
}[EqBoot4]
after contracting with half-projectors to convert from embedding space spinor indices to embedding space vector indices and rewriting the blocks directly in terms of \eqref{Eq4ptCBSoln}.  To obtain purely scalar conformal bootstrap equations, we can contract \eqref{EqBoot4} with the invariant tensors $\m{A}_{12}$,  $\epsilon_{12}$ and $\Gamma_{12}$, as well as the remaining embedding space coordinates $\bar{\eta}_3\cdot\m{A}_{12}$ and $\bar{\eta}_4\cdot\m{A}_{12}$, which yields the correct number of independent four-point conformal bootstrap equations in terms of objects that are functions of the conformal cross-ratios $u$ and $v$ only.

The general strategy outlined above thus generates conformal bootstrap equations in terms of the cross-ratios $u$ and $v$, as in position space, but directly in the embedding space.  Thus, in the context of the embedding space formalism, there is no need to project back to position space.

%%%%%%%%%%%%%%%%%%%%%%%%%%%%%%%%%%%%%%%%%%%%%%%%%%
%%%%%%%%%%%%%%%%%%%%%%%%%%%%%%%%%%%%%%%%%%%%%%%%%%

\section{Algorithm}\label{SecAlg}

In this section we expound our full general algorithm for computing the rotation matrices and generating the conformal bootstrap equations.  We illustrate each step of the algorithm in the context of the simplest nontrivial example, namely $\Vev{SSSV}$ (assuming $d$ large to simplify tensor product decompositions).  The algorithm is presented as a list of steps for the four-point function $\Vev{\m{O}_i\m{O}_j\m{O}_l\m{O}_k}$.  Since the irreducible representations of the external quasi-primary operators are fixed, we assume that their projection operators are known.  We give the input data for a more complicated example in Appendix \ref{SAppEx}.

To simplify the notation, the final results for $\Vev{SSSV}$ are written in terms of the usual scalar conformal block $\m{G}_{(\Delta,\ell)}^{(d,\Lambda,\Lambda')}\equiv\m{G}_{(\Delta,\ell)}^{(d,\Lambda,0,0,\Lambda',0,0)}$ as well as the normalization factor $\omega_{(\Delta,\ell)}^{(d,\Lambda,\Lambda')}\equiv\omega_{(\Delta,\ell)}^{(d,\Lambda,0,\Lambda',0)}$ with conformal dimension $\Delta\equiv\Delta_{m+\ell}$.  For convenience, we define the $s$- and $t$-channel quantities $a_s$, $b_s$ and $a_t$, $b_t$, respectively, to denote the differences of the conformal dimensions of the external operators.  These are $a_s=(\Delta_j-\Delta_i)/2$, $b_s=(\Delta_k-\Delta_l)/2$ and $a_t=(\Delta_l-\Delta_i)/2$, $b_t=(\Delta_k-\Delta_j)/2$ for the two channels.

%%%%%%%%%%%%%%%%%%%%%%%%%%%%%%%%%%%%%%%%%%%%%%%%%%

\subsection{Rotation Matrices}

\begin{enumerate}
\item\textit{Determine the possible infinite towers of exchanged quasi-primary operators appearing in $\Vev{\m{O}_i\m{O}_j\m{O}_l\m{O}_k}$ by identifying the overlapping set of exchanged operators $\m{O}_m$ for the two different three-point correlation functions $\Vev{\m{O}_i\m{O}_j\m{O}_m}$ and $\Vev{\m{O}_l\m{O}_k\m{O}_m}$.}

For $\Vev{SSSV}$, the two three-point correlation functions of interest are $\Vev{SS\m{O}_m}$ with $\m{O}_m\in\{\m{O}^{\ell\bs{e}_1}\}$ and $\Vev{SV\m{O}_m}$ with $\m{O}_m\in\{\m{O}^{\ell\bs{e}_1},\m{O}^{\bs{e}_2+\ell\bs{e}_1}\}$.  Since the only non-vanishing three-point functions with two scalars are those that contain a spin-$\ell$ quasi-primary operator and since such an exchanged operator leads to non-vanishing three-point functions with one scalar and one vector, it follows that there is only one infinite tower of exchanged quasi-primary operators with $\m{O}^{\ell\bs{e}_1}$ for which $n_v^m=0$ and $\xi_m=0$.
\item\textit{For each infinite tower of exchanged quasi-primary operators, find the corresponding projection operators using group theory and the complete set of allowed tensor structures in the OPE basis with the help of Table \ref{TabBoot} and the Fock conditions.}

The projection operator for $\ell\bs{e}_1$ is well-known and is given in \eqref{EqPle1}.

Since $\bs{0}\otimes\bs{0}\otimes\ell\bs{e}_1=\ell\bs{e}_1$, there is only one tensor structure for $\Vev{SS\m{O}^{\ell\bs{e}_1}}$ and it is given by
\eqn{(\tCF{(1|}{ij,m+\ell}{12})_{\{E\}}^{\phantom{\{E\}}\{F\}}=(\m{A}_{12E}^{\phantom{12E}F})^\ell\to\tCF{(1|}{ijm}{12}=1\quad\text{with}\quad i_1=0, n_1=\ell,}[EqTS-SSO]
where we have extracted the special part (which is trivial in this case).

Meanwhile, for the other three-point correlation function, namely $\Vev{SV\m{O}^{\ell\bs{e}_1}}$, one has $\bs{0}\otimes\bs{e}_1\otimes\ell\bs{e}_1=[\bs{e}_2+(\ell-1)\bs{e}_1]\oplus(\ell+1)\bs{e}_1\oplus(\ell-1)\bs{e}_1$, which leads to two tensor structures given by
\eqn{
\begin{gathered}
(\tCF{(1|}{ij,m+\ell}{12})_{B\{E\}}^{\phantom{B\{E\}}\{F\}}=\m{A}_{12B}^{\phantom{12B}F}(\m{A}_{12E}^{\phantom{12E}F})^\ell\to\tCF{(1|}{ijm}{12}=\m{A}_{12B}^{\phantom{12B}F}\quad\text{with}\quad i_1=0, n_1=\ell+1,\\
(\tCF{(2|}{ij,m+\ell}{12})_{B\{E\}}^{\phantom{B\{E\}}\{F\}}=\m{A}_{12BE}(\m{A}_{12E}^{\phantom{12E}F})^{\ell-1}\to\tCF{(2|}{ij,m+1}{12}=\m{A}_{12BE}\quad\text{with}\quad i_2=1, n_2=\ell-1.
\end{gathered}
}[EqTS-SVO]
We can straightforwardly identify the special parts and related input data from the form of the respective tensor structures.
\item\textit{Invoking \eqref{EqG3Soln} for the three-point tensorial block, use \eqref{Eq3ptCBSoln} to write down the explicit three-point conformal blocks for each tensor structure.}

For $\Vev{SS\m{O}^{\ell\bs{e}_1}}$, the sole three-point conformal block associated with the tensor structure \eqref{EqTS-SSO} is
\eqna{
(\mathscr{G}_{(1|}^{ij|m+\ell})_{\{E\}}(\eta_1,\eta_2,\eta_3)&=(\h{\m{P}}_{31}^{\ell\bs{e}_1})_{\{E\}}^{\phantom{\{E\}}\{E'\}}\frac{\m{G}_{(\chi_{m+\ell},\ell)\{E'\}}^{(d,\chi_{m+\ell}/2+h_{ij,m+\ell},0)}(\eta_1,\eta_2,\eta_3)}{R_{(\chi_{m+\ell},\ell)}^{(d,\chi_{m+\ell}/2+h_{ij,m+\ell},0)}}\\
&=\frac{1}{R_{(\chi_{m+\ell},\ell)}^{(d,\chi_{m+\ell}/2+h_{ij,m+\ell},0)}}[\h{\m{P}}_{31}^{\ell\bs{e}_1}(\m{A}_{12}\cdot\tilde{\eta}_3)^\ell]_{\{E\}},
}[Eq3ptCB-SSO]
where we have applied \eqref{Eq3ptCBSoln} with the appropriate input data and then used \eqref{EqG3Soln}.

The same procedure can be followed for the two three-point conformal blocks of $\Vev{SV\m{O}^{\ell\bs{e}_1}}$ as dictated by \eqref{EqTS-SVO}, leading to
\eqna{
&(\mathscr{G}_{(1|}^{ij|m+\ell})_{\{aA\}\{bB\}\{eE\}}(\eta_1,\eta_2,\eta_3)\\
&\qquad=(\h{\m{P}}_{21}^{\bs{N}_j})_B^{\phantom{B}B'}(\h{\m{P}}_{31}^{\ell\bs{e}_1})_{\{E\}}^{\phantom{\{E\}}\{E'\}}\m{A}_{12B'}^{\phantom{12B'}F}\frac{\m{G}_{(\chi_{m+\ell},\ell)\{E'\}\{F\}}^{(d,\chi_{m+\ell}/2+h_{ij,m+\ell},1)}(\eta_1,\eta_2,\eta_3)}{R_{(\chi_{m+\ell},\ell)}^{(d,\chi_{m+\ell}/2+h_{ij,m+\ell},1)}},
}[Eq3ptCBG-SVO1]
and
\eqna{
&(\mathscr{G}_{(2|}^{ij|m+\ell})_{\{aA\}\{bB\}\{eE\}}(\eta_1,\eta_2,\eta_3)\\
&\qquad=(\h{\m{P}}_{21}^{\bs{N}_j})_B^{\phantom{B}B'}(\h{\m{P}}_{31}^{\ell\bs{e}_1})_{\{E\}}^{\phantom{\{E\}}\{E'\}}\m{A}_{13B'}^{\phantom{13B'}E''}\\
&\qquad\phantom{=}\times\left[\frac{\m{G}_{(\chi_{m+\ell}+1,\ell-1)\{E'\}\{E'E''\}}^{(d,\chi_{m+\ell}/2+h_{ij,m+\ell}-1/2,2)}(\eta_1,\eta_2,\eta_3)}{R_{(\chi_{m+\ell}+1,\ell-1)}^{(d,\chi_{m+\ell}/2+h_{ij,m+\ell}-1/2,2)}}+g_{E'E''}\frac{\m{G}_{(\chi_{m+\ell},\ell-1)\{E'\}}^{(d,\chi_{m+\ell}/2+h_{ij,m+\ell},0)}(\eta_1,\eta_2,\eta_3)}{R_{(\chi_{m+\ell},\ell-1)}^{(d,\chi_{m+\ell}/2+h_{ij,m+\ell},0)}}\right].
}[Eq3ptCBG-SVO2]
For easier readability, we did not expand the blocks using \eqref{EqG3Soln} at this stage, as we will express them in the three-point basis in the next step by following the procedure described in Appendix \ref{SAppRM}.
\item\textit{Re-express the previous three-point conformal blocks, which are in the OPE basis, in the three-point basis \eqref{EqTS3pt} to extract the rotation matrix elements.}

The OPE tensor structure \eqref{EqTS-SSO} implies that the corresponding three-point conformal block in the three-point basis is
\eqn{(\mathscr{G}_{[1|}^{ij|m+\ell})_{\{E\}}(\eta_1,\eta_2,\eta_3)=(\h{\m{P}}_{31}^{\ell\bs{e}_1})_{\{E\}}^{\phantom{\{E\}}\{E'\}}[(\m{A}_{12}\cdot\tilde{\eta}_3)_{E'}]^\ell.}
Therefore, from the three-point conformal block \eqref{Eq3ptCB-SSO}, it is straightforward to conclude that the rotation matrix for $\Vev{SSO^{\ell\bs{e}_1}}$ is simply
\eqn{(R_{ij,m+\ell}^{-1})_{11}=\frac{1}{R_{(\Delta,\ell)}^{(d,a_s,0)}},}[EqRM-SSO]
by direct comparison with the result in the three-point basis.

For $\Vev{SV\m{O}^{\ell\bs{e}_1}}$, the OPE tensor structures \eqref{EqTS-SVO} lead to the associated three-point conformal blocks in the three-point basis given by
\eqna{
(\mathscr{G}_{[1|}^{ij|m+\ell})_{B\{E\}}(\eta_1,\eta_2,\eta_3)&=(\h{\m{P}}_{21}^{\bs{e}_1})_B^{\phantom{B}B'}(\h{\m{P}}_{31}^{\ell\bs{e}_1})_{\{E\}}^{\phantom{\{E\}}\{E'\}}(\m{A}_{12}\cdot\tilde{\eta}_3)_{B'}[(\m{A}_{12}\cdot\tilde{\eta}_3)_{E'}]^\ell,\\
(\mathscr{G}_{[2|}^{ij|m+\ell})_{B\{E\}}(\eta_1,\eta_2,\eta_3)&=(\h{\m{P}}_{21}^{\bs{e}_1})_B^{\phantom{B}B'}(\h{\m{P}}_{31}^{\ell\bs{e}_1})_{\{E\}}^{\phantom{\{E\}}\{E'\}}\m{A}_{123B'E'}[(\m{A}_{12}\cdot\tilde{\eta}_3)_{E'}]^{\ell-1}.
}
To rotate to the proper basis starting from \eqref{Eq3ptCBG-SVO1} and \eqref{Eq3ptCBG-SVO2}, one can expand using \eqref{EqG3Soln} as noted above.  But it is also possible to expand the special parts of the tensor structures in terms of \eqref{EqA123} with the identities \eqref{EqOPEtoTS} and then use the contiguous relations \eqref{EqG3CR} to reach contractions of the type \eqref{EqG3SolnA123}, which can then be easily expressed in terms of objects relevant to the three-point basis (see Appendix \ref{SAppRM}).

For \eqref{Eq3ptCBG-SVO1}, one first rewrites $\m{A}_{12B'}^{\phantom{12B'}F}$ in terms of $\m{A}_{123B'}^{\phantom{123B'}F}$ and $(\m{A}_{12}\cdot\tilde{\eta}_3)_{B'}(\m{A}_{12}\cdot\tilde{\eta}_3)^F$ which leads to
\eqna{
&(\mathscr{G}_{(1|}^{ij|m+\ell})_{\{aA\}\{bB\}\{eE\}}(\eta_1,\eta_2,\eta_3)\\
&\qquad=(\h{\m{P}}_{21}^{\bs{N}_j})_B^{\phantom{B}B'}(\h{\m{P}}_{31}^{\ell\bs{e}_1})_{\{E\}}^{\phantom{\{E\}}\{E'\}}\left\{(R_{ij,m+\ell}^{-1})_{11}(\m{A}_{12}\cdot\tilde{\eta}_3)_{B'}[(\m{A}_{12}\cdot\tilde{\eta}_3)_{E'}]^\ell\right.\\
&\qquad\phantom{=}\left.+(R_{ij,m+\ell}^{-1})_{12}\m{A}_{123B'E'}[(\m{A}_{12}\cdot\tilde{\eta}_3)_{E'}]^{\ell-1}\right\},
}[Eq3ptCB-SVO1]
such that two of the four rotation matrix elements are
\eqn{
\begin{gathered}
(R_{ij,m+\ell}^{-1})_{11}=\frac{1}{2}\tilde{\kappa}_{(\Delta,\ell)}^{(d,a_s+1/2,0)}(0)-\frac{1}{2}\tilde{\kappa}_{(\Delta-1,\ell)}^{(d,a_s,0)}(0)+\frac{1}{2}\rho^{(d,1;\Delta/2-a_s-\ell/2-1/2)}\tilde{\kappa}_{(\Delta,\ell)}^{(d,a_s-1/2,0)}(0),\\
(R_{ij,m+\ell}^{-1})_{12}=\tilde{\kappa}_{(\Delta,\ell)}^{(d,a_s,1)}(1).
\end{gathered}
}[EqRM-SVO1]

Similarly, the first term of \eqref{Eq3ptCBG-SVO2} can be rewritten, up to a prefactor, as the projection operators contracted with
\eqn{\m{A}_{13B'}^{\phantom{13B'}F}\m{A}_{13E'}^{\phantom{13E'}F}\m{G}_{(\chi_{m+\ell}+1,\ell-1)\{E'\}\{FF\}}^{(d,\chi_{m+\ell}/2+h_{ij,m+\ell}-1/2,2)}(\eta_1,\eta_2,\eta_3).}
Re-expressing both $\m{A}_{13}$ in terms of $\m{A}_{123}$ and $(\m{A}_{12}\cdot\tilde{\eta}_3)(\m{A}_{12}\cdot\tilde{\eta}_3)$ allows to proceed with the technique of Appendix \ref{SAppRM} [the second term of \eqref{Eq3ptCBG-SVO2} being trivial], which ultimately yields
\eqna{
&(\mathscr{G}_{(2|}^{ij|m+\ell})_{\{aA\}\{bB\}\{eE\}}(\eta_1,\eta_2,\eta_3)\\
&\qquad=(\h{\m{P}}_{21}^{\bs{N}_j})_B^{\phantom{B}B'}(\h{\m{P}}_{31}^{\ell\bs{e}_1})_{\{E\}}^{\phantom{\{E\}}\{E'\}}\left\{(R_{ij,m+\ell}^{-1})_{21}(\m{A}_{12}\cdot\tilde{\eta}_3)_{B'}[(\m{A}_{12}\cdot\tilde{\eta}_3)_{E'}]^\ell\right.\\
&\qquad\phantom{=}\left.+(R_{ij,m+\ell}^{-1})_{22}\m{A}_{123B'E'}[(\m{A}_{12}\cdot\tilde{\eta}_3)_{E'}]^{\ell-1}\right\},
}[Eq3ptCB-SVO2]
with the last two rotation matrix elements being given by
\eqna{
&(R_{ij,m+\ell}^{-1})_{21}\\
&\qquad=\frac{1}{4}\tilde{\kappa}_{(\Delta-1,\ell-1)}^{(d,a_s-1/2,0)}(0)-\frac{1}{4}\tilde{\kappa}_{(\Delta,\ell-1)}^{(d,a_s-1/2,1)}(1)-\frac{1}{4}\tilde{\kappa}_{(\Delta,\ell-1)}^{(d,a_s,1)}(1)+\frac{1}{4}\tilde{\kappa}_{(\Delta+1,\ell-1)}^{(d,a_s+1/2,0)}(0)\\
&\qquad\phantom{=}-\frac{1}{2}\rho^{(d,1;\Delta/2-a_s-\ell/2+1/2)}\tilde{\kappa}_{(\Delta,\ell-1)}^{(d,a_s-1,0)}(0)+\frac{1}{4}\rho^{(d,1;\Delta/2-a_s-\ell/2+1/2)}\tilde{\kappa}_{(\Delta+1,\ell-1)}^{(d,a_s-1,1)}(1)\\
&\qquad\phantom{=}-\frac{1}{2}\rho^{(d,1;\Delta/2-a_s-\ell/2+1/2)}\tilde{\kappa}_{(\Delta+1,\ell-1)}^{(d,a_s-1/2,0)}(0)+\frac{1}{4}\rho^{(d,2;\Delta/2-a_s-\ell/2+1/2)}\tilde{\kappa}_{(\Delta+1,\ell-1)}^{(d,a_s-3/2,0)}(0),\\
&(R_{ij,m+\ell}^{-1})_{22}\\
&\qquad=\tilde{\kappa}_{(\Delta,\ell-1)}^{(d,a_s,0)}(0)-\frac{1}{2}\tilde{\kappa}_{(\Delta,\ell-1)}^{(d,a_s-1/2,1)}(1)+\tilde{\kappa}_{(\Delta+1,\ell-1)}^{(d,a_s-1/2,2)}(0)-\frac{1}{2}\tilde{\kappa}_{(\Delta+1,\ell-1)}^{(d,a_s,1)}(1)\\
&\qquad\phantom{=}+\frac{1}{2}\tilde{\kappa}_{(\Delta+1,\ell-1)}^{(d,a_s-1/2,2)}(2)+\frac{1}{2}\rho^{(d,1;\Delta/2-a_s-\ell/2+1/2)}\tilde{\kappa}_{(\Delta+1,\ell-1)}^{(d,a_s-1,1)}(1).
}[EqRM-SVO2]
Therefore, in the $s$-channel the rotation matrix elements appearing in \eqref{EqBoot4} are \eqref{EqRM-SSO} for $\Vev{\m{O}_i\m{O}_j\m{O}_m}=\Vev{SS\m{O}^{\ell\bs{e}_1}}$ as well as \eqref{EqRM-SVO1} and \eqref{EqRM-SVO2} with $i\to l$ and $j\to k$ ($a_s\to b_s$) for $\Vev{\m{O}_l\m{O}_k\m{O}_m}=\Vev{SV\m{O}^{\ell\bs{e}_1}}$.
\item\textit{Repeat the steps above for the $t$-channel $\Vev{\m{O}_i\m{O}_l\m{O}_j\m{O}_k}$.}

Since $\Vev{SSSV}$ is invariant at the level of the irreducible representations of the external quasi-primary operators under permutation to the $t$-channel, we may directly adapt the above results to the $t$-channel case by making straightforward substitutions.  Hence, the rotation matrix elements in the $t$-channel relevant for the bootstrap equations \eqref{EqBoot4} are \eqref{EqRM-SSO} with $j\to l$ ($a_s\to a_t$) for $\Vev{\m{O}_i\m{O}_l\m{O}_m}=\Vev{SS\m{O}^{\ell\bs{e}_1}}$ as well as \eqref{EqRM-SVO1} and \eqref{EqRM-SVO2} with $i\to j$ and $j\to k$ ($a_s\to b_t$) for $\Vev{\m{O}_j\m{O}_k\m{O}_m}=\Vev{SV\m{O}^{\ell\bs{e}_1}}$.
\end{enumerate}

%%%%%%%%%%%%%%%%%%%%%%%%%%%%%%%%%%%%%%%%%%%%%%%%%%

\subsection{Bootstrap Equations}

\begin{enumerate}
\item\textit{For both the $s$-channel $\Vev{\m{O}_i\m{O}_j\m{O}_l\m{O}_k}$ and the $t$-channel $\Vev{\m{O}_i\m{O}_l\m{O}_j\m{O}_k}$, compute the four-point conformal blocks using \eqref{Eq4ptCBSoln} and the data previously found (projection operators and tensor structures) for each infinite tower of exchanged quasi-primary operators.}

Starting with the $s$-channel $\Vev{SSSV}$, the sole infinite tower of exchanged quasi-primary operators $\m{O}^{\ell\bs{e}_1}$ has the projection operators given by \eqref{EqPle1} which implies that the expansion \eqref{EqPExp} is trivial with only one term ($t=1$ and $\ell_1=0$) for which $\mathscr{A}_1(d,\ell)=1$ and $\h{\m{Q}}_{23|1}^{\bs{0}}=1$.  Therefore, the two different associated four-point conformal blocks are simply
\eqna{
&(\mathscr{G}_{(1|1)}^{ij|m+\ell|lk})_{\{aA\}\{bB\}\{dD\}\{cC\}}(\eta_1,\eta_2,\eta_4,\eta_3)\\
&\qquad=(\h{\m{P}}_{34}^{\bs{N}_k})_C^{\phantom{C}F'}\frac{\m{G}_{(\chi_{m+\ell},\ell)\{\}\{\}\{F'\}\{\}}^{(d,\chi_{m+\ell}/2+h_{ij,m+\ell},0,0,\chi_{m+\ell}/2+h_{lk,m+\ell},1,0)}}{c_{(d,\ell)}R_{(\chi_{m+\ell},\ell)}^{(d,\chi_{m+\ell}/2+h_{ij,m+\ell},0)}R_{(\chi_{m+\ell},\ell)}^{(d,\chi_{m+\ell}/2+h_{lk,m+\ell},1)}},
}[Eq4ptCBG-s1]
and
\eqna{
&(\mathscr{G}_{(1|2)}^{ij|m+\ell|lk})_{\{aA\}\{bB\}\{dD\}\{cC\}}(\eta_1,\eta_2,\eta_4,\eta_3)\\
&\qquad=(\h{\m{P}}_{34}^{\bs{N}_k})_C^{\phantom{C}E'}\frac{\m{G}_{(\chi_{m+\ell},\ell)\{\}\{\}\{\}\{E'\}}^{(d,\chi_{m+\ell}/2+h_{ij,m+\ell},0,0,\chi_{m+\ell}/2+h_{lk,m+\ell}-1/2,0,1)}}{c_{(d,\ell)}R_{(\chi_{m+\ell},\ell)}^{(d,\chi_{m+\ell}/2+h_{ij,m+\ell},0)}R_{(\chi_{m+\ell},\ell)}^{(d,\chi_{m+\ell}/2+h_{lk,m+\ell}-1/2,0)}},
}[Eq4ptCBG-s2]
when using the tensor structures \eqref{EqTS-SSO} and \eqref{EqTS-SVO} in the definition \eqref{Eq4ptCBSoln}.

Since the irreducible representations of the external quasi-primary operators in the $t$-channel are positioned exactly in the same order, the four-point conformal blocks are analogous with obvious substitutions, \textit{i.e.}\
\eqna{
&(\mathscr{G}_{(1|1)}^{il|m+\ell|jk})_{\{aA\}\{dD\}\{bB\}\{cC\}}(\eta_1,\eta_4,\eta_2,\eta_3)\\
&\qquad=(\h{\m{P}}_{32}^{\bs{N}_k})_C^{\phantom{C}F'}\frac{\m{G}_{(\chi_{m+\ell},\ell)\{\}\{\}\{F'\}\{\}}^{(d,\chi_{m+\ell}/2+h_{il,m+\ell},0,0,\chi_{m+\ell}/2+h_{jk,m+\ell},1,0)}}{c_{(d,\ell)}R_{(\chi_{m+\ell},\ell)}^{(d,\chi_{m+\ell}/2+h_{il,m+\ell},0)}R_{(\chi_{m+\ell},\ell)}^{(d,\chi_{m+\ell}/2+h_{jk,m+\ell},1)}},
}[Eq4ptCBG-t1]
and
\eqna{
&(\mathscr{G}_{(1|2)}^{il|m+\ell|jk})_{\{aA\}\{dD\}\{bB\}\{cC\}}(\eta_1,\eta_4,\eta_2,\eta_3)\\
&\qquad=(\h{\m{P}}_{32}^{\bs{N}_k})_C^{\phantom{C}E'}\frac{\m{G}_{(\chi_{m+\ell},\ell)\{\}\{\}\{\}\{E'\}}^{(d,\chi_{m+\ell}/2+h_{il,m+\ell},0,0,\chi_{m+\ell}/2+h_{jk,m+\ell}-1/2,0,1)}}{c_{(d,\ell)}R_{(\chi_{m+\ell},\ell)}^{(d,\chi_{m+\ell}/2+h_{il,m+\ell},0)}R_{(\chi_{m+\ell},\ell)}^{(d,\chi_{m+\ell}/2+h_{jk,m+\ell}-1/2,0)}}.
}[Eq4ptCBG-t2]
\item\textit{Perform the $\circ$-product with the help of \eqref{EqCircProd} to write each four-point conformal block (in both channels) as a linear combination of the four-point tensorial blocks \eqref{EqG4}.}

It is clear from the four-point conformal blocks in both channels \eqref{Eq4ptCBG-s1}, \eqref{Eq4ptCBG-s2}, \eqref{Eq4ptCBG-t1} and \eqref{Eq4ptCBG-t2} that no such $\circ$-product needs to be performed to express all conformal blocks in terms of the four-point tensorial blocks \eqref{EqG4}.  This occurs since the projection operators for $\ell\bs{e}_1$ are the base projection operators with trivial expansions \eqref{EqPExp}.
\item\textit{From the free embedding space indices of the external quasi-primary operators in the bootstrap equations \eqref{EqBoot4}, determine the independent set of $\m{A}_{12}$, $\epsilon_{12}$, $\Gamma_{12}$, $\bar{\eta}_3\cdot\m{A}_{12}$ and $\bar{\eta}_4\cdot\m{A}_{12}$ contractions which lead to fully scalar four-point bootstrap equations by referring to Table \ref{TabBoot}.}

For $\Vev{SSSV}$, we have $\bs{0}\otimes\bs{0}\otimes\bs{0}\otimes\bs{e}_1=\bs{e}_1$ which implies two independent contractions according to Table \ref{TabBoot}.  It is straightforward to see that the two contractions of the bootstrap equations \eqref{EqBoot4} that lead to two independent fully scalar four-point bootstrap equations are given by $(\bar{\eta}_3\cdot\m{A}_{12})^C$ and $(\bar{\eta}_4\cdot\m{A}_{12})^C$.  It is however more convenient to choose another basis for the contractions, namely $(\bar{\eta}_1\cdot\m{A}_{34})^C$ and $(\bar{\eta}_2\cdot\m{A}_{34})^C$, to simplify the computations below.
\item\textit{For each independent contraction found above, use the contiguous relations of Section \ref{SecTensorGeneralizations} and the action of the differential operators in Appendix \ref{SAppProofs} to rewrite the fully contracted four-point tensorial blocks appearing in the associated bootstrap equation \eqref{EqBoot4} in terms of standard four-point scalar blocks for spin-$\ell$ exchange.}

For the first contraction $(\bar{\eta}_1\cdot\m{A}_{34})^C$, the $s$-channel side of the bootstrap equation \eqref{EqBoot4} leads to
\eqna{
&(\bar{\eta}_1\cdot\m{A}_{34})^C(\mathscr{G}_{(1|1)}^{ij|m+\ell|lk})_{\{aA\}\{bB\}\{dD\}\{cC\}}(\eta_1,\eta_2,\eta_4,\eta_3)\\
&\qquad=(\bar{\eta}_1\cdot\m{A}_{34})^{F'}\frac{\m{G}_{(\chi_{m+\ell},\ell)\{\}\{\}\{F'\}\{\}}^{(d,\chi_{m+\ell}/2+h_{ij,m+\ell},0,0,\chi_{m+\ell}/2+h_{lk,m+\ell},1,0)}}{c_{(d,\ell)}R_{(\chi_{m+\ell},\ell)}^{(d,\chi_{m+\ell}/2+h_{ij,m+\ell},0)}R_{(\chi_{m+\ell},\ell)}^{(d,\chi_{m+\ell}/2+h_{lk,m+\ell},1)}}\\
&\qquad=-\frac{2b_s+1+a_s(1+u-v)}{(\Delta+1-d)u^{\frac{1}{2}}}\frac{\m{G}_{(\Delta,\ell)}^{(d,a_s,b_s+1/2)}}{\omega_{(\Delta,\ell)}^{(d,a_s,b_s+1/2)}}+\frac{1}{\Delta+1-d}\bar{\eta}_1\cdot\bar{\m{D}}_{34}\frac{\m{G}_{(\Delta,\ell)}^{(d,a_s,b_s+1/2)}}{\omega_{(\Delta,\ell)}^{(d,a_s,b_s+1/2)}}\\
&\qquad\phantom{=}+[\Delta-\ell+1-d-2d(\Delta+1-d)+2b_s(3\Delta+2b_s+\ell+2-3d)]\frac{(\Delta-2b_s-\ell-1)}{2(\Delta+1-d)u^{\frac{1}{2}}}\\
&\qquad\phantom{=}\times\frac{\m{G}_{(\Delta,\ell)}^{(d,a_s,b_s-1/2)}}{\omega_{(\Delta,\ell)}^{(d,a_s,b_s-1/2)}}-\frac{(\Delta-2b_s-\ell-1)(\Delta-2b_s-\ell+1-d)}{2(\Delta+1-d)}\bar{\eta}_1\cdot\bar{\m{D}}_{43}\frac{\m{G}_{(\Delta,\ell)}^{(d,a_s,b_s-1/2)}}{\omega_{(\Delta,\ell)}^{(d,a_s,b_s-1/2)}},
}[Eq4ptCB-s1C1]
and
\eqna{
&(\bar{\eta}_1\cdot\m{A}_{34})^C(\mathscr{G}_{(1|2)}^{ij|m+\ell|lk})_{\{aA\}\{bB\}\{dD\}\{cC\}}(\eta_1,\eta_2,\eta_4,\eta_3)\\
&\qquad=(\bar{\eta}_1\cdot\m{A}_{34})^{E'}\frac{\m{G}_{(\chi_{m+\ell},\ell)\{\}\{\}\{\}\{E'\}}^{(d,\chi_{m+\ell}/2+h_{ij,m+\ell},0,0,\chi_{m+\ell}/2+h_{lk,m+\ell}-1/2,0,1)}}{c_{(d,\ell)}R_{(\chi_{m+\ell},\ell)}^{(d,\chi_{m+\ell}/2+h_{ij,m+\ell},0)}R_{(\chi_{m+\ell},\ell)}^{(d,\chi_{m+\ell}/2+h_{lk,m+\ell}-1/2,0)}}\\
&\qquad=\frac{2b_s+1+a_s(1+u-v)}{\ell(\Delta-2b_s-\ell-1)(\Delta+1-d)u^{\frac{1}{2}}}\frac{\m{G}_{(\Delta,\ell)}^{(d,a_s,b_s+1/2)}}{\omega_{(\Delta,\ell)}^{(d,a_s,b_s+1/2)}}\\
&\qquad\phantom{=}-\frac{1}{\ell(\Delta-2b_s-\ell-1)(\Delta+1-d)}\bar{\eta}_1\cdot\bar{\m{D}}_{34}\frac{\m{G}_{(\Delta,\ell)}^{(d,a_s,b_s+1/2)}}{\omega_{(\Delta,\ell)}^{(d,a_s,b_s+1/2)}}\\
&\qquad\phantom{=}+\frac{(2b_s-1)(\Delta+2b_s+\ell+1-d)}{2\ell(\Delta+1-d)u^{\frac{1}{2}}}\frac{\m{G}_{(\Delta,\ell)}^{(d,a_s,b_s-1/2)}}{\omega_{(\Delta,\ell)}^{(d,a_s,b_s-1/2)}}\\
&\qquad\phantom{=}-\frac{\Delta+2b_s+\ell+1-d}{2\ell(\Delta+1-d)}\bar{\eta}_1\cdot\bar{\m{D}}_{43}\frac{\m{G}_{(\Delta,\ell)}^{(d,a_s,b_s-1/2)}}{\omega_{(\Delta,\ell)}^{(d,a_s,b_s-1/2)}},
}[Eq4ptCB-s2C1]
for the two four-point conformal blocks \eqref{Eq4ptCBG-s1} and \eqref{Eq4ptCBG-s2} respectively.  Meanwhile, upon including the proper embedding space metrics as in \eqref{EqBoot4} in the $t$-channel, we find
\eqna{
&(\bar{\eta}_1\cdot\m{A}_{34})^C\left((\m{A}_{34})^{n_v^k}\mathscr{G}_{(1|1)}^{il|m+\ell|jk}\right)_{\{aA\}\{dD\}\{bB\}\{cC\}}(\eta_1,\eta_4,\eta_2,\eta_3)\\
&\qquad=\left[(\bar{\eta}_1\cdot\m{A}_{23}\cdot\m{A}_{34})^{F'}\frac{\m{G}_{(\chi_{m+\ell},\ell)\{\}\{\}\{F'\}\{\}}^{(d,\chi_{m+\ell}/2+h_{il,m+\ell},0,0,\chi_{m+\ell}/2+h_{jk,m+\ell},1,0)}}{c_{(d,\ell)}R_{(\chi_{m+\ell},\ell)}^{(d,\chi_{m+\ell}/2+h_{il,m+\ell},0)}R_{(\chi_{m+\ell},\ell)}^{(d,\chi_{m+\ell}/2+h_{jk,m+\ell},1)}}\right]_{\eta_2\leftrightarrow\eta_4}\\
&\qquad=\left\{\left[(\bar{\eta}_1\cdot\m{A}_{34})^{F'}-\frac{1}{v}(\bar{\eta}_2\cdot\m{A}_{34})^{F'}\right]\frac{\m{G}_{(\chi_{m+\ell},\ell)\{\}\{\}\{F'\}\{\}}^{(d,\chi_{m+\ell}/2+h_{il,m+\ell},0,0,\chi_{m+\ell}/2+h_{jk,m+\ell},1,0)}}{c_{(d,\ell)}R_{(\chi_{m+\ell},\ell)}^{(d,\chi_{m+\ell}/2+h_{il,m+\ell},0)}R_{(\chi_{m+\ell},\ell)}^{(d,\chi_{m+\ell}/2+h_{jk,m+\ell},1)}}\right\}_{\eta_2\leftrightarrow\eta_4}\\
&\qquad=\left[(\ref{Eq4ptCB-s1C1})-\frac{1}{v}(\ref{Eq4ptCB-s1C2})\right]_{\substack{\eta_2\leftrightarrow\eta_4\\j\leftrightarrow l}},
}[Eq4ptCB-t1C1]
and
\eqna{
&(\bar{\eta}_1\cdot\m{A}_{34})^C\left((\m{A}_{34})^{n_v^k}\mathscr{G}_{(1|2)}^{il|m+\ell|jk}\right)_{\{aA\}\{dD\}\{bB\}\{cC\}}(\eta_1,\eta_4,\eta_2,\eta_3)\\
&\qquad=\left[(\bar{\eta}_1\cdot\m{A}_{23}\cdot\m{A}_{34})^{E'}\frac{\m{G}_{(\chi_{m+\ell},\ell)\{\}\{\}\{\}\{E'\}}^{(d,\chi_{m+\ell}/2+h_{il,m+\ell},0,0,\chi_{m+\ell}/2+h_{jk,m+\ell}-1/2,0,1)}}{c_{(d,\ell)}R_{(\chi_{m+\ell},\ell)}^{(d,\chi_{m+\ell}/2+h_{il,m+\ell},0)}R_{(\chi_{m+\ell},\ell)}^{(d,\chi_{m+\ell}/2+h_{jk,m+\ell}-1/2,0)}}\right]_{\eta_2\leftrightarrow\eta_4}\\
&\qquad=\left[(\ref{Eq4ptCB-s2C1})-\frac{1}{v}(\ref{Eq4ptCB-s2C2})\right]_{\substack{\eta_2\leftrightarrow\eta_4\\j\leftrightarrow l}},
}[Eq4ptCB-t2C1]
where the two contributions for each bootstrap equation in the $t$-channel correspond to the two possible contractions of the $s$-channel (see below).

For the second contraction $(\bar{\eta}_2\cdot\m{A}_{34})^C$, an equivalent argument implies
\eqna{
&(\bar{\eta}_2\cdot\m{A}_{34})^C(\mathscr{G}_{(1|1)}^{ij|m+\ell|lk})_{\{aA\}\{bB\}\{dD\}\{cC\}}(\eta_1,\eta_2,\eta_4,\eta_3)\\
&\qquad=(\bar{\eta}_2\cdot\m{A}_{34})^{F'}\frac{\m{G}_{(\chi_{m+\ell},\ell)\{\}\{\}\{F'\}\{\}}^{(d,\chi_{m+\ell}/2+h_{ij,m+\ell},0,0,\chi_{m+\ell}/2+h_{lk,m+\ell},1,0)}}{c_{(d,\ell)}R_{(\chi_{m+\ell},\ell)}^{(d,\chi_{m+\ell}/2+h_{ij,m+\ell},0)}R_{(\chi_{m+\ell},\ell)}^{(d,\chi_{m+\ell}/2+h_{lk,m+\ell},1)}}\\
&\qquad=-\frac{(2b_s+1)(1-u+v)+2a_s(1-u-v)}{2(\Delta+1-d)u^{\frac{1}{2}}}\frac{\m{G}_{(\Delta,\ell)}^{(d,a_s,b_s+1/2)}}{\omega_{(\Delta,\ell)}^{(d,a_s,b_s+1/2)}}\\
&\qquad\phantom{=}+\frac{1}{\Delta+1-d}\bar{\eta}_2\cdot\bar{\m{D}}_{34}\frac{\m{G}_{(\Delta,\ell)}^{(d,a_s,b_s+1/2)}}{\omega_{(\Delta,\ell)}^{(d,a_s,b_s+1/2,0)}}\\
&\qquad\phantom{=}+[4(d-2)(\Delta+1-d)-(2b_s-1)(3\Delta+2b_s+\ell+3-3d)(1-u+v)]\\
&\qquad\phantom{=}\times\frac{(\Delta-2b_s-\ell-1)}{4(\Delta+1-d)u^{\frac{1}{2}}}\frac{\m{G}_{(\Delta,\ell)}^{(d,a_s,b_s-1/2)}}{\omega_{(\Delta,\ell)}^{(d,a_s,b_s-1/2)}}\\
&\qquad\phantom{=}-\frac{(\Delta-2b_s-\ell-1)(\Delta-2b_s-\ell+1-d)}{2(\Delta+1-d)}\bar{\eta}_2\cdot\bar{\m{D}}_{43}\frac{\m{G}_{(\Delta,\ell)}^{(d,a_s,b_s-1/2)}}{\omega_{(\Delta,\ell)}^{(d,a_s,b_s-1/2)}},
}[Eq4ptCB-s1C2]
and
\eqna{
&(\bar{\eta}_2\cdot\m{A}_{34})^C(\mathscr{G}_{(1|2)}^{ij|m+\ell|lk})_{\{aA\}\{bB\}\{dD\}\{cC\}}(\eta_1,\eta_2,\eta_4,\eta_3)\\
&\qquad=(\bar{\eta}_2\cdot\m{A}_{34})^{E'}\frac{\m{G}_{(\chi_{m+\ell},\ell)\{\}\{\}\{\}\{E'\}}^{(d,\chi_{m+\ell}/2+h_{ij,m+\ell},0,0,\chi_{m+\ell}/2+h_{lk,m+\ell}-1/2,0,1)}}{c_{(d,\ell)}R_{(\chi_{m+\ell},\ell)}^{(d,\chi_{m+\ell}/2+h_{ij,m+\ell},0)}R_{(\chi_{m+\ell},\ell)}^{(d,\chi_{m+\ell}/2+h_{lk,m+\ell}-1/2,0)}}\\
&\qquad=\frac{(2b_s+1)(1-u+v)+2a_s(1-u-v)}{2\ell(\Delta-2b_s-\ell-1)(\Delta+1-d)u^{\frac{1}{2}}}\frac{\m{G}_{(\Delta,\ell)}^{(d,a_s,b_s+1/2)}}{\omega_{(\Delta,\ell)}^{(d,a_s,b_s+1/2)}}\\
&\qquad\phantom{=}-\frac{1}{\ell(\Delta-2b_s-\ell-1)(\Delta+1-d)}\bar{\eta}_2\cdot\bar{\m{D}}_{34}\frac{\m{G}_{(\Delta,\ell)}^{(d,a_s,b_s+1/2)}}{\omega_{(\Delta,\ell)}^{(d,a_s,b_s+1/2)}}\\
&\qquad\phantom{=}+\frac{(2b_s-1)(\Delta+2b_s+\ell+1-d)(1-u+v)}{4\ell(\Delta+1-d)u^{\frac{1}{2}}}\frac{\m{G}_{(\Delta,\ell)}^{(d,a_s,b_s-1/2)}}{\omega_{(\Delta,\ell)}^{(d,a_s,b_s-1/2)}}\\
&\qquad\phantom{=}-\frac{\Delta+2b_s+\ell+1-d}{2\ell(\Delta+1-d)}\bar{\eta}_2\cdot\bar{\m{D}}_{43}\frac{\m{G}_{(\Delta,\ell)}^{(d,a_s,b_s-1/2)}}{\omega_{(\Delta,\ell)}^{(d,a_s,b_s-1/2)}},
}[Eq4ptCB-s2C2]
for the $s$-channel and
\eqna{
&(\bar{\eta}_2\cdot\m{A}_{34})^C\left((\m{A}_{34})^{n_v^k}\mathscr{G}_{(1|1)}^{il|m+\ell|jk}\right)_{\{aA\}\{dD\}\{bB\}\{cC\}}(\eta_1,\eta_4,\eta_2,\eta_3)\\
&\qquad=\left[u^{\frac{1}{2}}(\bar{\eta}_4\cdot\m{A}_{23}\cdot\m{A}_{34})^{F'}\frac{\m{G}_{(\chi_{m+\ell},\ell)\{\}\{\}\{F'\}\{\}}^{(d,\chi_{m+\ell}/2+h_{il,m+\ell},0,0,\chi_{m+\ell}/2+h_{jk,m+\ell},1,0)}}{c_{(d,\ell)}R_{(\chi_{m+\ell},\ell)}^{(d,\chi_{m+\ell}/2+h_{il,m+\ell},0)}R_{(\chi_{m+\ell},\ell)}^{(d,\chi_{m+\ell}/2+h_{jk,m+\ell},1)}}\right]_{\eta_2\leftrightarrow\eta_4}\\
&\qquad=\left[-\frac{u}{v}(\bar{\eta}_2\cdot\m{A}_{34})^{F'}\frac{\m{G}_{(\chi_{m+\ell},\ell)\{\}\{\}\{F'\}\{\}}^{(d,\chi_{m+\ell}/2+h_{il,m+\ell},0,0,\chi_{m+\ell}/2+h_{jk,m+\ell},1,0)}}{c_{(d,\ell)}R_{(\chi_{m+\ell},\ell)}^{(d,\chi_{m+\ell}/2+h_{il,m+\ell},0)}R_{(\chi_{m+\ell},\ell)}^{(d,\chi_{m+\ell}/2+h_{jk,m+\ell},1)}}\right]_{\eta_2\leftrightarrow\eta_4}\\
&\qquad=\left[-\frac{u}{v}(\ref{Eq4ptCB-s1C2})\right]_{\substack{\eta_2\leftrightarrow\eta_4\\j\leftrightarrow l}},
}[Eq4ptCB-t1C2]
and
\eqna{
&(\bar{\eta}_2\cdot\m{A}_{34})^C\left((\m{A}_{34})^{n_v^k}\mathscr{G}_{(1|2)}^{il|m+\ell|jk}\right)_{\{aA\}\{dD\}\{bB\}\{cC\}}(\eta_1,\eta_4,\eta_2,\eta_3)\\
&\qquad=\left[u^{\frac{1}{2}}(\bar{\eta}_4\cdot\m{A}_{23}\cdot\m{A}_{34})^{E'}\frac{\m{G}_{(\chi_{m+\ell},\ell)\{\}\{\}\{\}\{E'\}}^{(d,\chi_{m+\ell}/2+h_{il,m+\ell},0,0,\chi_{m+\ell}/2+h_{jk,m+\ell}-1/2,0,1)}}{c_{(d,\ell)}R_{(\chi_{m+\ell},\ell)}^{(d,\chi_{m+\ell}/2+h_{il,m+\ell},0)}R_{(\chi_{m+\ell},\ell)}^{(d,\chi_{m+\ell}/2+h_{jk,m+\ell}-1/2,0)}}\right]_{\eta_2\leftrightarrow\eta_4}\\
&\qquad=\left[-\frac{u}{v}(\ref{Eq4ptCB-s2C2})\right]_{\substack{\eta_2\leftrightarrow\eta_4\\j\leftrightarrow l}},
}[Eq4ptCB-t2C2]
for the $t$-channel.

For readability, we performed all the contractions in \eqref{Eq4ptCB-s1C1}, \eqref{Eq4ptCB-s2C1}, \eqref{Eq4ptCB-t1C1}, \eqref{Eq4ptCB-t2C1}, \eqref{Eq4ptCB-s1C2}, \eqref{Eq4ptCB-s2C2}, \eqref{Eq4ptCB-t1C2} and \eqref{Eq4ptCB-t2C2} but have not expanded the action of the differential operators (see Appendix \ref{SAppProofs}).
\item\textit{Put constraints on the OPE coefficients with the help of the bootstrap equations, taking into account conserved current relations between the three-point basis OPE coefficients \cite{Fortin:2020des}.}

Assuming a diagonal metric $G^{nm}=\delta^{nm}$ and starting from \eqref{EqBoot4} contracted with $(\bar{\eta}_1\cdot\m{A}_{34})^C$, the first bootstrap equation is given by
\eqna{
&\sum_m\sum_{b=1}^2\cCF{[1|}{ijm}\cCF{[b|}{lkm}v^{\frac{1}{2}(\Delta_j+\Delta_k)}(R_{ijm})_{11}\left[(R_{lkm})_{b1}(\ref{Eq4ptCB-s1C1})+(R_{lkm})_{b2}(\ref{Eq4ptCB-s2C1})\right]\\
&\qquad=\sum_m\sum_{b=1}^2\cCF{[1|}{ilm}\cCF{[b|}{jkm}u^{\frac{1}{2}(\Delta_k+\Delta_l)}(R_{ilm})_{11}\left[(R_{jkm})_{b1}(\ref{Eq4ptCB-t1C1})+(R_{jkm})_{b2}(\ref{Eq4ptCB-t2C1})\right].
}[EqBoot4-1]
The second bootstrap equation, which results from the contraction with $(\bar{\eta}_2\cdot\m{A}_{34})^C$, is instead given by
\eqna{
&\sum_m\sum_{b=1}^2\cCF{[1|}{ijm}\cCF{[b|}{lkm}v^{\frac{1}{2}(\Delta_j+\Delta_k)}(R_{ijm})_{11}\left[(R_{lkm})_{b1}(\ref{Eq4ptCB-s1C2})+(R_{lkm})_{b2}(\ref{Eq4ptCB-s2C2})\right]\\
&\qquad=\sum_m\sum_{b=1}^2\cCF{[1|}{ilm}\cCF{[b|}{jkm}u^{\frac{1}{2}(\Delta_k+\Delta_l)}(R_{ilm})_{11}\left[(R_{jkm})_{b1}(\ref{Eq4ptCB-t1C2})+(R_{jkm})_{b2}(\ref{Eq4ptCB-t2C2})\right].
}[EqBoot4-2]
As is well-known, \eqref{EqBoot4-1} and \eqref{EqBoot4-2} can be used to put constraints on the OPE coefficients \cite{Rattazzi:2008pe}.
\end{enumerate}

%%%%%%%%%%%%%%%%%%%%%%%%%%%%%%%%%%%%%%%%%%%%%%%%%%
%%%%%%%%%%%%%%%%%%%%%%%%%%%%%%%%%%%%%%%%%%%%%%%%%%

\section{Conclusion}\label{SecConc}

In this work, we developed the conformal bootstrap approach in the context of the embedding space formalism with a modified uplift, originally introduced in \cite{Fortin:2019fvx,Fortin:2019dnq}.  Our primary intention was to generate conformal bootstrap equations expressed in terms of standard scalar conformal blocks for spin-$\ell$ exchange instead of the scalar conformal blocks for scalar exchange, which is computationally more cumbersome.  We accomplished this in Section \ref{SecCF} by expressing the three- and four-point correlation functions obtained from the embedding space OPE in terms of tensorial generalizations of the three- and four-point conformal blocks.  Using group theory, we then determined the set of available objects that can be contracted with the (naturally tensorial) bootstrap equations to generate independent scalar conformal bootstrap equations.  By deriving a complete set of contiguous relations (Section \ref{SecTensorGeneralizations}) in order to reduce the tensorial conformal blocks down to the conformal blocks for symmetric-traceless exchange, we therefore succeeded in expressing the scalar bootstrap equations in terms of standard conformal blocks for spin-$\ell$ exchange and functions of the conformal cross-ratios.  We also introduced a convenient basis of three-point conformal blocks, or tensor structures, such that the OPE coefficients have trivial symmetry properties under permutations of the three external quasi-primary operators.  In particular, in this basis the tensor structures do not mix (Section \ref{SecBootstrap}).  Finally, we laid out the complete algorithm leading to independent scalar conformal bootstrap equations directly from the embedding space formalism (\textit{i.e.}\ without projecting back to position space) in Section \ref{SecAlg}.  Overall, the algorithm reduces to the following: 
\begin{enumerate}
\item Rotation Matrices (Three-Point Correlation Functions)
\begin{enumerate}
\item Determine the infinite towers of exchanged quasi-primary operators from group theory;
\item Find the appropriate projection operators and tensor structures;
\item Use \eqref{Eq3ptCBSoln} to compute the three-point conformal blocks in terms of the three-point tensorial blocks \eqref{EqG3};
\item Re-express the three-point conformal blocks in the proper basis to determine the rotation matrix elements.
\end{enumerate}
\item Bootstrap Equations (Four-Point Correlation Functions)
\begin{enumerate}
\item With the information above, compute the four-point conformal blocks \eqref{Eq4ptCBSoln};
\item Perform some simplifications to express the results in terms of the four-point tensorial blocks \eqref{EqG4};
\item From group theory, determine the independent set of contractions leading to scalar bootstrap equations;
\item Use the contiguous relations to perform the appropriate contractions obtained previously and obtain independent scalar bootstrap equations in terms of conformal blocks for spin-$\ell$ exchange.
\end{enumerate}
\end{enumerate}
Clearly, the method described here requires as input data the projection operators and the tensor structures, which are obtained from group theory.

Compared with our previous technique and other relevant approaches, the main advantages of the present method lie in the final form of the conformal bootstrap equations, which depend only on standard spin-$\ell$ conformal blocks, as well as the three-point conformal block basis for which OPE coefficients transform trivially under permutations of the external quasi-primary operators.  A key highlight of the method is its built-in capacity to treat all Lorentz representations democratically.  This is due to the modified uplift to the embedding space that is at the heart of the approach.  It is this salient feature that ultimately allows us to write down the final form of the four-point conformal blocks given in \eqref{Eq4ptCBSoln} and \eqref{EqG4p}.  From the form of this result, it is evident that all blocks corresponding to arbitrary exchanged representations and external quasi-primary operators have the same form, up to a choice of input data, in particular the projection operators and tensor structures, as well as a handful of parameters that may be fixed from the input data, such as $\{n_v^m,i_a,i_b,\ell_t\}$.  One possible shortcoming is the form of the rotation matrices.  Although the computation of the rotation matrix is constructive and as such fairly straightforward, it does however leave us with potentially complicated rotation matrix elements.  This is reminiscent of other approaches, such as the weight-shifting method \cite{Karateev:2017jgd}, where potentially cumbersome coefficients arise for a similar reason, namely due to a change of the three-point basis.  Further, the explicit appearance of differential operators inside the contiguous relations make them more complicated in form.  Nevertheless, they do lead to bootstrap equations expressed in terms of standard spin-$\ell$-exchange blocks, which is not always straightforward to obtain from related approaches (\textit{e.g.}\ the Casimir equation method \cite{Kravchuk:2017dzd}).  Although the first-order differential operators that potentially appear in the final form of the conformal blocks may in principle be removed via some identities due to Dolan and Osborn (Appendix \ref{SAppProofs}), it is an open question whether the resulting linear combination of conformal blocks is easier to handle computationally as opposed to the direct action of the differential operators.  We hope to address some of these shortcomings and concerns in future work.

It is natural to attempt to harness the method presented here for the study of correlation functions of conserved currents or energy-momentum tensors.  Indeed, the energy-momentum tensor is the only non-trivial local quasi-primary operator that exists in all sufficiently local CFTs.  It is therefore expected that the conformal bootstrap of four energy-momentum tensors could lead to interesting results on the central charge.

%%%%%%%%%%%%%%%%%%%%%%%%%%%%%%%%%%%%%%%%%%%%%%%%%%

\ack{
The work of JFF is supported by NSERC.  VP is supported by the Perimeter Institute for Theoretical Physics.  Research at Perimeter Institute is supported in part by the Government of Canada through the Department of Innovation, Science and Economic Development Canada and by the Province of Ontario through the Ministry of Colleges and Universities.  The work of WS is supported in part by the US Department of Energy under grant DE-SC00-17660.
}

%%%%%%%%%%%%%%%%%%%%%%%%%%%%%%%%%%%%%%%%%%%%%%%%%%
%%%%%%%%%%%%%%%%%%%%%%%%%%%%%%%%%%%%%%%%%%%%%%%%%%

\setcounter{section}{0}
\renewcommand{\thesection}{\Alph{section}}

\section{Proofs}\label{SAppProofs}

In this appendix, we sketch out the proofs of the contiguous relations used in the main text.  We begin by introducing a number of identities that are then used to prove the contiguous relations found in Section~\ref{SecTensorGeneralizations}.

%%%%%%%%%%%%%%%%%%%%%%%%%%%%%%%%%%%%%%%%%%%%%%%%%%

\subsection{Identities}\label{SSAppIds}

The first identities we need are given by
\eqn{[\m{D}_{12}^A,\m{D}_{12\{F\}}^{(d,h,n)}]=2(h+n)\frac{\eta_1^A}{(\ee{1}{2})^{\frac{1}{2}}}\m{D}_{12\{F\}}^{(d,h,n)}+n\m{D}_{12\{(F\}}^{(d,h+1,n-1)}\m{A}_{12F)}^{\phantom{12F)}A},}[EqIdCom]
and
\eqna{
\m{D}_{12\{F\}}^{(d,h,n)}&=\frac{\eta_{2F}}{(\ee{1}{2})^{\frac{1}{2}}}\m{D}_{12\{F\}}^{(d,h+1,n-1)}+2(h+n)\m{D}_{12F}\m{D}_{12\{F\}}^{(d,h,n-1)}\\
&\phantom{=}\qquad-(h+n)(d+2h+2n-2)\frac{\eta_{1F}}{(\ee{1}{2})^{\frac{1}{2}}}\m{D}_{12\{F\}}^{(d,h,n-1)},
}[EqIdD]
and can be found in \cite{Fortin:2019dnq}.

We next need to compute
\eqn{
\begin{gathered}
\m{D}_{43}^{E'}[(\m{A}_{12})^\ell\h{\m{P}}_{23}^{\ell\bs{e}_1}(\m{A}_{34})^\ell]_{\{E\}\{E'\}}=(d+\ell-2)\frac{(\ee{3}{4})^{\frac{1}{2}}\eta_2^{E'}}{(\ee{2}{3})}[(\m{A}_{12})^\ell\h{\m{P}}_{23}^{\ell\bs{e}_1}(\m{A}_{34})^\ell]_{\{E\}\{E'\}},\\
\m{D}_{43}^{E'}\frac{1}{(\ee{2}{3})^{\Delta-1}}=-(\Delta-1)\frac{(\ee{3}{4})^{\frac{1}{2}}(\m{A}_{34}\cdot\eta_2)^{E'}}{(\ee{2}{3})^\Delta},
\end{gathered}
}[EqIdD43]
as well as
\eqn{
\begin{gathered}
g^{EE'}[(\m{A}_{12})^\ell\h{\m{P}}_{23}^{\ell\bs{e}_1}(\m{A}_{34})^\ell]_{\{E\}\{E'\}}=\left[\m{A}_{23}^{EE'}+\frac{\eta_3^E\eta_2^{E'}}{(\ee{2}{3})}\right][(\m{A}_{12})^\ell\h{\m{P}}_{23}^{\ell\bs{e}_1}(\m{A}_{34})^\ell]_{\{E\}\{E'\}},\\
\m{A}_{23}^{EE'}[(\m{A}_{12})^\ell\h{\m{P}}_{23}^{\ell\bs{e}_1}(\m{A}_{34})^\ell]_{\{E\}\{E'\}}=\frac{(d+\ell-3)(d+2\ell-2)}{\ell(d+2\ell-4)}[(\m{A}_{12})^{\ell-1}\h{\m{P}}_{23}^{(\ell-1)\bs{e}_1}(\m{A}_{34})^{\ell-1}]_{\{E\}\{E'\}}.
\end{gathered}
}[EqIdg]
These follow straightforwardly from their definitions.

Finally, for future convenience, we also provide the algebra of the differential operators,
\eqn{
\begin{gathered}
{}[\m{D}_{21}^A,\m{D}_{12}^B]=(\Theta_1-\Theta_2)\m{A}_{12AB}+\frac{1}{(\ee{1}{2})^{\frac{1}{2}}}(\eta_1^A\m{D}_{21}^B-\eta_2^B\m{D}_{12}^A),\\
[\m{D}_{21}^A,\m{D}_{12}^2]=\m{D}_{12}^A(2\Theta_1-2\Theta_2+2-d)+2\frac{\eta_1^A}{(\ee{1}{2})^{\frac{1}{2}}}\m{D}_{12}\cdot\m{D}_{21},\\
\m{A}_{12}^{AB}[\m{D}_{21B},\m{D}_{12\{F\}}^{(d,h,n)}]=2(h+n)\m{D}_{12}^A\m{D}_{12\{F\}}^{(d,h-1,n)}(\Theta_1-\Theta_2+h-d/2),
\end{gathered}
}[EqIdD21Com]
where the last identity is easily proven from the first two.  We require these identities, together with
\eqn{\m{D}_{21A}[(\m{A}_{12})^\ell\h{\m{P}}_{23}^{\ell\bs{e}_1}(\m{A}_{34})^\ell]_{\{E\}\{E'\}}=-\ell\frac{\eta_{2(E}}{(\ee{1}{2})^{\frac{1}{2}}}[(\m{A}_{12})^\ell\h{\m{P}}_{23}^{\ell\bs{e}_1}(\m{A}_{34})^\ell]_{\{E)A\}\{E'\}},}[EqIdD21]
in order to compute the full set of contiguous relations.

%%%%%%%%%%%%%%%%%%%%%%%%%%%%%%%%%%%%%%%%%%%%%%%%%%

\subsection{Sketches}

It is trivial to prove the contiguous relations \eqref{EqG4CRg1} directly from the definition \eqref{EqG4}.  This is however not the case for the contiguous relations \eqref{EqG4CRg2}

Let us consider the first of these relations.  Here it is clear that the metric $g^{FE'}$ implies a contraction of $\eta_2^{E'}$ with $[(\m{A}_{12})^\ell\h{\m{P}}_{23}^{\ell\bs{e}_1}(\m{A}_{34})^\ell]_{\{E\}\{E'\}}$ inside the OPE differential operators.  Since
\eqn{\eta_2^{E'}\frac{[(\m{A}_{12})^\ell\h{\m{P}}_{23}^{\ell\bs{e}_1}(\m{A}_{34})^\ell]_{\{E\}\{E'\}}}{(\ee{2}{3})^\Delta}=\frac{1}{d+\ell-\Delta-1}\frac{1}{(\ee{3}{4})^{\frac{1}{2}}}\m{D}_{43}^{E'}\frac{[(\m{A}_{12})^\ell\h{\m{P}}_{23}^{\ell\bs{e}_1}(\m{A}_{34})^\ell]_{\{E\}\{E'\}}}{(\ee{2}{3})^{\Delta-1}}}
we can then invoke \eqref{EqIdD43} to absorb the new differential operator $\m{D}_{43}^{E'}$ in the OPE differential operator with the help of the commutation relations \eqref{EqIdCom} and \eqref{EqIdD}.  This then yields the first contiguous relation of \eqref{EqG4CRg2}.  The second contiguous relation is proved analogously.  Lastly, the third relation in \eqref{EqG4CRg2} can be derived from an application of \eqref{EqIdg}.

Turning to the contiguous relations involving $\eta$'s, we find that we can readily prove the relations \eqref{EqG4CReta1} directly from the definition \eqref{EqG4}.  However, it turns out that proving the remaining relations [\eqref{EqG4CReta2}, \eqref{EqG4CReta3} and \eqref{EqG4CReta4}] is more elaborate.  For these, we proceed by applying the relation
\eqna{
&\bar{\m{D}}_{12}^X\m{G}_{(\Delta,\ell)\{F\}\{E\}\{F'\}\{E'\}}^{(d,\Lambda,n,m,\Lambda',n',m')}\\
&\qquad=(\Lambda+m/2)u^{\frac{1}{2}}\bar{\eta}_4^X\m{G}_{(\Delta,\ell)\{F\}\{E\}\{F'\}\{E'\}}^{(d,\Lambda,n,m,\Lambda',n',m')}-(\Lambda+m/2)\bar{\eta}_2^X\m{G}_{(\Delta,\ell)\{F\}\{E\}\{F'\}\{E'\}}^{(d,\Lambda,n,m,\Lambda',n',m')}\\
&\qquad\phantom{=}-\frac{1}{2(\Lambda-\Delta/2+\ell/2+n/2+1)}\frac{R_{(\Delta,\ell)}^{(d,\Lambda,n)}}{R_{(\Delta,\ell)}^{(d,\Lambda+1,n)}}\bar{\eta}_2^X\m{G}_{(\Delta,\ell)\{F\}\{E\}\{F'\}\{E'\}}^{(d,\Lambda+1,n,m,\Lambda',n',m')}\\
&\qquad\phantom{=}-(\Delta/2+\ell/2-n/2-m/2-d/2)\bar{\eta}_1^X\m{G}_{(\Delta,\ell)\{F\}\{E\}\{F'\}\{E'\}}^{(d,\Lambda,n,m,\Lambda',n',m')}\\
&\qquad\phantom{=}-(\ell-m)g^{XE}\m{G}_{(\Delta,\ell)\{F\}\{E\}\{F'\}\{E'\}}^{(d,\Lambda,n,m+1,\Lambda',n',m')}\\
&\qquad\phantom{=}+\frac{1}{2(\Lambda-\Delta/2+\ell/2+n/2+1)}\frac{R_{(\Delta,\ell)}^{(d,\Lambda,n)}}{R_{(\Delta,\ell)}^{(d,\Lambda+1/2,n+1)}}g^{XF}\m{G}_{(\Delta,\ell)\{F\}\{E\}\{F'\}\{E'\}}^{(d,\Lambda+1/2,n+1,m,\Lambda',n',m')},
}[EqD12G]
which arises from a direct application of the definition of the four-point tensorial blocks \eqref{EqG4} along with \eqref{EqIdD}.  Further, we also have the relation
\eqna{
&\bar{\m{D}}_{21}^X\m{G}_{(\Delta,\ell)\{F\}\{E\}\{F'\}\{E'\}}^{(d,\Lambda,n,m,\Lambda',n',m')}\\
&\qquad=(\Lambda'+m'/2)u^{\frac{1}{2}}(\m{A}_{12}\cdot\bar{\eta}_3)^X\m{G}_{(\Delta,\ell)\{F\}\{E\}\{F'\}\{E'\}}^{(d,\Lambda,n,m,\Lambda',n',m')}\\
&\qquad\phantom{=}-(\Lambda+m/2+\Lambda'+m'/2)u^{\frac{1}{2}}(\m{A}_{12}\cdot\bar{\eta}_4)^X\m{G}_{(\Delta,\ell)\{F\}\{E\}\{F'\}\{E'\}}^{(d,\Lambda,n,m,\Lambda',n',m')}\\
&\qquad\phantom{=}+2(h+n)(h+\Delta-d/2)\frac{R_{(\Delta,\ell)}^{(d,\Lambda,n)}}{R_{(\Delta,\ell)}^{(d,\Lambda-1,n)}}\bar{\m{D}}_{12}^X\m{G}_{(\Delta,\ell)\{F\}\{E\}\{F'\}\{E'\}}^{(d,\Lambda-1,n,m,\Lambda',n',m')}\\
&\qquad\phantom{=}+2(\Lambda+m/2-1)(h+n)(h+\Delta-d/2)\frac{R_{(\Delta,\ell)}^{(d,\Lambda,n)}}{R_{(\Delta,\ell)}^{(d,\Lambda-1,n)}}u^{\frac{1}{2}}(\m{A}_{12}\cdot\bar{\eta}_4)^X\m{G}_{(\Delta,\ell)\{F\}\{E\}\{F'\}\{E'\}}^{(d,\Lambda-1,n,m,\Lambda',n',m')}\\
&\qquad\phantom{=}+2(\ell-m)(h+n)(\Delta-n+1-d)\frac{R_{(\Delta,\ell)}^{(d,\Lambda,n)}}{R_{(\Delta,\ell)}^{(d,\Lambda-1,n)}}g^{XE}\m{G}_{(\Delta,\ell)\{F\}\{E\}\{F'\}\{E'\}}^{(d,\Lambda-1,n,m+1,\Lambda',n',m')}\\
&\qquad\phantom{=}+2(\ell-m)(h+n)(\Delta-n+1-d)\frac{R_{(\Delta,\ell)}^{(d,\Lambda,n)}}{R_{(\Delta,\ell)}^{(d,\Lambda-1,n)}}\bar{\eta}_1^X\m{G}_{(\Delta,\ell)\{F\}\{E\}\{F'\}\{E'\}}^{(d,\Lambda-1,n,m,\Lambda',n',m')}\\
&\qquad\phantom{=}-m\frac{R_{(\Delta,\ell)}^{(d,\Lambda,n)}}{R_{(\Delta,\ell)}^{(d,\Lambda-1/2,n+1)}}g^{XY}\m{G}_{(\Delta,\ell)\{FE\}\{EY\}\{F'\}\{E'\}}^{(d,\Lambda-1/2,n+1,m,\Lambda',n',m')}\\
&\qquad\phantom{=}-m\frac{R_{(\Delta,\ell)}^{(d,\Lambda,n)}}{R_{(\Delta,\ell)}^{(d,\Lambda-1/2,n+1)}}\bar{\eta}_1^X\m{G}_{(\Delta,\ell)\{FE\}\{E\}\{F'\}\{E'\}}^{(d,\Lambda-1/2,n+1,m-1,\Lambda',n',m')},
}[EqD21G]
which originates from the definition of the four-point tensorial blocks and the identities \eqref{EqIdD21Com} and \eqref{EqIdD21}.  Here, to simplify the notation, we introduced
\eqn{
\begin{gathered}
\bar{\m{D}}_{12}^A=\frac{(\ee{2}{4})^{\frac{1}{2}}}{(\ee{1}{4})^{\frac{1}{2}}}\m{D}_{12}^A,\qquad\qquad\bar{\m{D}}_{21}^A=\frac{(\ee{1}{4})^{\frac{1}{2}}}{(\ee{2}{4})^{\frac{1}{2}}}\m{D}_{21}^A,\\
h=\Lambda-\Delta/2+\ell/2-n/2.
\end{gathered}
}
We next contract \eqref{EqD12G} with $\bar{\eta}_{3X}$, which results in the contiguous relation \eqref{EqG4CReta2}, but this is a partial answer.  Contracting \eqref{EqD21G} with $\bar{\eta}_{3X}$ then leads to a recurrence relation that can be solved, proving \eqref{EqG4CReta3}, and the solution can then be used in \eqref{EqG4CReta2} to obtain a proper contiguous relation.  The same can be done for the remaining contiguous relations by contracting \eqref{EqD12G} and \eqref{EqD21G} with $\bar{\eta}_{4X}$ instead.

Let us next consider the differential operators which arise in the contiguous relations.  In this framework, the final conformal blocks (after the successive application of the contiguous relations, as needed) are, in general, linear combinations of four-point scalar blocks on which first-order differential operators act.  It is therefore possible to re-express these first-order differential operators purely in terms of differential operators involving the conformal cross-ratios $u$ and $v$.  In particular, we have the relations
\eqn{
\begin{gathered}
\bar{\eta}_3\cdot\bar{\m{D}}_{12}=\bar{\eta}_4\cdot\bar{\m{D}}_{21}=\bar{\eta}_2\cdot\bar{\m{D}}_{43}=\bar{\eta}_1\cdot\bar{\m{D}}_{34}=u^{\frac{1}{2}}(\m{F}_1^{(\Lambda,\Lambda')}+\Lambda+\Lambda')+v\m{H}_1,\\
\bar{\eta}_4\cdot\bar{\m{D}}_{12}=\bar{\eta}_1\cdot\bar{\m{D}}_{43}=\m{H}_1,\\
\bar{\eta}_3\cdot\bar{\m{D}}_{21}=\bar{\eta}_2\cdot\bar{\m{D}}_{34}=v\m{H}_1,
\end{gathered}
}
where
\eqn{
\begin{gathered}
\m{H}_1=2u^{\frac{1}{2}}\partial_u-\frac{1-u-v}{u^{\frac{1}{2}}}\partial_v,\qquad\qquad\m{F}_1^{(\Lambda,\Lambda')}=(1-u-v)\partial_u-2v\partial_v-(\Lambda+\Lambda').
\end{gathered}
}
Following \cite{Dolan:2011dv} with the normalizations \eqref{EqG3R} and
\eqn{c_{(d,\ell)}=\frac{(d/2-1)_\ell}{(d-2)_\ell},}
the action of the first-order differential operators on the four-point scalar conformal blocks are
\eqna{
&\m{H}_1\m{G}_{(\Delta,\ell)}^{(d,\Lambda,0,0,\Lambda',0,0)}=\Delta\m{G}_{(\Delta-1,\ell)}^{(d,\Lambda+1/2,0,0,\Lambda'+1/2,0,0)}\\
&\qquad+\frac{(\ell-1)(d+\ell-2)(2\Lambda+\Delta+\ell)(2\Lambda'+\Delta+\ell)}{2(d+2\ell-2)(\Delta+\ell)(\Delta+\ell-1)}\m{G}_{(\Delta,\ell+1)}^{(d,\Lambda+1/2,0,0,\Lambda'+1/2,0,0)}\\
&\qquad-\frac{\ell(d+\ell-1)(2\Lambda+\Delta-\ell+2-d)(2\Lambda'+\Delta-\ell+2-d)}{2(d+2\ell-2)(\Delta-\ell+1-d)(\Delta-\ell+2-d)}\m{G}_{(\Delta,\ell-1)}^{(d,\Lambda+1/2,0,0,\Lambda'+1/2,0,0)}\\
&\qquad-\frac{(2\Lambda+\Delta+\ell)(2\Lambda'+\Delta+\ell)(2\Lambda+\Delta-\ell+2-d)(2\Lambda'+\Delta-\ell+2-d)}{4(2\Delta-d)(2\Delta+2-d)(\Delta+\ell-1)(\Delta+\ell)(\Delta-\ell+1-d)(\Delta-\ell+2-d)}\\
&\qquad\times(\Delta-1)(\Delta-d)(\Delta-d+2)\m{G}_{(\Delta+1,\ell)}^{(d,\Lambda+1/2,0,0,\Lambda'+1/2,0,0)},\\
&\m{F}_1^{(\Lambda,\Lambda')}\m{G}_{(\Delta,\ell)}^{(d,\Lambda,0,0,\Lambda',0,0)}=\frac{(d+\ell-2)(\Delta-\ell)}{d+2\ell-2}\m{G}_{(\Delta-1,\ell+1)}^{(d,\Lambda,0,0,\Lambda',0,0)}+\frac{\ell(\Delta+\ell-2+d)}{d+2\ell-2}\m{G}_{(\Delta-1,\ell-1)}^{(d,\Lambda,0,0,\Lambda',0,0)}\\
&\qquad-\frac{(2\Lambda-\Delta-\ell)(2\Lambda'-\Delta-\ell)(2\Lambda+\Delta+\ell)(2\Lambda'+\Delta+\ell)}{4(d+2\ell-2)(2\Delta-d)(2\Delta+2-d)(\Delta+\ell-1)(\Delta+\ell)^2(\Delta+\ell+1)}\\
&\qquad\times(\Delta-1)(\Delta+\ell-d)(\Delta+2-d)(d+\ell-2)\m{G}_{(\Delta+1,\ell+1)}^{(d,\Lambda,0,0,\Lambda',0,0)}\\
&\qquad-\frac{(2\Lambda+\Delta-\ell+2-d)(2\Lambda'+\Delta-\ell+2-d)(2\Lambda-\Delta+\ell-2+d)(2\Lambda'-\Delta+\ell-2+d)}{4(d+2\ell-2)(2\Delta-d)(2\Delta+2-d)(\Delta-\ell+1-d)(\Delta-\ell+2-d)^2(\Delta-\ell+3-d)}\\
&\qquad\times\ell(\Delta-1)(\Delta-\ell+2-2d)(\Delta+2-d)\m{G}_{(\Delta+1,\ell-1)}^{(d,\Lambda,0,0,\Lambda',0,0)}\\
&\qquad-\frac{2d\Lambda\Lambda'[\Delta(\Delta-d)+\ell(d+\ell-2)+2d-4]}{(\Delta+\ell-2)(\Delta+\ell)(\Delta-\ell-d)(\Delta-\ell+2-d)}\m{G}_{(\Delta,\ell)}^{(d,\Lambda,0,0,\Lambda',0,0)}.
}
The above identities encode the action of the differential operators on standard four-point scalar conformal blocks.

Consequently, once all contractions have been performed, we may then act with the various differential operators explicitly by invoking the above results to ultimately generate linear combinations of four-point scalar conformal blocks with coefficients that are functions of $u$ and $v$.

%%%%%%%%%%%%%%%%%%%%%%%%%%%%%%%%%%%%%%%%%%%%%%%%%%
%%%%%%%%%%%%%%%%%%%%%%%%%%%%%%%%%%%%%%%%%%%%%%%%%%

\section{Rotation Matrices}\label{SAppRM}

In this appendix, we streamline the computation of the rotation matrix starting from the explicit form of the three-point conformal blocks in the OPE basis.

%%%%%%%%%%%%%%%%%%%%%%%%%%%%%%%%%%%%%%%%%%%%%%%%%%

\subsection{Change of Bases}

It is always possible to rewrite the special part of the OPE tensor structures appearing in \eqref{Eq3ptCBSoln} directly in terms of the embedding space invariant tensors \eqref{EqA123}, which are the building blocks of the three-point conformal blocks in the tensor three-point basis \eqref{EqTS3pt}.  Indeed, from their definitions \eqref{EqA123} or using \eqref{Eqepsilon}, it is easy to show that
\eqn{
\begin{gathered}
\m{A}_{12}^{AB}=\m{A}_{123}^{AB}-\frac{1}{2}(\m{A}_{12}\cdot\tilde{\eta}_3)^A(\m{A}_{12}\cdot\tilde{\eta}_3)^B,\\
\epsilon_{12}^{A_1\cdots A_d}=-\frac{d}{2}\epsilon_{123}^{[A_1\cdots A_{d-1}}(\m{A}_{12}\cdot\tilde{\eta}_3)^{A_d]},\\
\Gamma_{12}^{A_1\cdots A_n}=\Gamma_{123}^{A_1\cdots A_n}-\frac{n}{2}\Gamma_{123}^{[A_1\cdots A_{n-1}}\,\tilde{\eta}_3\cdot\Gamma_{12}\,(\m{A}_{12}\cdot\tilde{\eta}_3)^{A_n]}.
\end{gathered}
}[EqOPEtoTS]
Moreover, since
\eqn{
\begin{gathered}
\Gamma^F=\Gamma_{123}^F-\frac{1}{2}\tilde{\eta}_3\cdot\Gamma_{12}\,(\m{A}_{12}\cdot\tilde{\eta}_3)^F+\tilde{\eta}_1\cdot\Gamma\,\tilde{\eta}_2^F+\tilde{\eta}_2\cdot\Gamma\,\tilde{\eta}_1^F,\\
\m{A}_{13}^{AB}=\m{A}_{123}^{AB}-\frac{1}{2}(\m{A}_{13}\cdot\tilde{\eta}_2)^A(\m{A}_{13}\cdot\tilde{\eta}_2)^B,
\end{gathered}
}[EqGammaFA13]
the three-point conformal blocks \eqref{Eq3ptCBSoln} lead to contractions of the three-point tensorial blocks \eqref{EqG3Soln} with embedding space coordinates which can be fully performed with the help of \eqref{EqG3CR}, and further contractions with the objects appearing in \eqref{EqA123} which are transverse with respect to the three embedding space coordinates.  In other words, expanding \eqref{Eq3ptCBSoln} using \eqref{EqOPEtoTS} and \eqref{EqGammaFA13} leads to three-point conformal blocks expressed as linear combinations of terms of the form
\eqna{
&(\m{A}_{123X}^{\phantom{123X}F})^n(\tilde{\eta}_1^F)^{t_1}(\tilde{\eta}_2^F)^{t_2}(\tilde{\eta}_3^F)^{t_3}\m{G}_{(\Delta,\ell)\{E\}\{F\}}^{(d,\Lambda,n+t_1+t_2+t_3)}(\eta_1,\eta_2,\eta_3)\\
&\qquad=\rho^{(d,t_2;-\Lambda+\Delta/2-\ell/2-n/2-t_1/2-t_2/2-t_3/2)}\frac{R_{(\Delta,\ell)}^{(d,\Lambda,n+t_1+t_2+t_3)}}{R_{(\Delta-t_3,\ell)}^{(d,\Lambda+t_1/2-t_2/2,n)}}\\
&\qquad\phantom{=}\times(\m{A}_{123X}^{\phantom{123X}F})^n\m{G}_{(\Delta-t_3,\ell)\{E\}\{F\}}^{(d,\Lambda+t_1/2-t_2/2,n)}(\eta_1,\eta_2,\eta_3),
}
where the contractions with $\tilde{\eta}_1^F$, $\tilde{\eta}_2^F$ and $\tilde{\eta}_3^F$ were performed using the contiguous relations \eqref{EqG3CR}.  Determining the rotation matrix thus relies on computing
\eqna{
&(\m{A}_{123X}^{\phantom{123X}F})^n\m{G}_{(\Delta,\ell)\{E\}\{F\}}^{(d,\Lambda,n)}(\eta_1,\eta_2,\eta_3)\\
&\qquad=(\m{A}_{123X}^{\phantom{123X}F})^n\sum_{\substack{s\geq0\\(n-s)\,\text{mod}\,2=0}}\kappa_{(\Delta,\ell)}^{(d,\Lambda,n)}(s)[\h{\m{P}}_{31}^{\ell\bs{e}_1}(\m{A}_{12}\cdot\tilde{\eta}_3)^{\ell-s}(g)^s]_{\{E\}\{(F\}}(g_{FF)})^{n/2-s/2}\\
&\qquad=\sum_{\substack{s\geq0\\(n-s)\,\text{mod}\,2=0}}\kappa_{(\Delta,\ell)}^{(d,\Lambda,n)}(s)[\h{\m{P}}_{31}^{\ell\bs{e}_1}(\m{A}_{12}\cdot\tilde{\eta}_3)^{\ell-s}(\m{A}_{123})^s]_{\{E\}\{(X\}}(\m{A}_{123XX)})^{n/2-s/2},
}[EqG3SolnA123]
where
\eqna{
\kappa_{(\Delta,\ell)}^{(d,\Lambda,n)}(s)&=\frac{(-1)^{n/2-s/2}(-2)^{n/2+s/2}n!(-\ell)_s}{(n/2-s/2)!s!}(-\Lambda+\Delta/2-\ell/2-n/2+1-d/2)_s\\
&\phantom{=}\qquad\times\frac{(\Lambda+\Delta/2+\ell/2-n/2)_{n/2-s/2}(-\Lambda+\Delta/2+\ell/2-n/2)_{n/2+s/2}}{(\Delta+1-d/2)_{-n/2+s/2}(\Lambda-\ell-n+2-d)_s}\\
&\phantom{=}\qquad\times{}_3F_2\left[\left.\begin{array}{c}-s,\Lambda+\Delta/2+\ell/2-s/2,\Lambda+\Delta/2-\ell/2-n/2+1-d/2\\\Lambda-\Delta/2-\ell/2-s/2+1,\Lambda-\Delta/2+\ell/2+n/2-s+d/2\end{array}\right|1\right].
}[Eqkappa]
Due to the full transversality of the metrics $\m{A}_{123X}^{\phantom{123X}F}$, we can obtain the expression \eqref{EqG3SolnA123} for the contracted three-point tensorial blocks from \eqref{EqG3Soln} by simply removing all embedding space coordinates carrying $F$-indices.  This implies that the $s_3$ sum in \eqref{EqG3Soln} disappears, along with all the sums in $\bar{I}_{12\{F\}}^{(d,h,n;p)}$ since its only non-vanishing contribution comes from the sole term with $g_{FF}$ (see \cite{Fortin:2019dnq}).  The two remaining sums are the $s_0$ sum of \eqref{EqG3Soln} (expressed here in terms of $s$) and the $t$ sum of \eqref{EqG3Soln} [generating the hypergeometric function in \eqref{Eqkappa}].

Finally, it is useful to introduce
\eqn{\tilde{\kappa}_{(\Delta,\ell)}^{(d,\Lambda,n)}(s)=\frac{\kappa_{(\Delta,\ell)}^{(d,\Lambda,n)}(s)}{R_{(\Delta,\ell)}^{(d,\Lambda,n)}},}[Eqkappat]
to simplify the notation of the rotation matrix elements.

%%%%%%%%%%%%%%%%%%%%%%%%%%%%%%%%%%%%%%%%%%%%%%%%%%
%%%%%%%%%%%%%%%%%%%%%%%%%%%%%%%%%%%%%%%%%%%%%%%%%%

\section{Projection Operators}\label{SAppProj}

In this appendix we give two hatted projection operators relevant for the example of Appendix \ref{SAppEx}, \textit{i.e.}\ $\Vev{SSVV}$ and $\Vev{SVSV}$.

%%%%%%%%%%%%%%%%%%%%%%%%%%%%%%%%%%%%%%%%%%%%%%%%%%

\subsection{\texorpdfstring{$\ell\bs{e}_1$}{le1}}

The first hatted projection operator of interest is the symmetric-traceless projector onto the irreducible representation $\ell\bs{e}_1$.  Its explicit form is well-known and is given by
\eqn{(\h{\m{P}}^{\ell\bs{e}_1})_{\mu^\ell}^{\phantom{\mu^\ell}\mu'^\ell}=\sum_{i=0}^{\lfloor\ell/2\rfloor}a_i(d,\ell)g_{(\mu_1\mu_2}g^{(\mu'_1\mu'_2}\cdots g_{\mu_{2i-1}\mu_{2i}}g^{\mu'_{2i-1}\mu'_{2i}}\delta_{\mu_{2i+1}}^{\phantom{\mu_{2i+1}}\mu'_{2i+1}}\cdots\delta_{\mu_\ell)}^{\phantom{\mu_\ell)}\mu'_\ell)},}[EqPle1]
where the coefficients are
\eqn{a_i(d,\ell)=\frac{(-\ell)_{2i}}{2^{2i}i!(-\ell+2-d/2)_i}.}
We note that technically, the explicit form \eqref{EqPle1} of the hatted projection operator is not necessary for determining the conformal bootstrap equations, as they are ultimately all expressed in terms of the usual spin-$\ell$ conformal blocks.

%%%%%%%%%%%%%%%%%%%%%%%%%%%%%%%%%%%%%%%%%%%%%%%%%%

\subsection{\texorpdfstring{$\bs{e}_m+\ell\bs{e}_1$}{em+le1}}

The second hatted projection operator relevant for the example of the next section is given by
\eqna{
&(\h{\m{P}}^{\bs{e}_m+\ell\bs{e}_1})_{\nu_m\cdots\nu_1\mu^\ell}^{\phantom{\nu_m\cdots\nu_1\mu^\ell}\mu'^\ell\nu_1'\cdots\nu_m'}\\
&\qquad=\Sym_{\{\mu\},\{\mu'\}}\Asym_{\{\nu\},\{\nu'\}}\left[\frac{m}{\ell+m}(\h{\m{P}}^{\bs{e}_m})_{\nu_m\cdots\nu_1}^{\phantom{\nu_m\cdots\nu_1}\nu'_1\cdots\nu'_m}(\h{\m{P}}^{\ell\bs{e}_1})_{\mu^\ell}^{\phantom{\mu^\ell}\mu'^\ell}\right.\\
&\qquad\phantom{=}+\frac{m\ell}{\ell+m}(\h{\m{P}}^{\bs{e}_m})_{\nu_m\cdots\nu_2\mu}^{\phantom{\nu_m\cdots\nu_2\mu}\nu'_1\cdots\nu'_m}(\h{\m{P}}^{\ell\bs{e}_1})_{\mu^{\ell-1}\nu_1}^{\phantom{\mu^{\ell-1})\nu_1]}\mu'^\ell}\\
&\qquad\phantom{=}-\frac{m^2\ell\,k(d,\ell,m)\,g_{\nu_1\lambda'}g^{\mu'\lambda}}{(\ell+m)(d+\ell-m)}(\h{\m{P}}^{\bs{e}_m})_{\nu_m\cdots\nu_2\lambda}^{\phantom{\nu_m\cdots\nu_2\lambda}\nu'_1\cdots\nu'_m}(\h{\m{P}}^{\ell\bs{e}_1})_{\mu^\ell}^{\phantom{\mu^\ell}\lambda'\mu'^{\ell-1}}\\
&\qquad\phantom{=}-\frac{m\ell^2\,k(d,\ell,m)\,g_{\mu\lambda'}g^{\mu'\lambda}}{(\ell+m)(d+\ell-m)}(\h{\m{P}}^{\bs{e}_m})_{\nu_m\cdots\nu_2\lambda}^{\phantom{\nu_m\cdots\nu_2\lambda}\nu'_1\cdots\nu'_m}(\h{\m{P}}^{\ell\bs{e}_1})_{\mu^{\ell-1}\nu_1}^{\phantom{\mu^{\ell-1}\nu_1}\lambda'\mu'^{\ell-1}}\\
&\qquad\phantom{=}+\frac{m^2\ell(\ell-1)\,g_{\nu_1\lambda'}g^{\mu'\mu'}}{(\ell+m)(d+2\ell-2)(d+\ell-m-2)}(\h{\m{P}}^{\bs{e}_{m-1}})_{\nu_m\cdots\nu_2}^{\phantom{\nu_m\cdots\nu_2}\nu'_2\cdots\nu'_m}(\h{\m{P}}^{\ell\bs{e}_1})_{\mu^\ell}^{\phantom{\mu^\ell}\nu'_1\lambda'\mu'^{\ell-2}}\\
&\qquad\phantom{=}+\frac{m\ell^2(\ell-1)\,g_{\mu\lambda'}g^{\mu'\mu'}}{(\ell+m)(d+2\ell-2)(d+\ell-m-2)}(\h{\m{P}}^{\bs{e}_{m-1}})_{\nu_m\cdots\nu_2}^{\phantom{\nu_m\cdots\nu_2}\nu'_2\cdots\nu'_m}(\h{\m{P}}^{\ell\bs{e}_1})_{\mu^{\ell-1}\nu_1}^{\phantom{\mu^{\ell-1}\nu_1}\nu'_1\lambda'\mu'^{\ell-2}}\\
&\qquad\phantom{=}+\frac{m(m-1)\ell^2(\ell-1)\,g_{\nu_1\lambda'}g^{\mu'\mu'}}{(\ell+m)(d+2\ell-2)(d+\ell-m-2)}(\h{\m{P}}^{\bs{e}_{m-1}})_{\nu_m\cdots\nu_3\mu}^{\phantom{\nu_m\cdots\nu_3\mu}\nu'_2\cdots\nu'_m}(\h{\m{P}}^{\ell\bs{e}_1})_{\mu^{\ell-1}\nu_2}^{\phantom{\mu^{\ell-1}\nu_2}\nu'_1\lambda'\mu'^{\ell-2}}\\
&\qquad\phantom{=}-\frac{2m^2(m-1)\ell(\ell-1)\,g_{\nu_1\lambda'}g^{\mu'\nu'_2}}{(\ell+m)(d+2\ell-2)(d+\ell-m)(d+\ell-m-2)}(\h{\m{P}}^{\bs{e}_{m-1}})_{\nu_m\cdots\nu_2}^{\phantom{\nu_m\cdots\nu_2}\mu'\nu'_3\cdots\nu'_m}(\h{\m{P}}^{\ell\bs{e}_1})_{\mu^\ell}^{\phantom{\mu^\ell}\nu'_1\lambda'\mu'^{\ell-2}}\\
&\qquad\phantom{=}-\frac{2m(m-1)\ell^2(\ell-1)\,g_{\mu\lambda'}g^{\mu'\nu'_2}}{(\ell+m)(d+2\ell-2)(d+\ell-m)(d+\ell-m-2)}(\h{\m{P}}^{\bs{e}_{m-1}})_{\nu_m\cdots\nu_2}^{\phantom{\nu_m\cdots\nu_2}\mu'\nu'_3\cdots\nu'_m}(\h{\m{P}}^{\ell\bs{e}_1})_{\mu^{\ell-1}\nu_1}^{\phantom{\mu^{\ell-1}\nu_1}\nu'_1\lambda'\mu'^{\ell-2}}\\
&\qquad\phantom{=}-\frac{2m(m-1)\ell^2(\ell-1)\,g_{\nu_1\lambda'}g^{\mu'\lambda}\delta_\mu^{\mu'}}{(\ell+m)(d+2\ell-2)(d+\ell-m)(d+\ell-m-2)}(\h{\m{P}}^{\bs{e}_{m-1}})_{\nu_m\cdots\nu_3\lambda}^{\phantom{\nu_m\cdots\nu_3\lambda}\nu'_2\cdots\nu'_m}(\h{\m{P}}^{\ell\bs{e}_1})_{\mu^{\ell-1}\nu_2}^{\phantom{\mu^{\ell-1}\nu_2}\nu'_1\lambda'\mu'^{\ell-2}}\\
&\qquad\phantom{=}-\frac{m(m-1)\ell^2\,k(d,\ell,m)\,g_{\nu_1\lambda'}g^{\mu'\nu'_1}}{(\ell+m)(d+\ell-m)}(\h{\m{P}}^{\bs{e}_{m-1}})_{\nu_m\cdots\nu_3\mu}^{\phantom{\nu_m\cdots\nu_3\mu}\nu'_2\cdots\nu'_m}(\h{\m{P}}^{\ell\bs{e}_1})_{\mu^{\ell-1}\nu_2}^{\phantom{\mu^{\ell-1}\nu_2}\lambda'\mu'^{\ell-1}}\\
&\qquad\phantom{=}\left.+\frac{2m(m-1)(m-2)\ell^2(\ell-1)\,g_{\nu_1\lambda'}g^{\mu'\nu'_2}\delta_\mu^{\nu'_3}}{(\ell+m)(d+2\ell-2)(d+\ell-m)(d+\ell-m-2)}(\h{\m{P}}^{\bs{e}_{m-2}})_{\nu_m\cdots\nu_3}^{\phantom{\nu_m\cdots\nu_3}\mu'\nu'_4\cdots\nu'_m}(\h{\m{P}}^{\ell\bs{e}_1})_{\mu^{\ell-1}\nu_2}^{\phantom{\mu^{\ell-1}\nu_2}\nu'_1\lambda'\mu'^{\ell-2}}\right],
}[EqPemle1]
where
\eqn{k(d,\ell,m)=1+\frac{2(\ell-1)}{(d+2\ell-2)(d+\ell-m-2)}.}

The irreducible representation $\bs{e}_m+\ell\bs{e}_1$ with $m=2$ appears in the correlation function $\Vev{SVSV}$.  The case $m=2$ was found in \cite{Rejon-Barrera:2015bpa} in another form reminiscent of the shifted projection operators of \cite{Fortin:2020ncr}.  Here it is expressed in terms of the standard projection operator \eqref{EqPle1}, as is required for generating (relatively) simple conformal bootstrap equations.

%%%%%%%%%%%%%%%%%%%%%%%%%%%%%%%%%%%%%%%%%%%%%%%%%%
%%%%%%%%%%%%%%%%%%%%%%%%%%%%%%%%%%%%%%%%%%%%%%%%%%

\section{Example}\label{SAppEx}

This appendix provides the input data necessary to write down the conformal bootstrap equations for $\Vev{SSVV}$.  Since the resulting bootstrap equations (rotation matrices and conformal blocks) are lengthy, we do not write them down explicitly.

%%%%%%%%%%%%%%%%%%%%%%%%%%%%%%%%%%%%%%%%%%%%%%%%%%

\subsection{s-channel}

In the $s$-channel $\Vev{SSVV}$ the two three-point correlation functions of interest are $\Vev{SS\m{O}_m}$ and $\Vev{VV\m{O}_m}$, respectively.

Since $\Vev{SS\m{O}_m}$ is non-trivial only for $\m{O}_m\in\{\m{O}^{\ell\bs{e}_1}\}$, the unique infinite tower of exchanged quasi-primary operators consists of $\m{O}^{\ell\bs{e}_1}$.  The relevant projection operators are given by \eqref{EqPle1} with $n_v^m=0$ and $\xi_m=0$.

The sole tensor structure for $\Vev{SS\m{O}^{\ell\bs{e}_1}}$ and its corresponding rotation matrix were already given in the main text (see Section \ref{SecAlg}).  Since
\eqn{\bs{e}_1\otimes\bs{e}_1\otimes\ell\bs{e}_1=(\ell+2)\bs{e}_1\oplus\ell\bs{e}_1\oplus\ell\bs{e}_1\oplus\ell\bs{e}_1\oplus(\ell-2)\bs{e}_1\oplus\cdots,}
where the remaining irreducible representations are not symmetric-traceless, following Table \ref{TabBoot} we find that there are $5$ tensor structures for $\Vev{VV\m{O}^{\ell\bs{e}_1}}$.  These are given explicitly by
\eqna{
&(\tCF{(1|}{ij,m+\ell}{12})_{AB\{E\}}^{\phantom{AB\{E\}}\{F\}}=\m{A}_{12A}^{\phantom{12A}F}\m{A}_{12B}^{\phantom{12B}F}(\m{A}_{12E}^{\phantom{12E}F})^\ell\\
&\qquad\to\tCF{(1|}{ijm}{12}=\m{A}_{12A}^{\phantom{12A}F}\m{A}_{12B}^{\phantom{12B}F}\quad\text{with}\quad i_1=0, n_1=\ell+2,\\
&(\tCF{(2|}{ij,m+\ell}{12})_{AB\{E\}}^{\phantom{AB\{E\}}\{F\}}=\m{A}_{12A}^{\phantom{12A}F}\m{A}_{12BE}(\m{A}_{12E}^{\phantom{12E}F})^{\ell-1}\\
&\qquad\to\tCF{(2|}{ij,m+1}{12}=\m{A}_{12A}^{\phantom{12A}F}\m{A}_{12BE}\quad\text{with}\quad i_2=1, n_2=\ell,\\
&(\tCF{(3|}{ij,m+\ell}{12})_{AB\{E\}}^{\phantom{AB\{E\}}\{F\}}=\m{A}_{12AE}\m{A}_{12B}^{\phantom{12B}F}(\m{A}_{12E}^{\phantom{12E}F})^{\ell-1}\\
&\qquad\to\tCF{(3|}{ij,m+1}{12}=\m{A}_{12AE}\m{A}_{12B}^{\phantom{12B}F}\quad\text{with}\quad i_3=1, n_3=\ell,\\
&(\tCF{(4|}{ij,m+\ell}{12})_{AB\{E\}}^{\phantom{AB\{E\}}\{F\}}=\m{A}_{12AB}(\m{A}_{12E}^{\phantom{12E}F})^\ell\\
&\qquad\to\tCF{(4|}{ijm}{12}=\m{A}_{12AB}\quad\text{with}\quad i_4=0, n_4=\ell,\\
&(\tCF{(5|}{ij,m+\ell}{12})_{AB\{E\}}^{\phantom{AB\{E\}}\{F\}}=\m{A}_{12AE}\m{A}_{12BE}(\m{A}_{12E}^{\phantom{12E}F})^{\ell-2}\\
&\qquad\to\tCF{(5|}{ij,m+2}{12}=\m{A}_{12AE}\m{A}_{12BE}\quad\text{with}\quad i_5=2, n_5=\ell-2.
}
The $\Vev{VV\m{O}^{\ell\bs{e}_1}}$ rotation matrix is thus a $5\times5$ matrix and there are $1\times5=5$ different four-point conformal blocks in the $s$-channel.

%%%%%%%%%%%%%%%%%%%%%%%%%%%%%%%%%%%%%%%%%%%%%%%%%%

\subsection{t-channel}

In the $t$-channel $\Vev{SVSV}$, we are interested in the possible irreducible representations of the exchanged quasi-primary operators of the same type of three-point correlation functions, namely two copies of $\Vev{SV\m{O}_m}$, for which $\m{O}_m\in\{\m{O}^{\ell\bs{e}_1},\m{O}^{\bs{e}_2+\ell\bs{e}_1}\}$.  Hence there are two distinct infinite towers of exchanged quasi-primary operators, with the projection operators for $\m{O}^{\ell\bs{e}_1}$ given in \eqref{EqPle1} ($n_v^m=0$ and $\xi_m$) and the projection operators for $\m{O}^{\bs{e}_2+\ell\bs{e}_1}$ given in \eqref{EqPemle1} (with $m=2$) ($n_v^m=2$ and $\xi_m=0$).

The relevant tensor structures and rotation matrix for $\Vev{SV\m{O}^{\ell\bs{e}_1}}$ are explicitly given in the main text (see Section \ref{SecAlg}).  In the case of $\Vev{SV\m{O}^{\bs{e}_2+\ell\bs{e}_1}}$, we see that Table \ref{TabBoot} states that there is only one tensor structure,
\eqn{\bs{0}\otimes\bs{e}_1\otimes(\bs{e}_2+\ell\bs{e}_1)=(\bs{e}_3+\ell\bs{e}_1)\oplus[2\bs{e}_2+(\ell-1)\bs{e}_1]\oplus[\bs{e}_2+(\ell+1)\bs{e}_1]\oplus[\bs{e}_2+(\ell-1)\bs{e}_1]\oplus(\ell+1)\bs{e}_1.}
In the basis of interest, the tensor structure is given by
\eqn{(\tCF{(1|}{ij,m+\ell}{12})_{B\{E\}}^{\phantom{B\{E\}}\{F\}}=\m{A}_{12BE_1}\m{A}_{12E_2}^{\phantom{12E_2}F}(\m{A}_{12E}^{\phantom{12E}F})^\ell\to\tCF{(1|}{ijm}{12}=\m{A}_{12BE_1}\m{A}_{12E_2}^{\phantom{12E_2}F}\quad\text{with}\quad i_1=0, n_1=\ell+1.}
where $E_1$ and $E_2$ are the two antisymmetric indices of the exchanged quasi-primary operator.

It follows that there are $2\times2+1\times1=5$ different four-point conformal blocks in the $t$-channel.

%%%%%%%%%%%%%%%%%%%%%%%%%%%%%%%%%%%%%%%%%%%%%%%%%%

\subsection{Bootstrap Equations}

Finally, to determine the fully-scalar conformal bootstrap equations, we refer to Table \ref{TabBoot} and note that
\eqn{\bs{0}\otimes\bs{e}_1\otimes\bs{0}\otimes\bs{e}_1=\bs{e}_2\oplus2\bs{e}_1\oplus\bs{0},}
which implies that there are $1+3+1=5$ independent contractions to be performed on \eqref{EqBoot4}.  A good choice of independent contractions is
\eqn{
\begin{gathered}
\m{A}_{34}^{CD},\qquad(\bar{\eta}_1\cdot\m{A}_{34})^C(\bar{\eta}_1\cdot\m{A}_{34})^D,\qquad(\bar{\eta}_2\cdot\m{A}_{34})^C(\bar{\eta}_2\cdot\m{A}_{34})^D,\\
(\bar{\eta}_1\cdot\m{A}_{34})^C(\bar{\eta}_2\cdot\m{A}_{34})^D,\qquad(\bar{\eta}_2\cdot\m{A}_{34})^C(\bar{\eta}_1\cdot\m{A}_{34})^D,
\end{gathered}
}
where the three contractions on the first line correspond to $\bs{0}$ and two of the three $2\bs{e}_1$, while standard symmetric and antisymmetric linear combinations of the two contractions on the second line correspond to the last $2\bs{e}_1$ (symmetric) and the $\bs{e}_2$ (antisymmetric), respectively.

%%%%%%%%%%%%%%%%%%%%%%%%%%%%%%%%%%%%%%%%%%%%%%%%%%
%%%%%%%%%%%%%%%%%%%%%%%%%%%%%%%%%%%%%%%%%%%%%%%%%%

\bibliography{Bootstrap}

\end{document}